\documentclass[revtex4]{emulateapj}
\usepackage{natbib}
\usepackage{lscape}
\usepackage{pdflscape,textcomp}
\usepackage{apjfonts}
\usepackage{color}
\usepackage{graphicx}
\usepackage{amsmath,amsthm,amssymb}
\usepackage{epsfig}
\usepackage{subfigure}
\usepackage{mathrsfs}
\usepackage{url}
\newcommand{\Msun}{\mbox{$M_{\odot}$}}
\newcommand{\Mjup}{\mbox{$M_{Jup}$}}
\newcommand{\gps}{\ensuremath{g_{\rm P1}}}
\newcommand{\rps}{\ensuremath{r_{\rm P1}}}
\newcommand{\ips}{\ensuremath{i_{\rm P1}}}
\newcommand{\zps}{\ensuremath{z_{\rm P1}}}
\newcommand{\yps}{\ensuremath{y_{\rm P1}}}

\slugcomment{To appear in ApJ.}

\begin{document}
\title{Brown Dwarfs in Young Moving Groups from Pan-STARRS1. I. AB Doradus}
\author{Kimberly M. Aller\altaffilmark{1,4}, Michael
  C. Liu\altaffilmark{1}, Eugene A. Magnier\altaffilmark{1}, William
  M. J. Best\altaffilmark{1}, Michael C. Kotson\altaffilmark{1,2},
  William S. Burgett\altaffilmark{1}, Kenneth
  C. Chambers\altaffilmark{1}, Klaus W. Hodapp\altaffilmark{1},
  Heather Flewelling\altaffilmark{1}, Nick Kaiser\altaffilmark{1},
  Nigel Metcalf\altaffilmark{3}, John L. Tonry\altaffilmark{1},
  Richard J. Wainscoat\altaffilmark{1}, Christopher
  Waters\altaffilmark{1}}

\affil{\altaffilmark{1} University of Hawaii, Institute of Astronomy, 2860
  Woodlawn Drive, Honolulu, HI 96822}
\affil{\altaffilmark{2} Lincoln Laboratory, Massachusetts Institute of Technology}
\affil{\altaffilmark{3} Department of Physics, Durham University, South Road,
  Durham DH1 3LE, UK}
\altaffiltext{4}{Visiting Astronomer at the Infrared Telescope
  Facility, which is operated by the University of Hawaii under
  Cooperative Agreement no. NNX-08AE38A with the National Aeronautics
  and Space Administration, Science Mission Directorate, Planetary
  Astronomy Program.}


\begin{abstract}
Substellar members of young ($\lesssim$150~Myr) moving groups are
valuable benchmarks to empirically define brown dwarf evolution with
age and to study the low-mass end of the initial mass function. We
have combined Pan-STARRS1 (PS1) proper motions with optical--IR
photometry from PS1, 2MASS and {\it WISE} to search for substellar
members of the AB~Dor Moving Group within $\approx$50~pc and with
spectral types of late-M to early-L, corresponding to masses down to
$\approx$30~$\Mjup$ at the age of the group ($\approx$125~Myr). Including
both photometry and proper motions allows us to better select
candidates by excluding field dwarfs whose colors are similar to young
AB~Dor Moving Group members. Our near-IR spectroscopy has identified
six ultracool dwarfs (M6--L4; $\approx$30--100~$\Mjup$) with
intermediate surface gravities ({\sc int-g}) as candidate members of
the AB~Dor Moving Group. We find another two candidate members with
spectra showing hints of youth but consistent with field gravities. We
also find four field brown dwarfs unassociated with the AB~Dor Moving
Group, three of which have {\sc int-g} gravity classification. While
signatures of youth are present in the spectra of our $\approx$125~Myr
objects, neither their $J$--$K$ nor $W1$--$W2$ colors are
significantly redder than field dwarfs with the same spectral types,
unlike younger ultracool dwarfs. We also determined PS1 parallaxes for
eight of our candidates and one previously identified AB~Dor Moving
Group candidate. Although radial velocities (and parallaxes, for some)
are still needed to fully assess membership, these new objects provide
valuable insight into the spectral characteristics and evolution of
young brown dwarfs.

\end{abstract}


\section{Introduction}

Young moving groups (YMGs) are coeval associations of stars with
similar space motions and with ages ranging from $\sim$10--100~Myr. It
is believed that these groups have left their natal molecular cloud
after formation and dispersed into the field
\citep{2004ARA&A..42..685Z}. As such, YMGs link stars in molecular
clouds ($\sim$1~Myr) to field stars ($\gtrsim$1~Gyr) no longer
affiliated with their birthsites. Thus YMGs are valuable laboratories
for studying recent star formation in the solar neighborhood. Because
of the proximity of the known YMGs ($\lesssim$100~pc), they are ideal
candidates for characterizing the initial mass function (IMF) down to
substellar masses. Although substellar objects are generally very
faint, younger brown dwarfs are more luminous
\citep{2000ApJ...542..464C} thus more readily detected.

Characterization of young brown dwarfs and directly imaged planets
have revealed that their spectral properties differ from those of
their old field counterparts
\citep[e.g.][]{2005A&A...438L..25C,2008Sci...322.1348M,2010ApJ...723..850B,2013ApJ...774...55B,2010A&A...517A..76P}. Young
brown dwarfs have redder NIR colors and have spectra distinct from
field objects. Studies of brown dwarfs in young clusters and moving
groups have begun to delineate the brown dwarf spectral evolution, due
to the lower surface gravity of younger objects
\citep[e.g.][]{2007ApJ...657..511A,2013ApJ...772...79A}. In order to
further characterize this evolution, we need to identify a larger
sample of substellar objects at various young ages
($\sim$10--100~Myr). Determining the substellar spectral sequence in
YMGs with different ages would be a key step towards better
understanding substellar evolution and benchmarking spectral
indicators of youth.

The AB~Doradus (AB~Dor) Moving Group was first recognized as a sparse,
comoving group of stars in the Local Association by
\citet{2004ApJ...613L..65Z}. The age estimates for the AB~Dor Moving
Group vary substantially depending on the method, ranging from
50 to 150~Myr. Initially, \citet{2004ApJ...613L..65Z} estimated an age
of $\sim$50~Myr from color-magnitude diagrams. Analysis using
evolutionary tracks and dynamical masses to study AB~Dor~C, a member
of a quadruple system in the AB~Dor Moving Group, yielded an age for
the system of $\sim$75~Myr \citep{2007ApJ...665..736C}. However,
color-magnitude diagram comparisons of the lower-mass AB~Dor Moving
Group members with the Pleiades \citep[$\approx$125~Myr;
  e.g.][]{1996ApJ...458..600B,1998ApJ...499L..61M} and IC~2391
\citep[$\sim$35--50~Myr;
  e.g.][]{1999ApJ...522L..53B,2004ApJ...614..386B} suggested that the
AB~Dor Moving Group is roughly coeval with the Pleiades and older than
IC~2391 \citep{2005ApJ...628L..69L}. Traceback of the AB~Dor Moving
Group kinematics have also concluded that the group and the Pleiades
likely formed from the same large-scale star formation event, and thus
should be nearly same age
\citep{2005ApJ...628L..69L,2007MNRAS.377..441O}. By combining chemical
and kinematic analysis of the AB~Dor Moving Group members,
\citet{2013ApJ...766....6B} have also constrained the group to be
approximately the age of the Pleiades with a lower age limit of 110~Myr.

Currently, the AB~Dor Moving Group has one of the largest number of
stellar members of the known YMGs, with $\approx$50 confirmed members
with parallaxes
\citep{2004ARA&A..42..685Z,2008hsf2.book..757T,2011ApJ...732...61Z}. However,
the lack of low-mass stars ($\lesssim$0.5~$\Msun$) in the known
membership has prompted several recent surveys aimed at discovering
these missing members. By using photometry and/or kinematics, several
additional low mass stellar candidate members (late-K--mid-M dwarfs;
$\sim$0.1--0.5~$\Msun$) have emerged
\citep[e.g.][]{2009ApJ...699..649S,2012AJ....143...80S,2013ApJ...762...88M}. However,
these surveys were less sensitive to the cooler, fainter substellar
members. Currently, only CD-35~2722~B \citep[L3;
][]{2011ApJ...729..139W}, 2MASS~J1425$-$3650
\citep[L4;][]{2015ApJ...798...73G}, 2MASS~J0355+1133
\citep[L5;][]{2013AN....334...85L,2013AJ....145....2F},
WISEP~J0047+6803 \citep[L7;][]{2015ApJ...799..203G} and
SDSS~J1110+0116 \citep[T5.5;][]{2015ApJ...808L..20G} have been
confirmed as bona fide substellar members of the AB~Dor Moving
Group. \citet{2014ApJ...783..121G,2015ApJ...798...73G} have begun a
systematic search to identify lower mass stellar and substellar
candidate members using Bayesian inference to calculate their YMG
membership probabilities from proper motion, photometry, and, if
available, radial velocities and distances.

In order to further the search for substellar YMG members, we are
conducting a deep, wide-field search based on optical imaging data
from Pan-STARRS1. Pan-STARRS1 (PS1) is a multi-wavelength,
multi-epoch, optical imaging survey which covers $\approx$75\% of the
sky. PS1 goes $\sim$1~mag fainter than SDSS in the $z$ band
\citep{2000AJ....120.1579Y}. Also, PS1 has a novel
\yps~(0.918--1.001~$\mu$m) filter, which extends the wavelength
coverage further into the near-infrared than past optical surveys,
such as SDSS. Compared to previous optical surveys, the PS1 red
optical filters, \zps\, and \yps, allow for more sensitivity and
better characterization at redder wavelengths, both of which are
advantageous for identifying substellar objects. We also combine PS1
and 2MASS astrometry to compute proper motions for our search (the
addition of 2MASS astrometry increases our time baseline by a factor
of $\sim$3). Precise proper motions significantly increases our
ability to distinguish faint, substellar candidate AB~Dor Moving Group
members from field interlopers.

We use 2MASS, PS1, and {\it WISE} to select substellar AB~Dor Moving
Group candidates. In Section~\ref{sec:data} we discuss the PS1
photometry and proper motion precision as well as the addition of
2MASS and {\it WISE} data. In Section~\ref{sec:searchmethod} we
describe our search method, which uses photometrically determined
spectral types and proper motion analysis to select candidates. In
Section~\ref{sec:observations} we describe our spectroscopic followup
and reduction. In Section~\ref{sec:results} we determine spectral
types, determine parallactic and photometric distances, assess the
youth of our candidates, consider their membership in the AB~Dor
Moving Group, and estimate their physical properties. Our discussion
is in Section~\ref{sec:discussion}, and our conclusions are in
Section~\ref{sec:conclusions}.


\section{Survey Data} \label{sec:data}

PS1 is a 1.8~m, wide-field telescope located on Haleakal\={a} on the
island of Maui. The PS1 3$\pi$~Survey covers the sky north of
$-$30$^{\circ}$ decl., $\approx$75$\%$ of the sky. The survey's
five optical filters, \gps, \rps, \ips, \zps, and \yps, are described
in \citet{2010ApJS..191..376S} and \citet{2012ApJ...750...99T}. At
each epoch a single field is exposed for 60~s in \gps, 38~s in \rps,
and 30~s in \ips, \zps, and \yps. The photometry and astrometry from
each epoch have been combined to obtain average magnitudes and proper
motions.

We used data from the PS1 3$\pi$~survey (which began in 2010) to
construct spectral energy distributions (SEDs) and determine proper
motions of our candidates. We chose good quality data according to the
photometric quality flags set in the PS1 Desktop Virtual Observatory
(DVO) database \citep{2006amos.confE..50M}. Specifically, we select
objects characterized by all of the following attributes: fits a
point-spread function (PSF) model (not extended), is not saturated,
has good sky measurement, is not likely a cosmic ray, a diffraction
spike, a ghost or a glint, and does not lie between the image chips
(i.e. to choose \yps~ data we require (\yps:flags \& 0x0000.0300)!=0
and (\yps:flags \& 0x0000.1000)=0); and has the quality flag
psf\_qf~$\ge$~0.9 to ensure that at least 90\% of the object is
unmasked. Furthermore, we require objects to be detected at least
twice in a single night in at least one of the five filters (\gps,
\rps, \ips, \zps, \yps) to remove potential spurious sources that
would only appear as single detections. Finally, we require that a
single bandpass measurement error be $\le$~0.1~mag in order to use
that bandpass.

We then matched PS1 objects with their 2MASS counterparts using the
2MASS Point Source Catalog \citep{2006AJ....131.1163S}. We chose good
$J$, $H$, and $K$ photometry based on the following requirements:
measurement error $\le$~0.2~mag and cc\_flg~=~0 (i.e. no
confusion). We also matched PS1 objects with their {\it WISE}
counterparts from the {\it WISE} Point Source Catalog
\citep{2010AJ....140.1868W}, which was available when we began our
search. We chose objects with good photometry in $W1$ and $W2$ with
measurement errors $\le$~0.2~mag and cc\_flg~=~0 (i.e. no
confusion). The final {\it WISE} photometry for our objects presented
here is from the updated AllWISE Catalog \citep{2014ApJ...783..122K},
which was released after our initial search.

We also used astrometric data from PS1 and 2MASS. The PS1 proper
motions have a 2--3~year time baseline. To increase the time baseline
from 2--3~years to $\sim$10~years, we combined astrometry from PS1 and
2MASS. For our dataset, the typical 2MASS astrometric uncertainties
are $\approx$70~mas, far larger than our typical PS1 uncertainties of
$\approx$15~mas. By extending our time-baseline, we improved our
typical proper motion precision from $\approx$7~mas~yr$^{-1}$ to
$\approx$5~mas~yr$^{-1}$.

When calculating the proper motion for our objects, we used an
outlier-resistant fitting method which calculates bisquare weights
(Tukey's biweight) and iteratively fits the data to minimize the
residuals. This method reduces the effects that potentially spurious
data points have on our final proper motion fit. We determined our
proper motion uncertainties by using bootstrap resampling of our
data. We discuss the quality of our PS1+2MASS proper motions in the
Appendix.

Finally, we constructed an initial catalog of objects with good
quality PS1+2MASS+{\it WISE} photometry and PS1+2MASS proper motions
signal-to-noise ratio (S/N) $\ge$10, corresponding to a typical proper
motion of $\ge$140~mas~yr$^{-1}$. Although by requiring the proper
motions to have a high S/N we will miss slower-moving candidate
members, only $\approx$13\% of the confirmed AB~Dor Moving Group
members tabulated in \citet{2013ApJ...762...88M} have proper motions
below 140~mas~yr$^{-1}$ and are within 50~pc (the approximate
photometric distance limit of our search).

\section{Candidate Selection} \label{sec:searchmethod}

After combining PS1, 2MASS, and {\it WISE} photometry to create an
initial catalog of objects with good quality proper motions and
photometry (Section~\ref{sec:data}), we screened our initial
catalog for probable late-M and L dwarfs using the following color
cuts: $y-J \ge 1.4$, $z-y \ge 0.5$, and $W1-W2 \ge 0$. In addition, we
selected for objects with \ips--\zps~$\ge$~0.9 if the \ips~ photometry
met our quality requirements (Section~\ref{sec:data}). In order to
remove potential galaxies from our sample we chose objects with $W2-W3
\le 3$ \citep{2010AJ....140.1868W}.

Next, we constructed SEDs and estimated spectral types and photometric
distances using the template-fitting method of
\citet{2013ApJ...773...63A}. We first created template SEDs from known
ultracool dwarfs with spectral types of M and L based on the
compilations by \citet{2007AJ....134.2340K},
\citet{2009AJ....137....1F}, \citet{2010ApJ...710.1627L}, and
DwarfArchives.org. Then we estimated the spectral type of our
candidates by determining the best-matched template SED using a
chi-squared fit. Our estimated photometric distances ($d_{phot}$) were
based on the \citet{2012ApJS..201...19D} average $J_{2MASS}$ absolute
magnitudes as a function of spectral type
(Section~\ref{sec:dphot}). We required our candidates to have
estimated spectral types from our SED fit later than M5, which
corresponds to the stellar/substellar boundary at the age of the
AB~Dor Moving Group.

In addition, we limited our search to candidates with proper motions
between 40 and 1000~mas~yr$^{-1}$. The lower proper motion limit is
set because we only chose candidates with proper motion S/N~$\ge$~10
and our minimum proper motion uncertainty was
$\approx$4~mas~yr$^{-1}$. Also, because we only matched PS1 sources
with 2MASS counterparts within a 10$\arcsec$ radius, our proper motion
upper limit was approximately 1000~mas~yr$^{-1}$.

\begin{figure*}
  \begin{center}
    \includegraphics[width=0.45\textwidth]{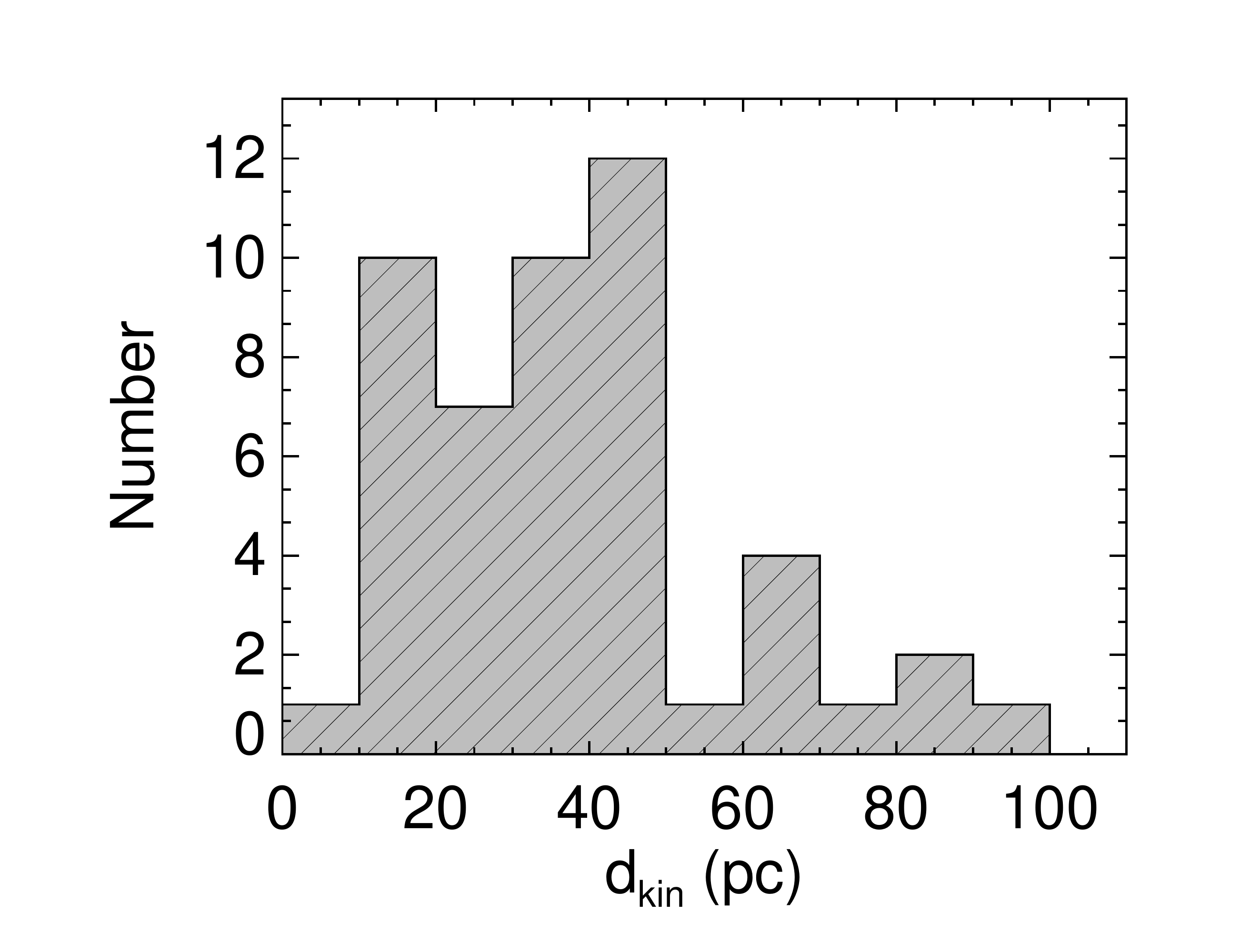}
    \includegraphics[width=0.45\textwidth]{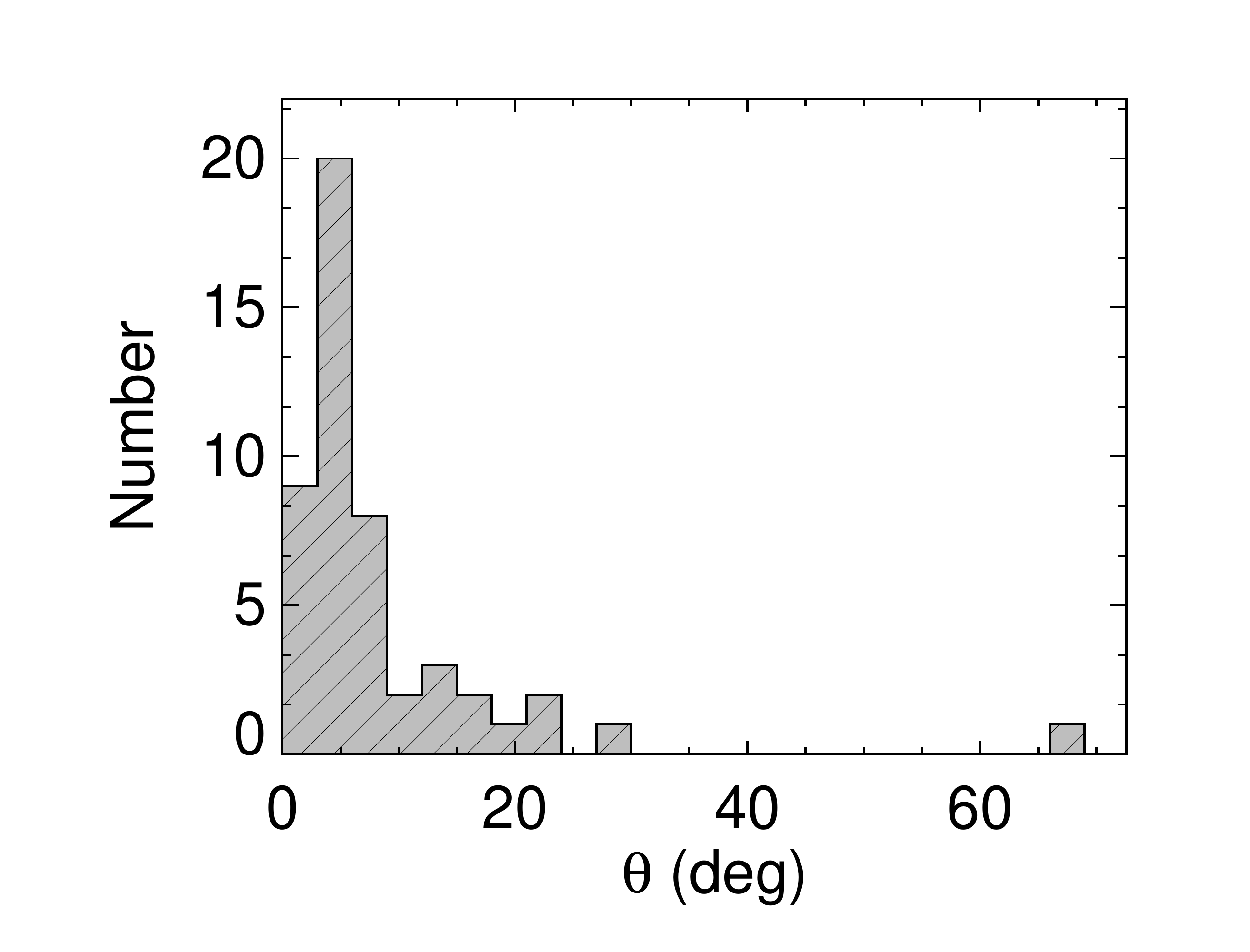}
    \caption{The kinematic distance ($d_{kin}$) and $\theta$ for the
      known AB~Dor Moving Group members with membership probability
      greater than 90\% and parallaxes \citep{2008hsf2.book..757T}. We
      required our candidate AB~Dor Moving Group members to have
      $\theta \le 40^{\circ}$ because the majority of the known members
      also satisfy this requirement. \label{fig:abdtheta}}
  \end{center}
\end{figure*}

Because moving group members have space motions with a common
characteristic direction and amplitude, we further refined our
candidate selection using proper motion and sky position. Following
\citet{2012AJ....143...80S}, we screened for candidates with space
motions consistent with the AB~Dor Moving Group. Specifically, we used
proper motions to calculate the angle ($\theta$) projected onto the
plane of the sky between our candidates' proper motions and the
average space motion vector of the known members of the AB~Dor Moving
Group \citep{2008hsf2.book..757T}. We also determined the kinematic
distance ($d_{kin}$), namely a candidate's distance if it were a
member of the AB~Dor Moving Group with the same absolute proper motion
velocity (i.e. velocity in km~s$^{-1}$) as the average of the known
members. In order to determine our selection criteria, we calculated
$\theta$ and $d_{kin}$ for the known members of the AB~Dor Moving
Group with parallaxes from \citet{2008hsf2.book..757T} and determined
that these members mainly have $\theta$~$\lesssim$~40$^{\circ}$
(Figure~\ref{fig:abdtheta}). In addition, the 1$\sigma$ distance range
for known AB~Dor Moving Group members is within 50~pc
\citep{2014ApJ...783..121G}. Therefore, we required our candidate
AB~Dor Moving Group members to have
$\theta-\sigma_{theta}$~$\le$~40$^{\circ}$ and
$d_{kin}-\sigma_{d_{kin}}$~$\le$~50~pc. Finally, we also chose objects
with $d_{phot}$ less than 50~pc
($d_{phot}-\sigma_{d_{phot}}$~$\le$~50~pc) and consistent with their
d$_{kin}$ within the uncertainties in both d$_{kin}$ and
d$_{phot}$. We allowed the d$_{phot}$ to vary within 50\% to allow for
uncertainties in determining photometric distance and in the absolute
magnitudes of our candidates based on estimated spectral type.

In addition to photometric and kinematic information, we limited our
search for AB~Dor Moving Group members by sky positions. Based on the
positions of known members, all candidates were selected to have
Galactic latitude below 60$^{\circ}$ and declinations below
70$^{\circ}$. We also ignored objects within 3$^{\circ}$ of the
Galactic plane because of crowding.

Given the photometric distance limit of our search, we could detect
objects with spectral types of $\approx$~L4 at 50~pc, which
corresponds to a mass of $\approx$~30~\Mjup~ given the age of the
AB~Dor Moving Group. Thus our search is sensitive to candidate
substellar AB~Dor Moving Group members with masses from the
substellar-stellar boundary down to $\approx$~30~\Mjup~ out to
50~pc. We tabulate the photometry, proper motions, d$_{phot}$,
d$_{kin}$, and $\theta$ for our new candidate AB~Dor Moving Group
members in Table~\ref{table:obsprop}.

\section{Observations} \label{sec:observations}

Field M and L dwarfs have similar colors and are more numerous than M
and L dwarf members of the AB~Dor Moving Group. Therefore we require
spectroscopy to determine whether our candidates are young brown
dwarfs, and thus potential members of this young ($\approx$125~Myr old) moving
group. Young brown dwarfs are spectroscopically distinguishable from
field ultracool dwarfs because their lower surface gravity affects the
depths of absorption lines and the overall continuum shape in the NIR
\citep[e.g.][]{2007ApJ...657..511A,2013ApJ...772...79A}.

We obtained spectroscopic followup using SpeX
\citep{2003PASP..115..362R}, the near-IR (0.8--2.5~$\mu m$)
spectrograph on the 3~m NASA Infrared Telescope Facility (IRTF) on
Maunakea. We used the low-resolution (LowRes15) prism mode with a
0.5$\arcsec$ slit width (R$\sim$130) for dwarfs with estimated
spectral types from SED fitting later than M8. For earlier M~dwarfs,
we used the moderate-resolution cross-dispersed (SXD) mode
(R$\sim$750). These resolutions are sufficient to determine spectral
type and assess youth using the \citet{2013ApJ...772...79A}
classification methods. Note that all spectra taken after 2014 August
were observed using the upgraded version of SpeX (uSpeX), which has
slightly larger wavelength coverage, 0.7--2.5~$\mu$m.

We also obtained spectroscopic followup using GNIRS, the near-IR
(0.8--2.5~$\mu$m) spectrograph on the 8~m {\it Gemini} telescope on
Maunakea. We used the moderate-resolution (R$\sim$1700)
cross-dispersed (SXD) mode with the 32~l~mm$^{-1}$ grating with the
0.15$\arcsec$/pix camera and the 0.3$\arcsec$ slit.

Our observations were obtained using a standard ABBA nod pattern for
sky subtraction. We observed an A0V standard star following each
candidate and then took wavelength and flatfield calibrations
immediately afterward. All SpeX spectra were reduced using version 3.4
(version 4.0 for uSpeX data) of the SpeXtool package
\citep{2003PASP..115..389V,2004PASP..116..362C}. We reduced the GNIRS
spectra using a custom version of SpeXtool
\citep{2013ApJ...777L..20L}. Table~\ref{table:obs} summarizes the
observation details.

\section{Results} \label{sec:results}

\subsection{Spectral Analysis \label{sec:specanalysis}}

We determined the NIR spectral type of our objects using both the
index-based and visual methods of \citet{2013ApJ...772...79A}. First,
the quantitative method combines the spectral-type sensitive indices
from \citet{2007ApJ...657..511A}, \citet{2004ApJ...610.1045S}, and
\citet{2003ApJ...596..561M} to calculate the average spectral
type. All of these spectral type indices are valid across the spectral
type range of our candidates except the \citet{2003ApJ...596..561M}
H$_{2}$O--D index, which is only valid for L dwarfs. We also performed
a Monte Carlo simulation to propagate measurement errors of our
reduced spectra into the index calculations in order to determine the
spectral type uncertainties derived from each index.

\begin{figure*}
  \begin{center}
    \centering
    \includegraphics[width=0.9\textwidth]{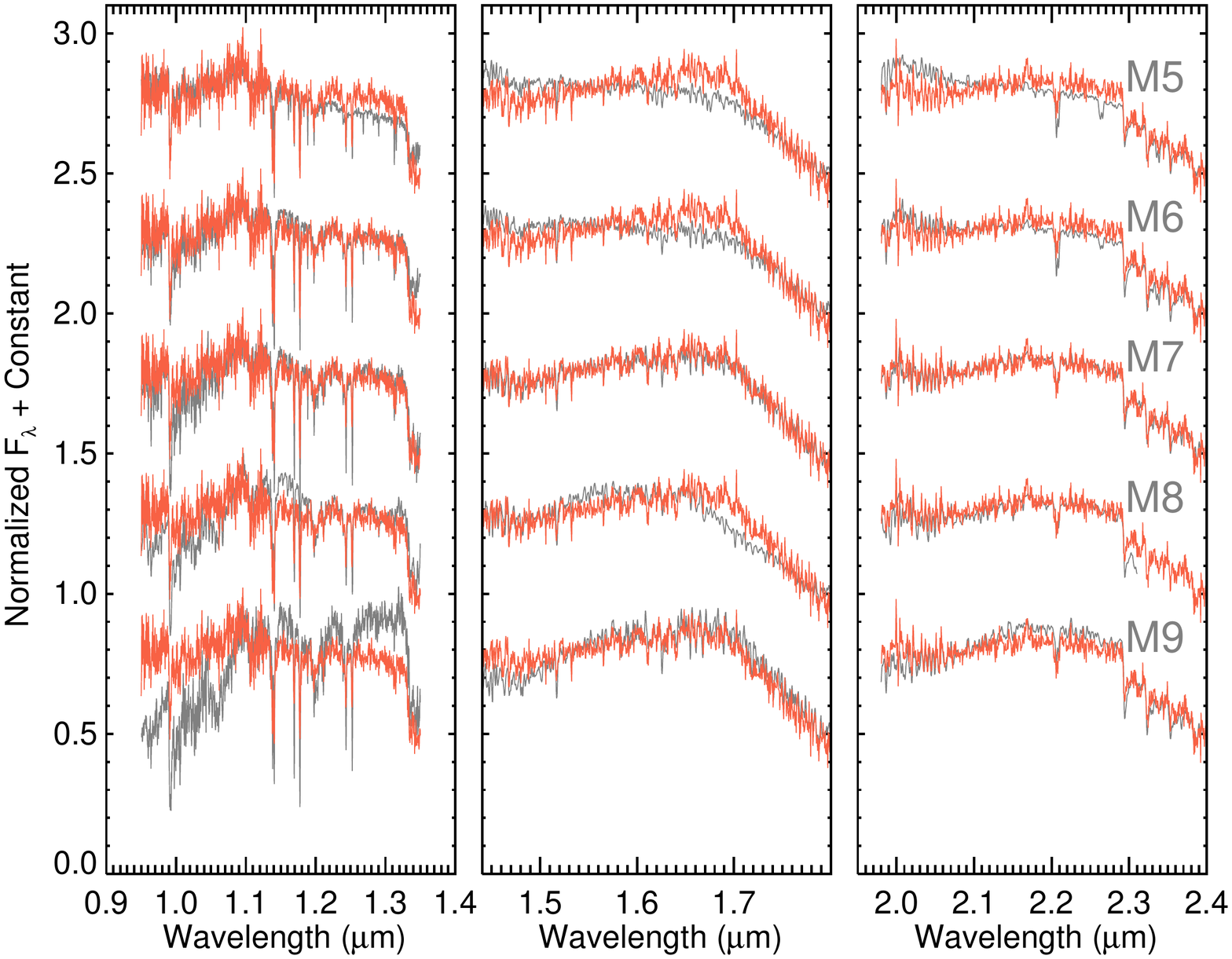}
    \caption{An example of visual classification of one of our
      candidate AB~Dor Moving Group members, PSO~J035.8$-$15
      (\emph{dark orange}), by comparing to the
      \citet{2010ApJS..190..100K} spectral standard M~dwarfs
      (\emph{gray}). The standards were taken from the IRTF Spectral
      Library \citep{2005ApJ...623.1115C} and smoothed to the
      resolution of our candidate spectrum (in this case,
      R$\sim$750). The standards are, from top to bottom: Gl~51 (M5),
      Gl~406 (M6), Gl~644C (M7), VB~10 (M8), and LHS~2924 (M9). The
      spectral type determined from visual classification for this
      object is M7$\pm$1 in both the $J$ and $K$
      band. \label{fig:visualclass}}
  \end{center}
\end{figure*}

Second, in addition to measuring indices to determine spectral type,
we also visually compared our objects to M and L~dwarf spectroscopic
standards defined in \citet{2010ApJS..190..100K}. We used standard
spectra taken from the IRTF Spectral Library
\citep{2005ApJ...623.1115C} and the SpeX Prism
Library\footnote{\url{http://pono.ucsd.edu/\textasciitilde adam/browndwarfs/spexprism}}.
Following the visual classification methods for young and intermediate
age objects of \citet{2013ApJ...772...79A}, we normalize both our
candidates and the standard template in each NIR band separately (see
Figure~\ref{fig:visualclass} for an example). However, because the
$H$~band of young/intermediate-aged brown dwarfs often have a
distinctly different shape from old objects, the $H$~band is not used
for visual spectral type classification. Because selecting a standard
with 1 subtype difference compared to our best-fitting standard
produced a noticeably poorer fit, we assumed an uncertainty of 1
subtype for our visual classification \citep[consistent
  with][]{2013ApJ...772...79A}. 

\begin{figure*}
  \begin{center}
    \includegraphics[width=.8\textwidth]{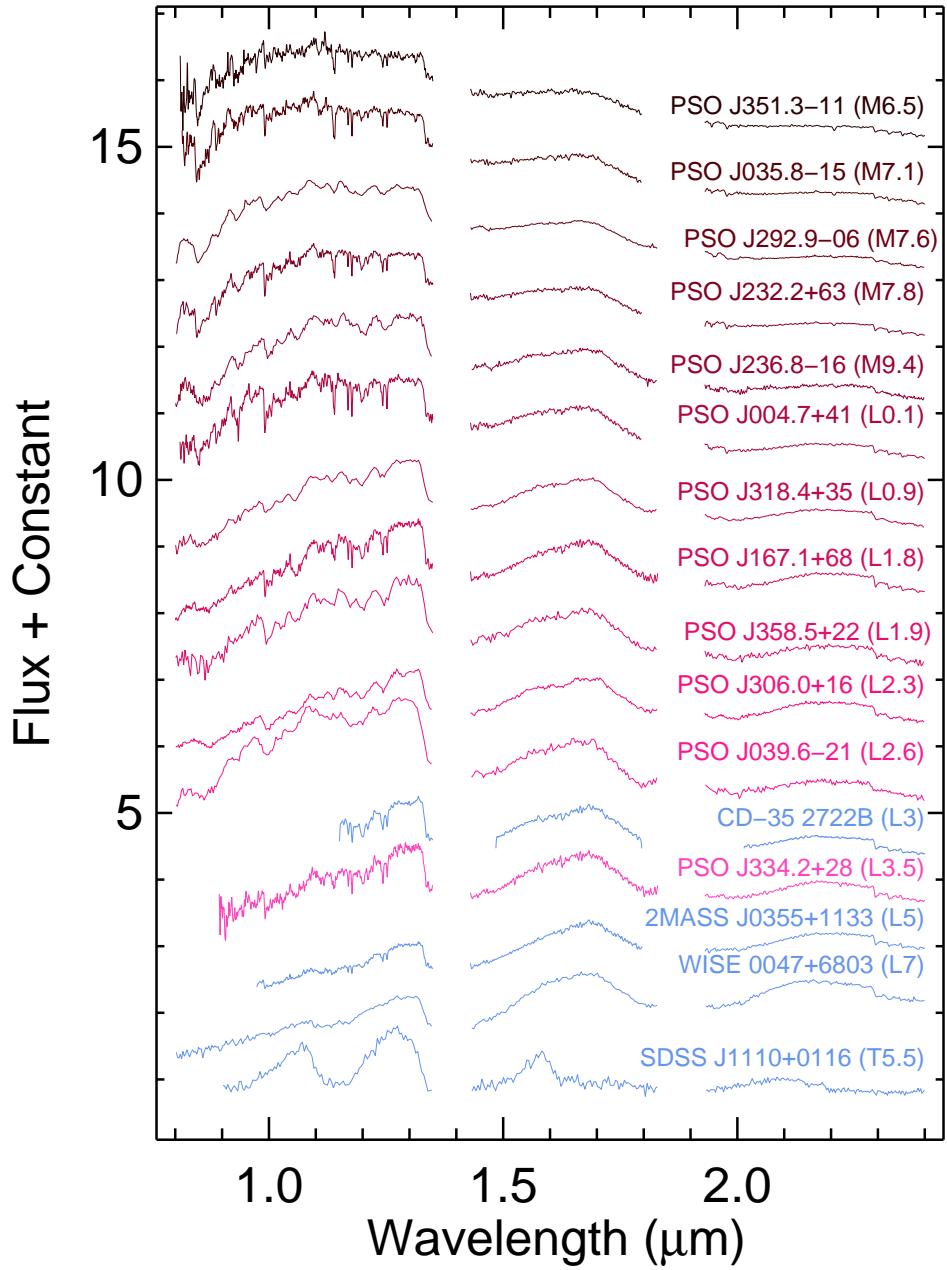}
    \caption{NIR spectra of our confirmed-young brown dwarf AB~Dor
      Moving Group candidates (\emph{shades of red}) in addition to
      the four known brown dwarf members (\emph{blue}) with publicly
      available spectra: CD-35~2722~B \citep{2011ApJ...729..139W},
      2MASS~J0355+1133
      \citep{2013AN....334...85L,2013AJ....145....2F},
      WISEP~J0047+6803 \citep{2012AJ....144...94G,2015ApJ...799..203G}
      and SDSS~J1110+0116 \citep{2015ApJ...808L..20G}. The spectra are
      ordered from earliest spectral type (highest mass) to latest
      spectral type (lowest mass).\label{fig:sptabd}}
  \end{center}
\end{figure*}

Our final spectral type is the weighted mean of the index-based and
visual spectral types as in \citet{2013ApJ...772...79A}. As we use the
\citet{2013ApJ...772...79A} method, we also adopt a conservative
spectral type uncertainty of 1 subtype. The spectral types for our
candidates are tabulated in Table~\ref{table:spt}. In
Figure~\ref{fig:sptabd} we show the spectra of our candidates in
addition to the four known brown dwarf members, CD-35~2722~B
\citep{2011ApJ...729..139W}, 2MASS~J0355+1133
\citep{2013AN....334...85L,2013AJ....145....2F}, WISEP~J0047+6803
\citep{2015ApJ...799..203G} and SDSS~J1101+0116
\citep{2015ApJ...808L..20G}.

We then assessed the gravity classification of our objects using
spectral indices defined in \citet{2013ApJ...772...79A}. Under their
classification scheme, several indices are measured in the $J$ and
$H$~bands and then are each assigned a score (0, 1, or 2) according to
the index value and the spectral type of the object, with higher
numbers indicating lower gravity. These scores are combined into a
final 4-number gravity score that represent the FeH, VO, alkali lines,
and $H$-band continuum indices (e.g. 0110, 2110, etc.). Finally, this
gravity score is used to determine the overall gravity classification
for the object: field gravity ({\sc fld-g}), intermediate gravity
({\sc int-g}), or very low gravity ({\sc vl-g}). We describe the
method in more detail below and also describe some modifications we
have made to account for the modest S/N of some of our spectra.

\subsubsection{Gravity Index Calculations and Uncertainties}

Depending on the spectral resolution, there is a specific set of
gravity indices used to assess the overall gravity classification of
an object. The FeH$_{z}$, VO$_{z}$ and KI$_{J}$ indices from
\citet{2013ApJ...772...79A} are tailored to assess gravity in low
resolution (R$\sim$130) spectra. In order to measure these indices
from our moderate-resolution SXD spectra (R$\sim$750), we smoothed
those spectra to R=130. The H-cont index from
\citet{2013ApJ...772...79A}, is used to assess gravity in either low
or moderate-resolution spectra by measuring the shape of the $H$-band,
specifically how close the blue end of the $H$-band continuum is to a
straight line. For our moderate resolution spectra, following
\citet{2013ApJ...772...79A}, we also used the FeH$_{J}$ index and the
alkali line indices in the $J$~band (NaI~[1.138~$\mu$m],
KI~[1.169~$\mu$m], KI~[1.177~$\mu$m], and KI~[1.253~$\mu$m]) to assess
gravity with the continuum used to compute pseudo-equivalent widths
defined by a linear fit.

For each of these gravity indices, \citet{2013ApJ...772...79A}
estimate the flux uncertainties from the rms scatter about a linear
fit to the continuum window around each index. In the case of high S/N
spectra this uncertainty in fitting the continuum is the dominant
source of error to measure pseudo-equivalent widths. However, this
approach underestimates the uncertainties for our lower S/N spectra
with S/N$\sim$40--100. Therefore, in our method, we determined the
uncertainty in each index ($\sigma$) using a Monte Carlo simulation to
propagate the spectrum measurement errors.

\begin{figure*}
  \begin{center}
    \centering
    \includegraphics[width=0.8\textwidth]{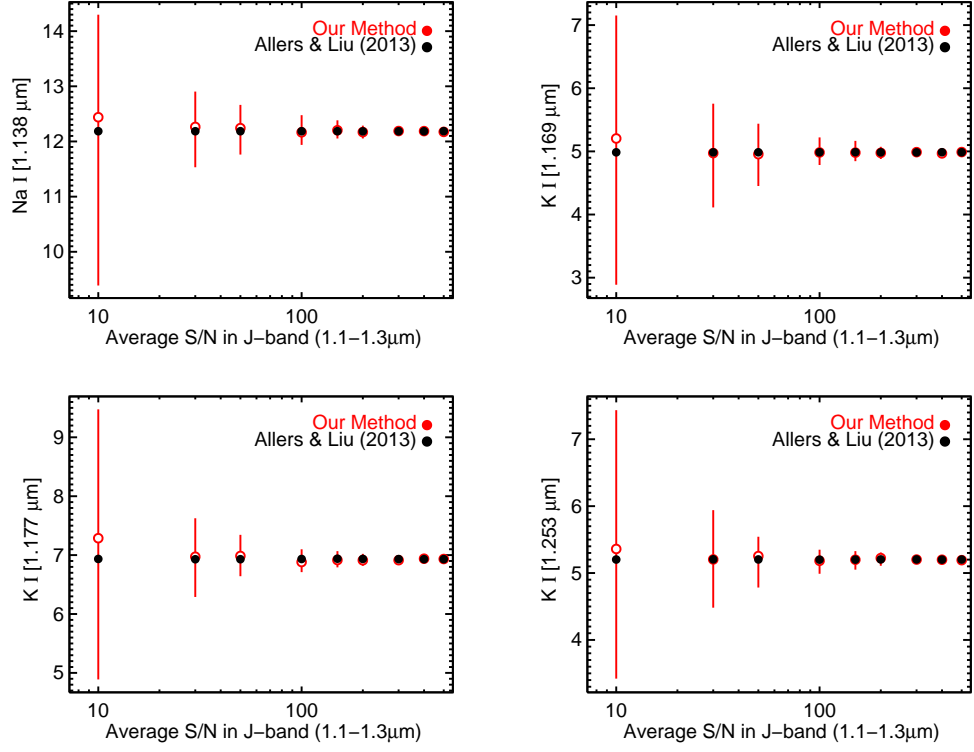}
    \caption{Comparison of the pseudo-equivalent width measurement
      uncertainties using our method (\emph{red open circles}) and the
      \citet{2013ApJ...772...79A} method (\emph{black circles}). We
      include measurement uncertainties, whereas
      \citet{2013ApJ...772...79A} only consider uncertainties in
      measuring continuum, which would dominate for high S/N spectra
      (S/N$\gtrsim$200). For our relatively lower S/N objects
      (S/N$\approx$40--100), we need to also include measurement
      errors when computing the uncertainties in the pseudo-equivalent
      widths. Although our measurement uncertainties are larger, our
      pseudo-equivalent widths are consistent with
      \citet{2013ApJ...772...79A} \label{fig:ewerrors}}
  \end{center}
\end{figure*}

In order to examine the effects of low S/N in calculating the
uncertainties for the gravity indices, we simulated spectra with a
range of S/N and determined the alkali line gravity index values and
uncertainties. We degraded the S/N of Gl~752B, an M8 from the IRTF
Spectral Library \citep{2009ApJS..185..289R}, from S/N of 500 down to
10. Then we calculated the measurement errors for the alkali line
gravity indices (NaI~[1.138~$\mu$m], KI~[1.169~$\mu$m],
KI~[1.177~$\mu$m], and KI~[1.253~$\mu$m]) using both a Monte Carlo
simulation to propagate the measurement errors and the original method
of \citet{2013ApJ...772...79A}. Our simulation shows that although
both methods are consistent within uncertainties for these indices
(Figure~\ref{fig:ewerrors}), the measurement errors are the main
source of error for spectra with modest S/N ($\lesssim$200).

\subsubsection{Gravity Index Scores} \label{sec:gravityindexscores}

After computing the index values and their uncertainties, we determine
gravity scores for each of these indices, namely 0, 1 or 2. Following
\citet{2013ApJ...772...79A}, indices that are undefined for an object,
because of spectral type and/or resolution, are given a score of
"n". The \citet{2013ApJ...772...79A} method also gives indices with
values that are within 1$\sigma$ (the index uncertainty) from the
field sequence values a score of "?". We handle such objects slightly
differently and instead do not give indices a score of "?" in order to
better identify borderline objects between the {\sc int-g} and {\sc
  fld-g} values.

For objects with low-resolution spectra (R$\sim$130), only the
FeH$_{z}$, VO$_{z}$, KI$_{J}$ and H-cont indices are used. As an
example, for PSO~J039.6$-$21, the FeH$_{z}$, VO$_{z}$, KI$_{J}$ and
H-cont scores are: 1021. But strictly following
\citet{2013ApJ...772...79A}, the final scores would instead be 102?
because the H-cont index value is within 1$\sigma$ of the {\sc fld-g}
value.

For objects with moderate-resolution spectra (R$\sim$750), the
\citet{2013ApJ...772...79A} method uses a similar set of four
measurements to assess gravity: one based on FeH, VO$_{z}$, the alkali
lines, and H-cont. The index scores for VO$_{z}$ and H-cont are
computed in the same manner as for the low-resolution
spectra. However, unlike for the low-resolution spectra, the final FeH
and alkali line scores are determined by combining the scores from
multiple indices. The final alkali line score is the mean (rounded up
to the nearest integer) of the individual scores from the
NaI~[1.138~$\mu$m], KI~[1.169~$\mu$m], KI~[1.177~$\mu$m], and
KI~[1.253~$\mu$m] indices. As an example, for PSO~J035.8$-$15, the
scores for NaI~[1.138~$\mu$m], KI~[1.169~$\mu$m], KI~[1.177~$\mu$m],
and KI~[1.253~$\mu$m] are 0101, thus the final alkali line index score
is 1. The final FeH index score is the larger of the FeH$_{z}$ and
FeH$_{J}$ scores. For 2MASS~J0233--15, the scores for FeH$_{z}$ and
FeH$_{J}$ are 11, and thus the final FeH score is 1. After
amalgamating the indices used to compute the final FeH and alkali line
scores, the four scores (FeH, VO$_{z}$, alkali lines, and H-cont) of
our example object, PSO~J035.8$-$15, are 1n12. In this example, the
scores are the same when strictly following the
\citet{2013ApJ...772...79A} method, since none of the index scores are
"?".

\subsubsection{Gravity Classification}

Following \citet{2013ApJ...772...79A}, after determining the scores
for the FeH, VO$_{z}$, alkali lines (KI$_{J}$ for the low-resolution
spectra or a combination of four pseudo-equivalent widths for
moderate-resolution spectra), and $H$-band continuum indices, we reduce
these four scores into a single value to represent the overall gravity
classification. This overall gravity classification value is the
median of these final four scores. If there was an even number of
defined indices, we take the average of the two scores straddling the
median. Objects with an overall gravity classification value $\le$~0.5
are classified as {\sc fld-g}. Those objects with 0.5~$<$~gravity
classification value~$<$~1.5 are classified as {\sc int-g}, and those
with a gravity classification value $\ge$~1.5 are classified as {\sc
  vl-g}.

Because our method uses a Monte Carlo simulation to propagate
measurement uncertainties into the index values, our overall gravity
classification also has uncertainties. Each Monte Carlo realization of
an object's spectrum produces an overall gravity classification value,
and we report the median value from all the realizations as the final
result and the 68\% confidence limits as the uncertainty. For
instance, 2MASS~J0223--15 has an overall gravity classification value
of 1.0$^{+1.0}_{-0.0}$, which corresponds to {\sc int-g}. For
comparison, using the original \citet{2013ApJ...772...79A} method,
2MASS~J0223--15 has gravity scores of 1n12, so the overall gravity
classification value would be 1, which also corresponds to {\sc
  int-g} but without any uncertainties.

As mentioned in Section~\ref{sec:gravityindexscores}, unlike
\citet{2013ApJ...772...79A}, we do not ignore indices with values
straddling the intermediate gravity and field gravity values. As a
result, we may be able to identify borderline objects that have index
values that hint at intermediate gravities but are not clearly
separated from the field values. Therefore, we choose to classify
objects with overall gravity classification values within 1$\sigma$ of
{\sc int-g} as borderline intermediate gravity ({\sc int-g?}). For
example, an object with an overall gravity classification value of
0.5$^{+0.5}_{-0.5}$ would be classified as {\sc int-g?}.

\begin{figure*}
  \begin{center}
    \centering
    \includegraphics[width=0.7\textwidth]{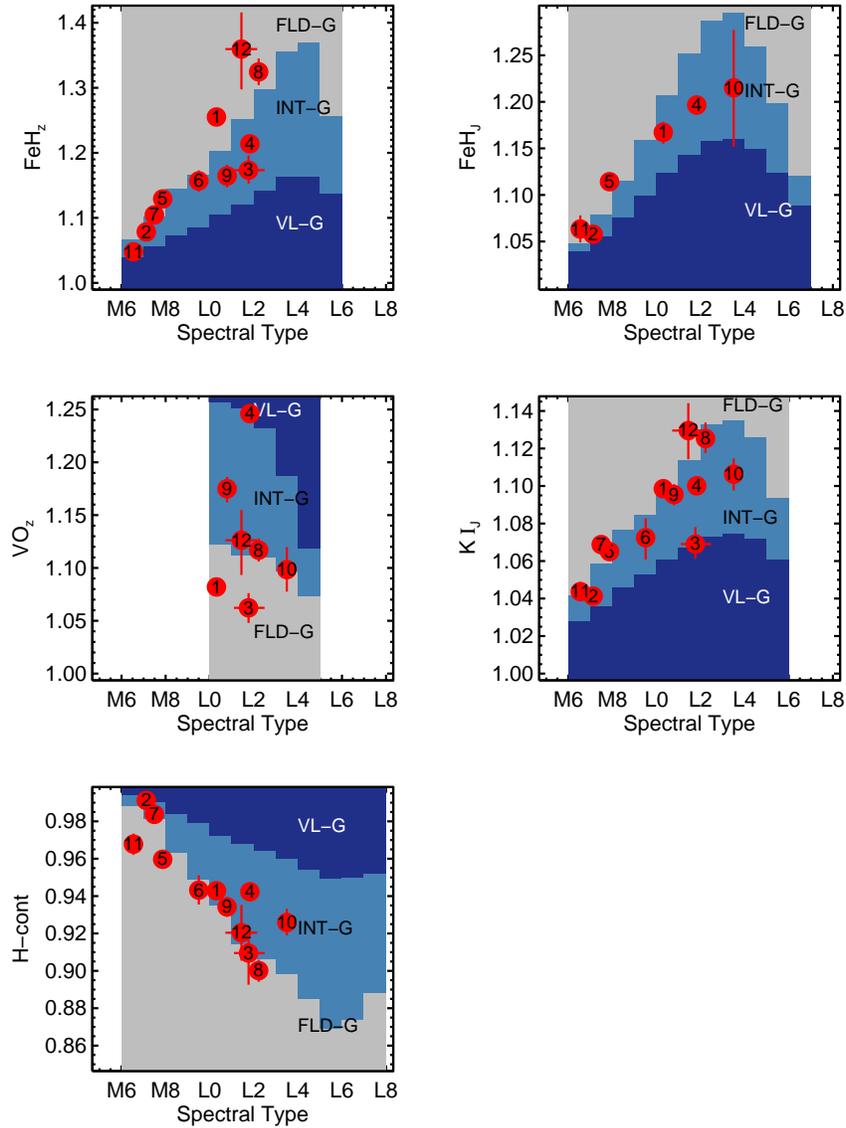}
    \caption{NIR spectral indices from all of our candidate AB~Dor
      Moving Group members. For the FeH$_{z}$, VO$_{z}$ and KI$_{J}$
      indices, we smoothed the spectra to prism resolution
      (R$\sim$130) because the indices are tailored for
      prism-resolution spectra. We compare the indices of our numbered
      candidates (\emph{red circles}) with the defining values for
      very low gravity ({\sc vl-g}), intermediate gravity ({\sc
        int-g}) and field gravity ({\sc fld-g}) taken from
      \citet{2013ApJ...772...79A}. Our objects are numbered as
      follows: (1)~PSO~J004.7+41, (2)~PSO~J035.8$-$15, (3)~PSO~J039.6$-$21,
      (4)~PSO~J167.1+68, (5)~PSO~J232.2+63, (6)~PSO~J236.8$-$16,
      (7)~PSO~J292.9$-$06, (8)~PSO~J306.0+16, (9)~PSO~J318.4+35,
      (10)~PSO~J334.2+28, (11)~PSO~J351.3$-$11, (12)~PSO~J358.5+22. The
      shaded regions define the gravity classification of {\sc fld-g}
      (\emph{gray}), {\sc int-g} (\emph{gray blue}) and {\sc vl-g}
      (\emph{dark blue}) for each index. Note that some indices only
      can be used for overall gravity classification within a range of
      spectral types or resolution (i.e. FeH$_{J}$ is only for
      R$\gtrsim$500). In these cases, we do not use the index value to
      determine the object's gravity score. We expect our AB~Dor
      Moving Group candidates to have {\sc int-g} given the age of the
      group ($\approx$125~Myr). \label{fig:prismind}}
  \end{center}
\end{figure*}

\begin{figure*}
  \begin{center}
    \centering
    \includegraphics[width=0.8\textwidth]{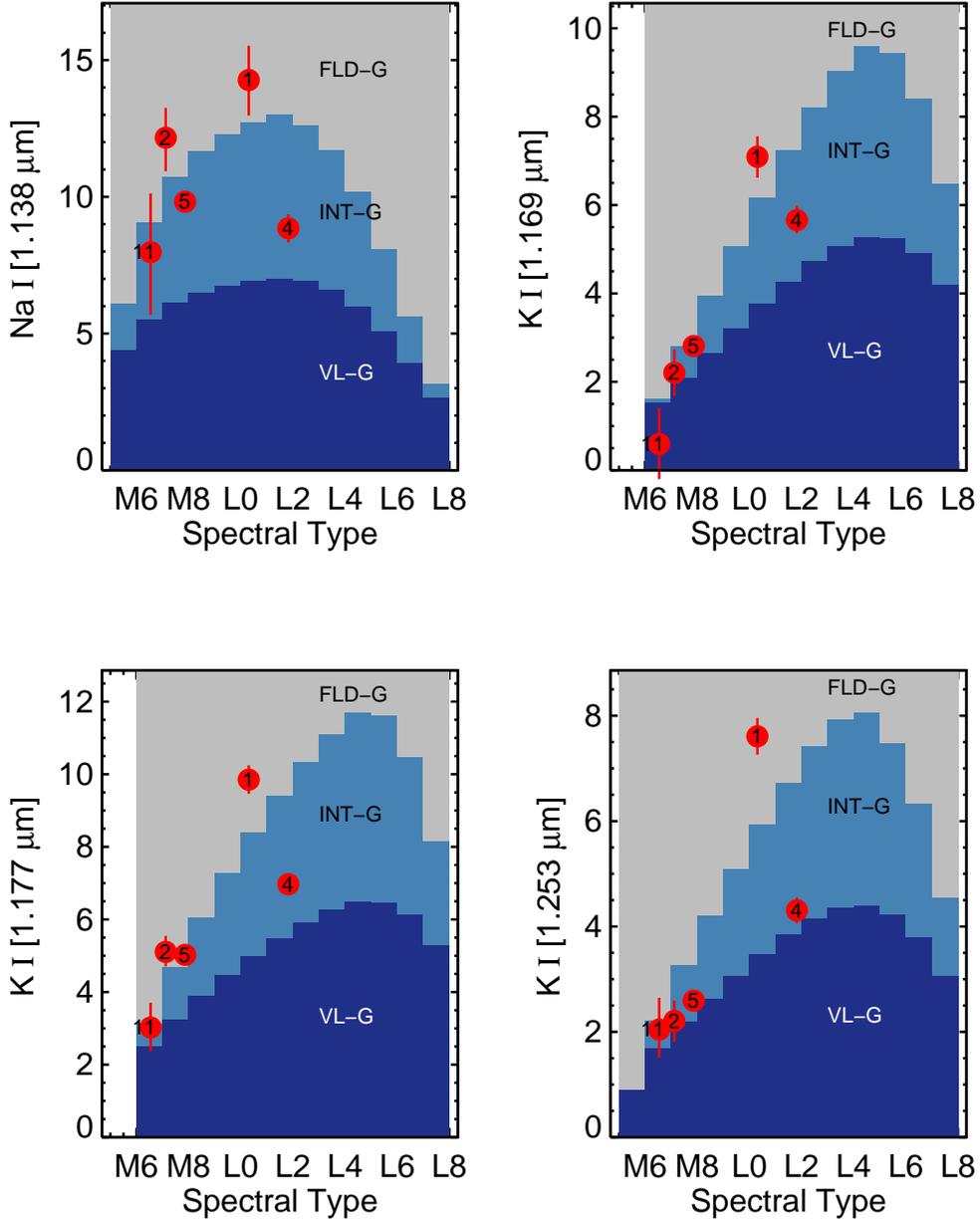}
    \caption{NIR spectral indices of our candidate AB~Dor Moving Group
      members taken in the cross-dispersed mode in IRTF/SpeX. We
      compare the indices of our candidates with the defining values
      for very low gravity ({\sc vl-g}), intermediate gravity ({\sc
      int-g}), and field gravity ({\sc fld-g}) taken from
      \citet{2013ApJ...772...79A}. The shaded regions define the
      gravity classification of {\sc fld-g} (\emph{gray}), {\sc int-g}
      (\emph{gray blue}) and {\sc vl-g} (\emph{dark blue}) for each
      index. We expect our AB~Dor Moving Group candidates to have
      intermediate gravities. Our objects have the same number labels as
      Figure~\ref{fig:prismind}. \label{fig:sxdind}}
  \end{center}
\end{figure*}

\begin{figure*}
  \begin{center}
    \centering
    \includegraphics[width=.7\textwidth]{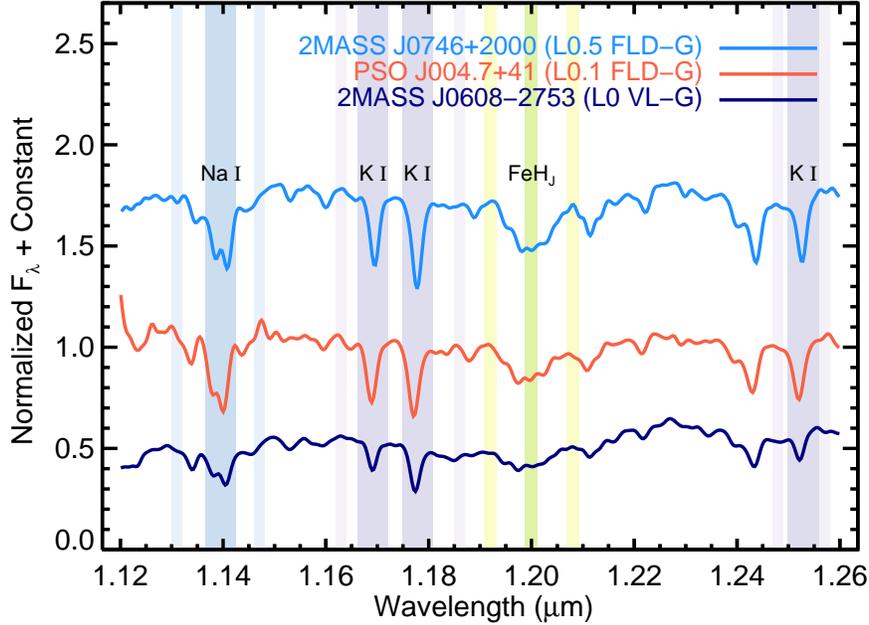}
    \caption{NIR moderate-resolution (R$\sim$750) $J$-band spectra
      from IRTF/SpeX of our AB~Dor Moving Group candidate,
      PSO~J004.7+41 (\emph{red}), compared with a field gravity ({\sc
        fld-g}, \emph{light blue}) and young, very low gravity ({\sc
        vl-g}, \emph{dark blue}) dwarf of similar spectral type
      (within half a spectral type). For our comparison spectra, we
      used the field spectral type standard objects from
      \citet{2010ApJS..190..100K} if there was available SXD spectra
      in the SpeX Library. If not, we used the non-standard field
      object taken in SXD mode with the closest spectral type to our
      object. Young dwarfs were taken from the list of standard young
      objects in \citet{2013ApJ...772...79A} if there was available
      moderate-resolution spectra. If not, we used a non-standard {\sc
        vl-g} object from \citet{2013ApJ...772...79A} with a spectral
      type within a half a spectral type of our
      object. Gravity-sensitive features in the $J$ band from
      \citet{2013ApJ...772...79A} are labeled and the wavelength
      ranges used to calculate gravity indices are highlighted for
      Na~I (\emph{blue}), K~I (\emph{purple}), and FeH$_{J}$
      (\emph{yellow-green}). Members of the AB~Dor Moving Group have
      an age of $\approx$125~Myr, thus are expected to have intermediate
      gravities lying between the field dwarfs ({\sc fld-g}) and young
      dwarfs ({\sc vl-g}). Although this object, PSO~J004.7+41 has a
      gravity classification of {\sc fld-g}, the gravity indices show
      minor hints of intermediate gravity. \label{fig:sxd1}}
  \end{center}
\end{figure*}

\begin{figure*}
  \vspace{-1 cm}
  \begin{center}
    \includegraphics[width=.7\textwidth]{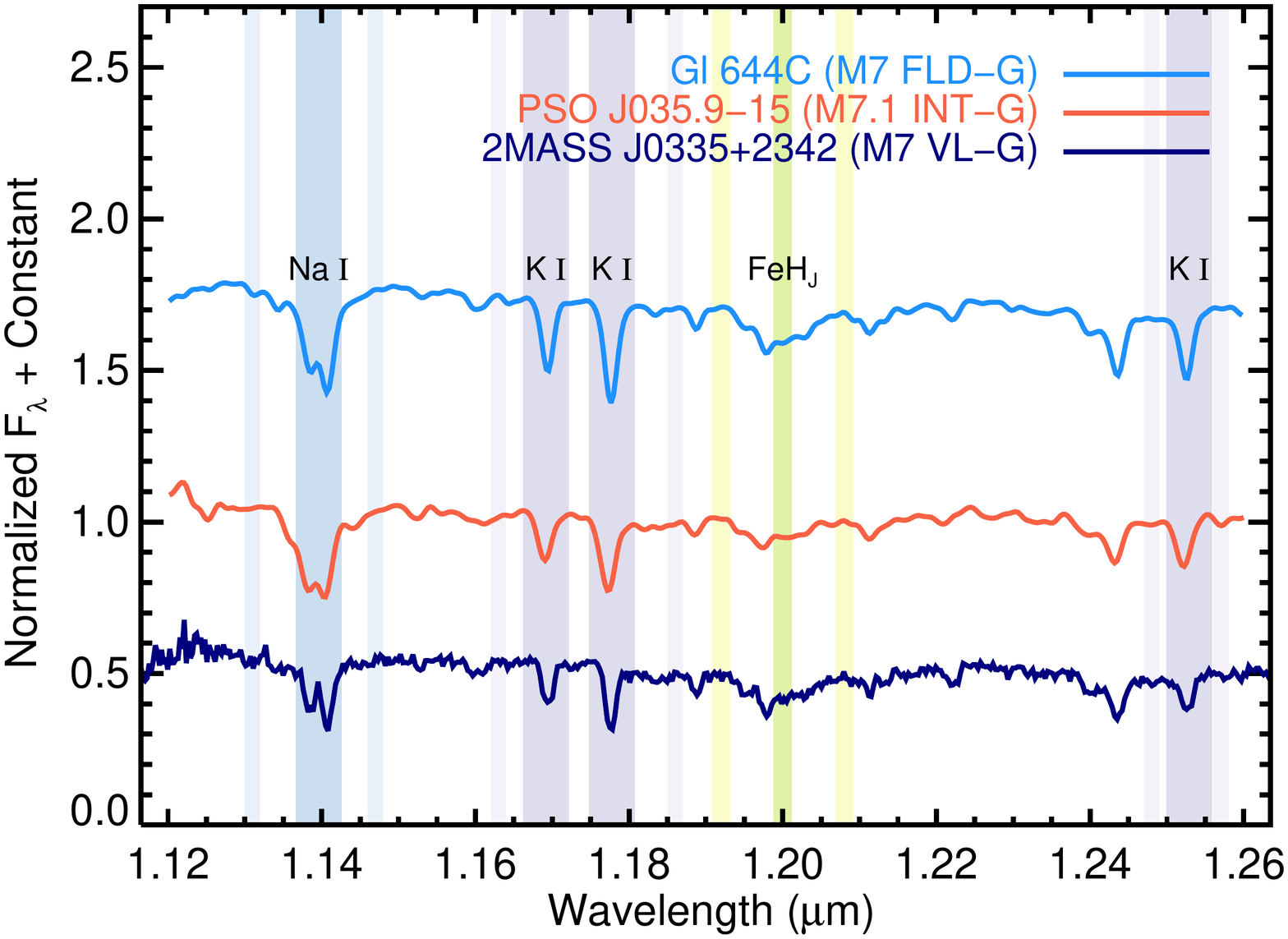}
    \caption{NIR moderate-resolution (R$\sim$750) spectra from
      IRTF/SpeX of our AB~Dor Moving Group candidate, PSO~J035.8$-$15
      (\emph{red}), compared with a {\sc fld-g} (\emph{light blue}) and
      {\sc vl-g} (\emph{dark blue}) dwarf of similar spectral type (within
      half a spectral type). We choose the comparison spectra as
      described in Figure~\ref{fig:sxd1}. \label{fig:sxd2}}
  \end{center}
\end{figure*}

\begin{figure*}
  \begin{center}
    \includegraphics[width=.7\textwidth]{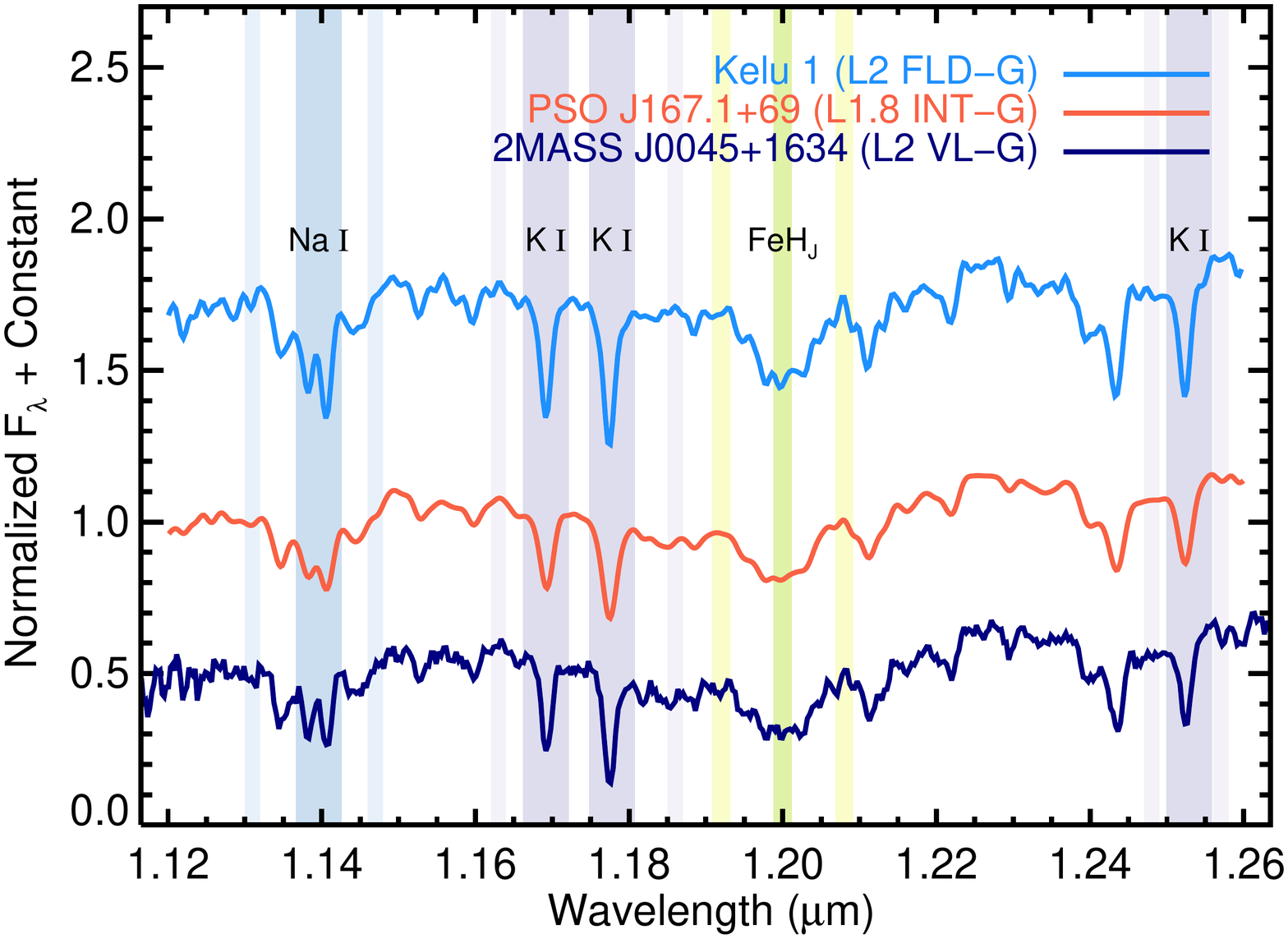}
    \caption{NIR moderate-resolution (R$\sim$750) spectra from
      IRTF/SpeX of one of our AB~Dor Moving Group candidates,
      PSO~J167.1+68 (\emph{red}), compared with a {\sc fld-g}
      (\emph{light blue}) and {\sc vl-g} (\emph{dark blue}) dwarf of
      similar spectral type (within half a spectral type). We choose
      the comparison spectra as described in
      Figure~\ref{fig:sxd1}. \label{fig:sxd3}}
  \end{center}
\end{figure*}

\begin{figure*}
  \vspace{-1 cm}
  \begin{center}
    \includegraphics[width=.7\textwidth]{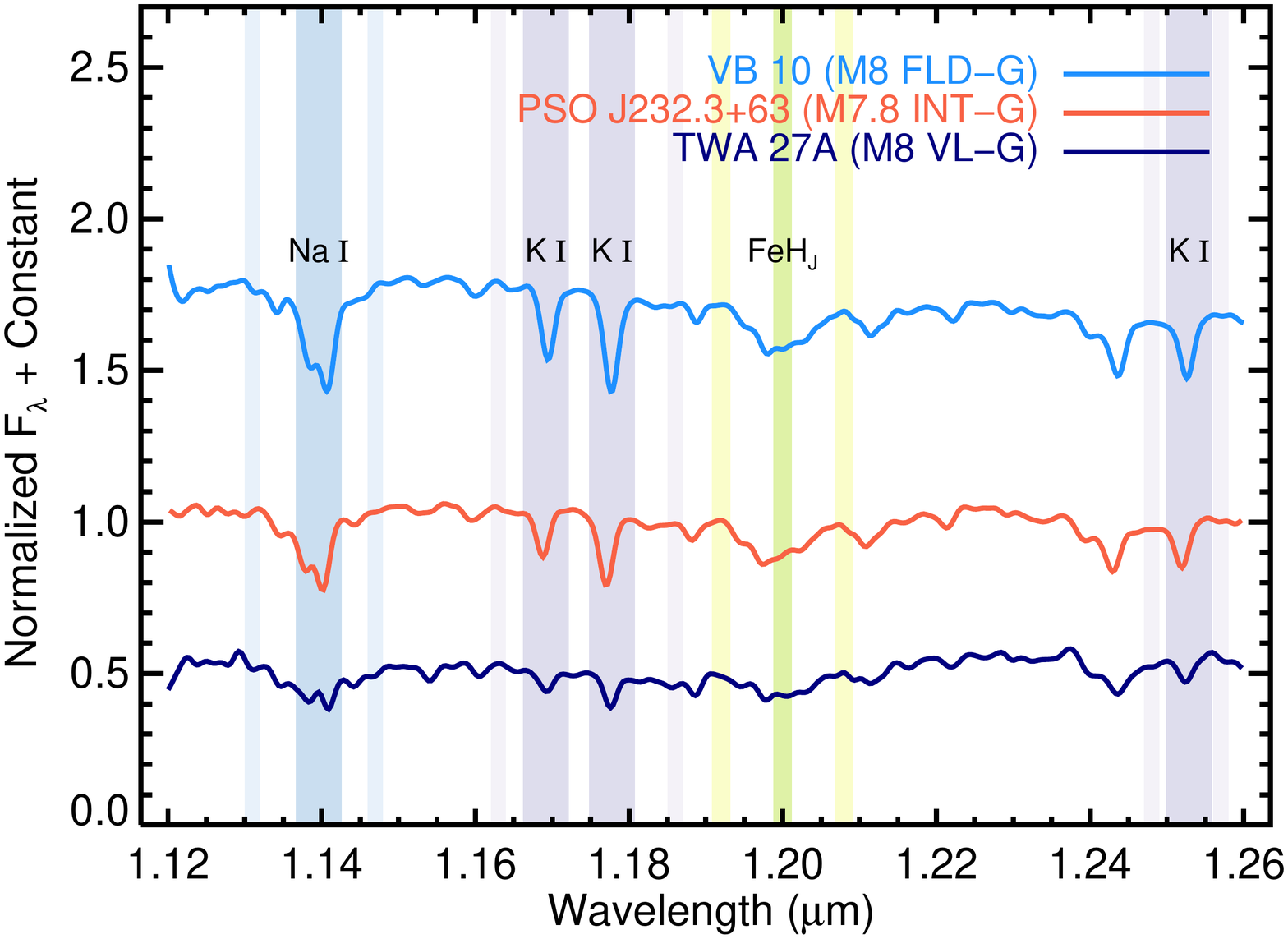}
    \caption{NIR moderate-resolution (R$\sim$750) spectra from
      IRTF/SpeX of one of our AB~Dor Moving Group candidates,
      PSO~J232.2+63 (\emph{red}), compared with a {\sc fld-g}
      (\emph{light blue}) and {\sc vl-g} (\emph{dark blue}) dwarf of
      similar spectral type (within half a spectral type). We choose
      the comparison spectra as described in
      Figure~\ref{fig:sxd1}. \label{fig:sxd4}}
  \end{center}
\end{figure*}

\begin{figure*}
  \begin{center}
    \includegraphics[width=.7\textwidth]{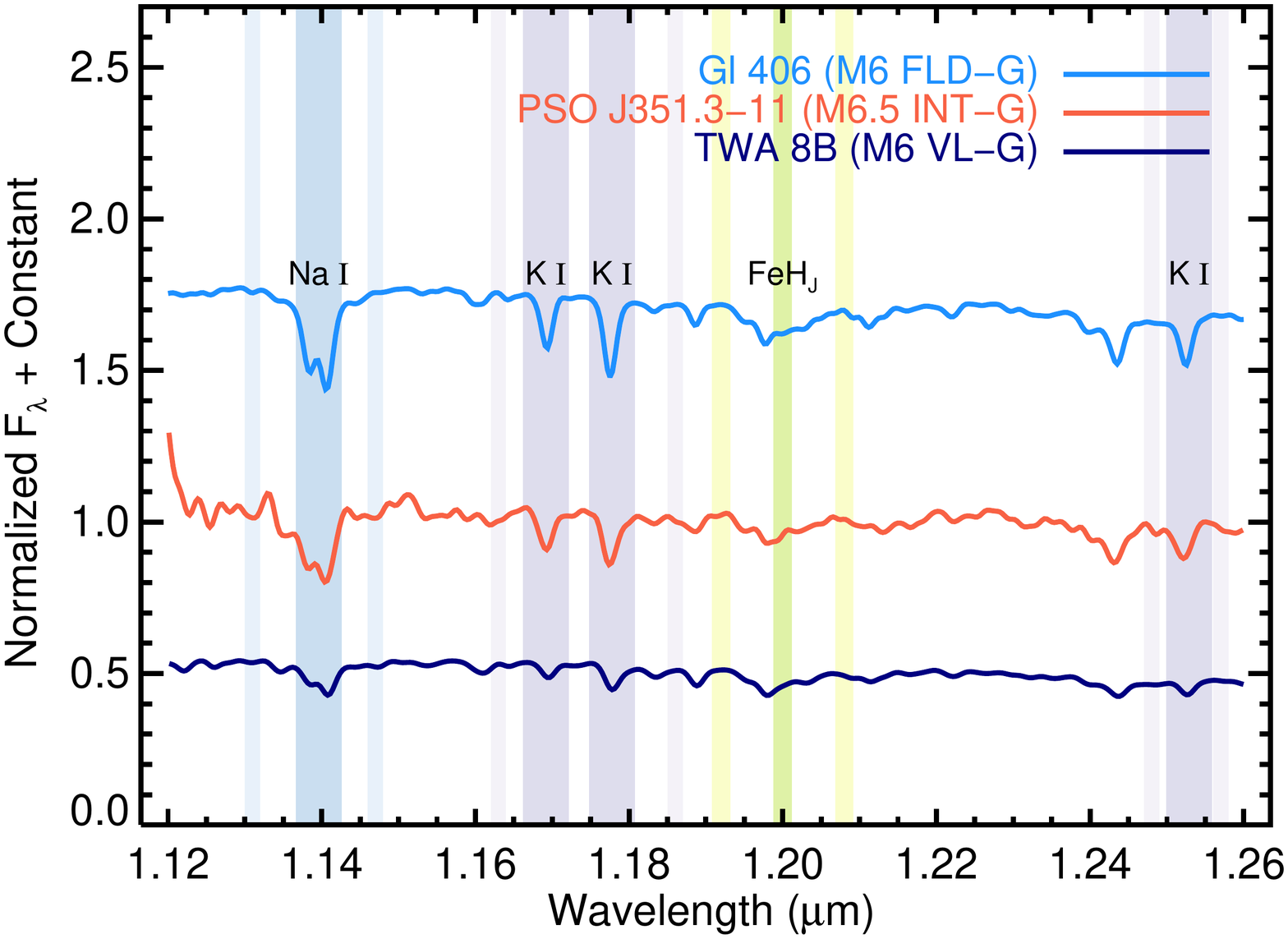}
    \caption{NIR moderate-resolution (R$\sim$750) spectra from
      IRTF/SpeX of one of our AB~Dor Moving Group candidates,
      PSO~J351.3$-$11 (\emph{red}), compared with a {\sc fld-g}
      (\emph{light blue}) and {\sc vl-g} (\emph{dark blue}) dwarf of
      similar spectral type (within half a spectral type). We choose
      the comparison spectra as described in
      Figure~\ref{fig:sxd1}. \label{fig:sxd5}}
  \end{center}
\end{figure*}

\begin{figure*}
  \vspace{-1 cm}
  \begin{center}
    \includegraphics[width=.7\textwidth]{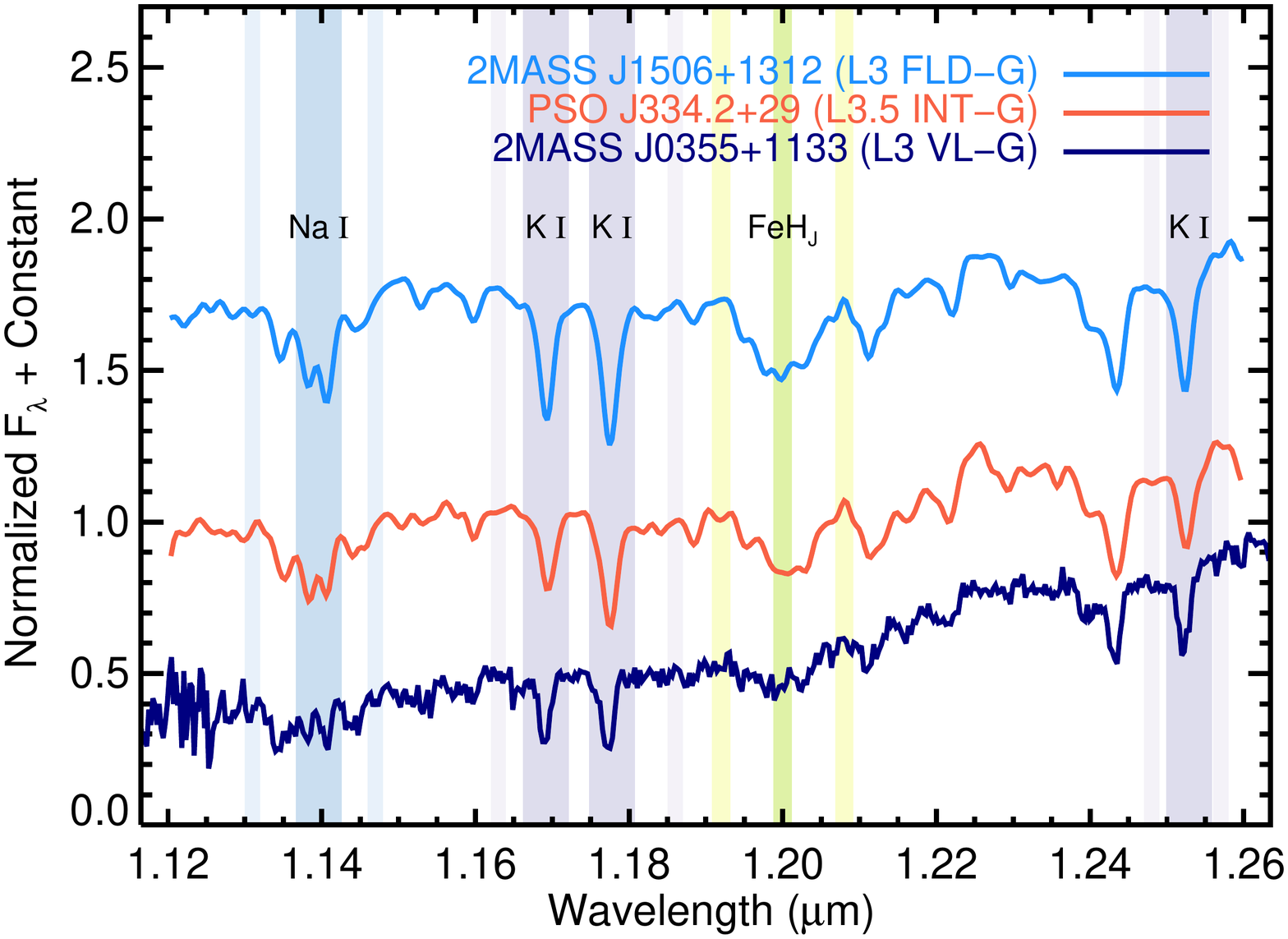}
    \caption{NIR moderate-resolution (R$\sim$1700) spectra from {\it
        Gemini}/GNIRS of one of our AB~Dor Moving Group candidates,
      PSO~J334.2+28 (\emph{red}), compared with a {\sc fld-g}
      (\emph{light blue}) and {\sc vl-g} (\emph{dark blue}) dwarf of
      similar spectral type (within half a spectral type). We choose
      the comparison spectra as described in
      Figure~\ref{fig:sxd1}. \label{fig:sxd6}}
  \end{center}
\end{figure*}

\begin{figure*}
  \begin{center}
    \includegraphics[width=.45\textwidth]{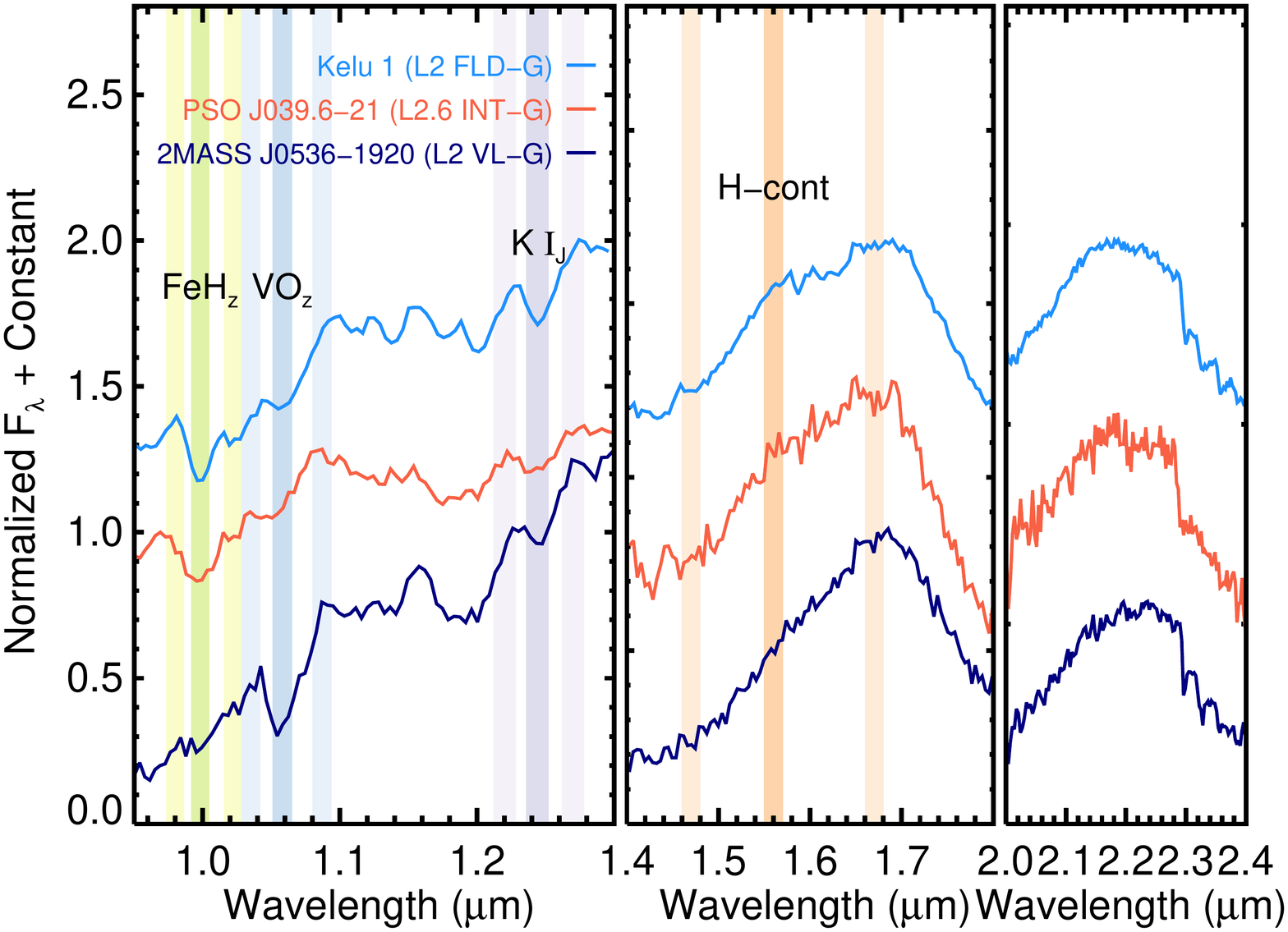}
    \includegraphics[width=.45\textwidth]{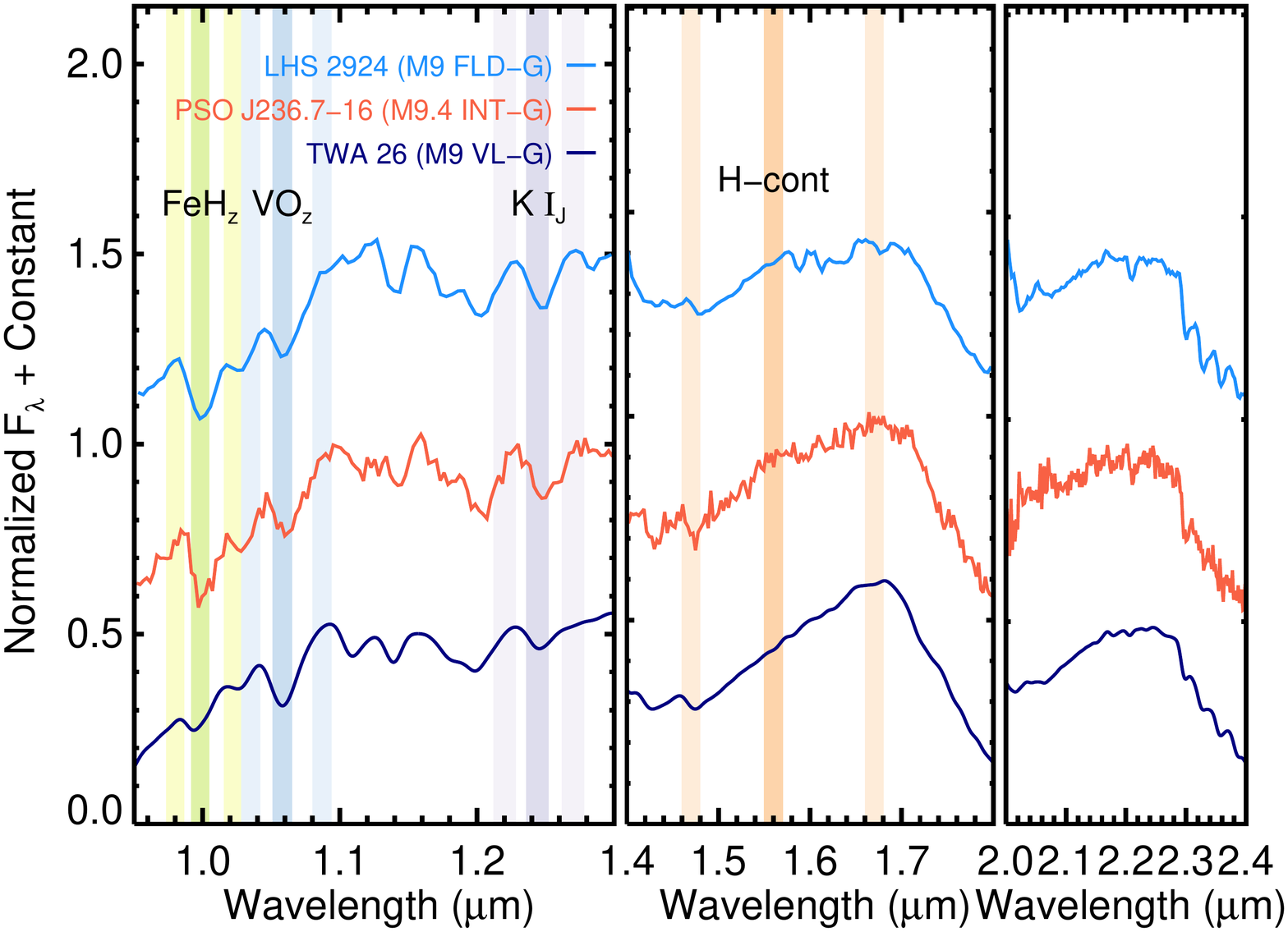}
    \includegraphics[width=.45\textwidth]{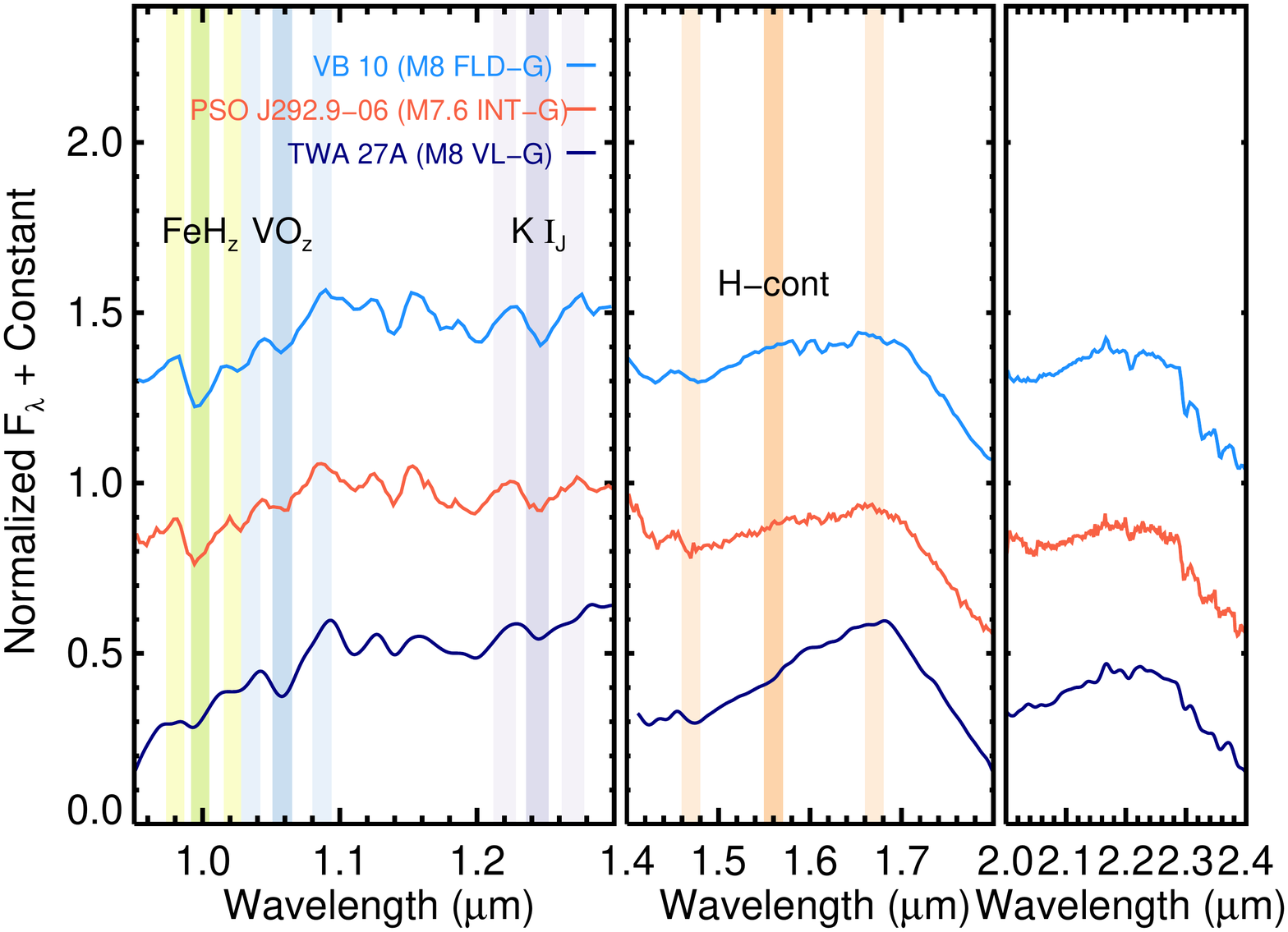}
    \includegraphics[width=.45\textwidth]{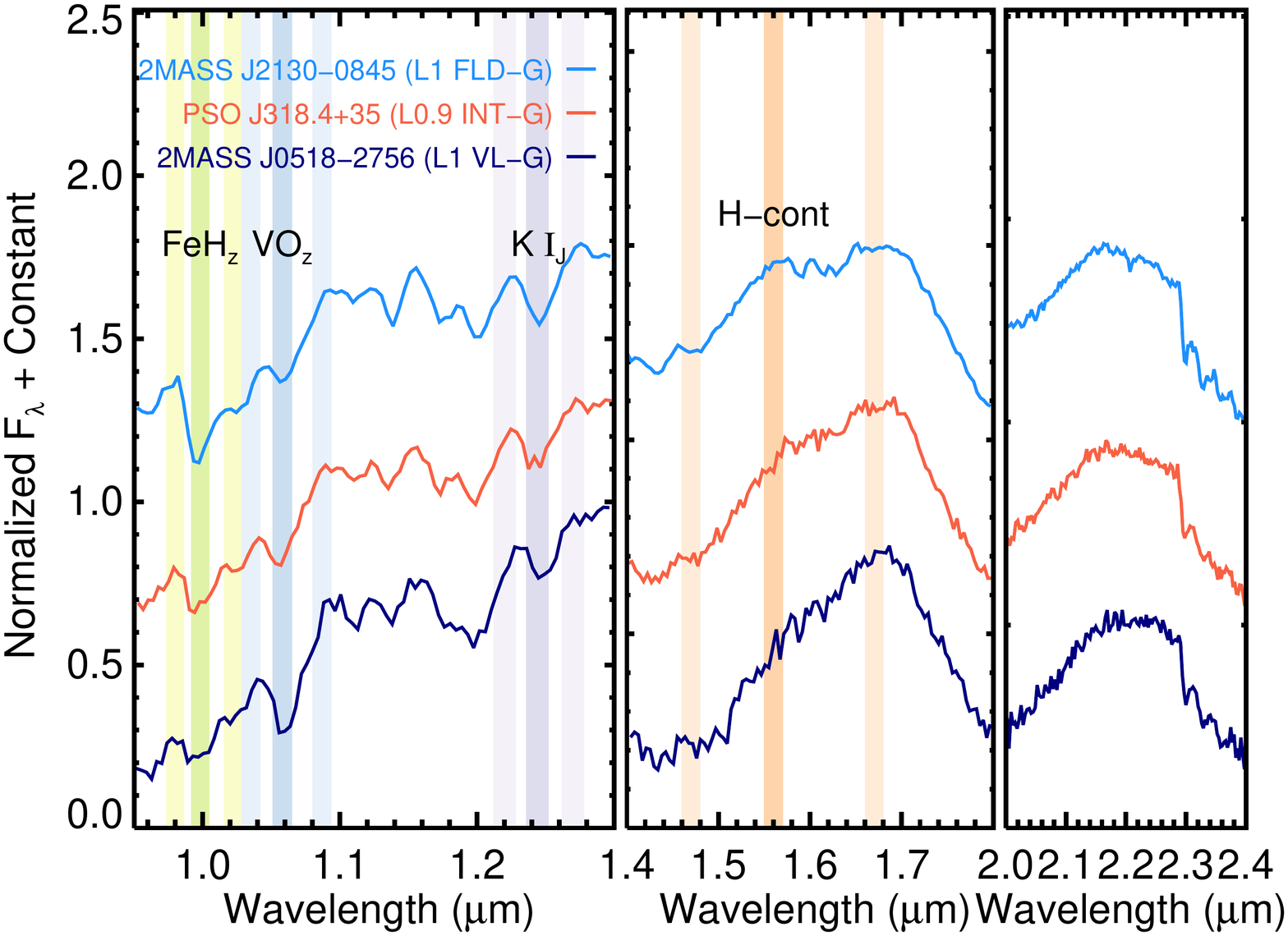}
    \caption{NIR low-resolution (R$\sim$130) spectra from IRTF/SpeX of
      our candidates with overall gravity classifications of {\sc
        int-g} compared with a {\sc fld-g} (\emph{light blue}) and a
      young, very low gravity (\emph{dark blue}) dwarf of similar
      spectral type (within half a spectral type). We used field
      spectral standards from \citet{2010ApJS..190..100K} if there was
      available low-resolution or moderate-resolution spectra. Young
      dwarfs were taken from the list of standard {\sc vl-g} objects
      in \citet{2013ApJ...772...79A} with publicly available
      spectra. All comparison spectra have been smoothed to
      R$\sim$130. Gravity-sensitive features from
      \citet{2013ApJ...772...79A} are labeled and the wavelength
      ranges used to calculate the gravity indices are highlighted for
      FeH$_{z}$ (\emph{yellow-green}), VO$_{z}$ (\emph{blue}),
      KI$_{J}$ (\emph{purple}), and H-cont
      (\emph{orange}). \label{fig:prism}}
\end{center}
\end{figure*}

\begin{figure*}
  \begin{center}
    \includegraphics[width=.45\textwidth]{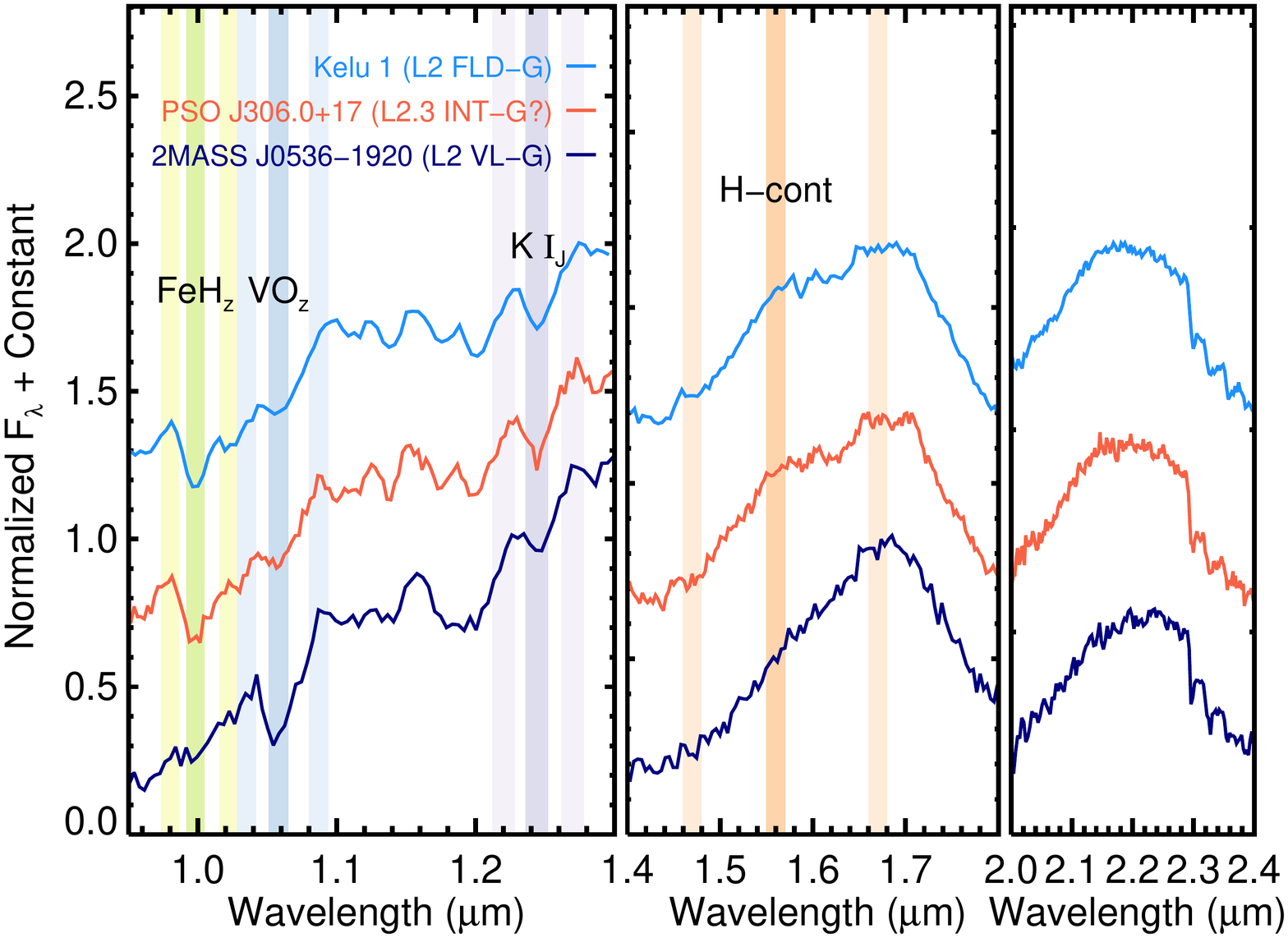}
    \includegraphics[width=.45\textwidth]{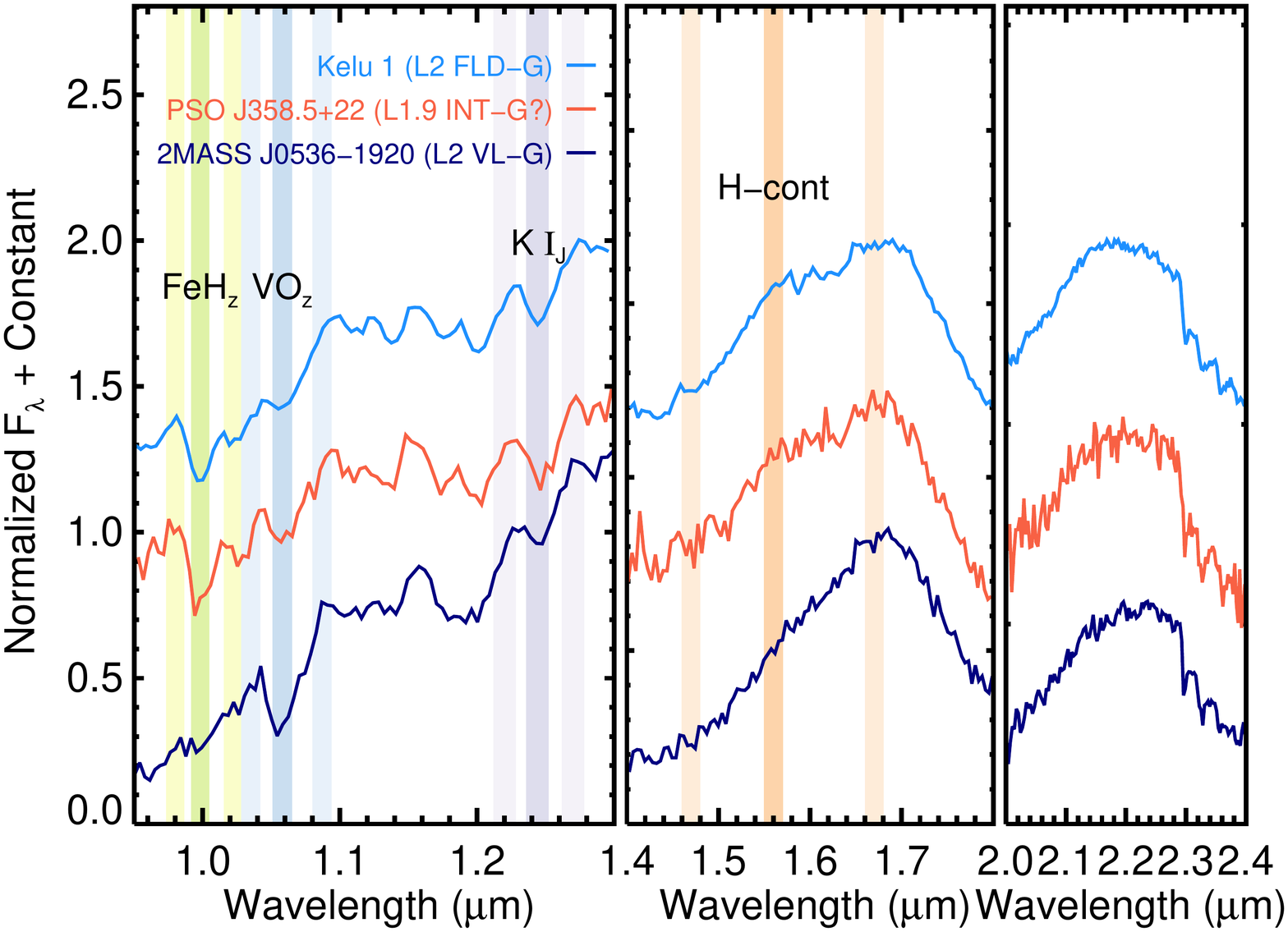}    
    \caption{NIR low-resolution (R$\sim$130) spectra from IRTF/SpeX of
      our candidates with overall gravity classifications of {\sc
        int-g?} compared with a {\sc fld-g} (\emph{light blue}) and a
      young, very low gravity (\emph{dark blue}) dwarf of similar
      spectral type (within half a spectral type). All comparison
      spectra have been chosen as described in
      Figure~\ref{fig:prismind} and smoothed to
      R$\sim$130. Gravity-sensitive features from
      \citep{2013ApJ...772...79A} are labeled and the wavelength
      ranges used to calculate the gravity indices are highlighted for
      FeH$_{z}$ (\emph{yellow-green}), VO$_{z}$ (\emph{blue}),
      KI$_{J}$ (\emph{purple}), and H-cont
      (\emph{orange}). \label{fig:prism2}}
  \end{center}
\end{figure*}

Figures~\ref{fig:prismind} and \ref{fig:sxdind} compare the index
values for each of our objects with the {\sc fld-g}, {\sc int-g}, and
{\sc vl-g} regions. In addition, as a visual check on the overall
gravity classification, we compare our spectra to known old field
dwarfs ({\sc fld-g}) and young dwarfs ({\sc vl-g}) in
Figures~\ref{fig:sxd1}--\ref{fig:prism2}. Table~\ref{table:gravity}
tabulates the gravity indices and classification for our objects.

\subsection{PS1 Parallaxes, Photometric Distances and Absolute Magnitudes \label{sec:dphot}}

PS1 has observed the field of each target repeatedly over several
years, allowing us to measure the parallax and proper motions of our
targets using relative astrometry techniques. We use the image
calibrations calculated by the standard Pan-STARRS analysis, but we
re-fit the parallax and proper motions for each object with our own
re-analysis of these calibrated coordinates (Magnier et al. 2016, in
preparation).

The standard astrometric analysis of the PS1 images uses a set of
low-order polynomial transformations to correct for the distortion
introduced by the optical system and the atmosphere, along with
individual corrections for each of the 60 CCDs.  The camera-level
polynomials are of the form $\sum a_{i,j} x^iy^j$ with $i + j \leq
3$. The individual chip corrections consist of a linear transformation
(to account for the chip location and rotation) plus a grid of fine
corrections across the chip, with up to 6$\times$6 correction cells
per chip. The astrometric transformations are determined by an
iteratively calculation to minimize the scatter of the adopted
reference stars in the database. Mean per-epoch residuals for
moderately bright stars range from 10 to 25~mas depending on the
Galactic latitude: regions of higher stellar density allow for a
better correction.

The standard astrometry analysis also fits each star for proper motion
and parallax, but it does not currently use a sufficiently robust
outlier rejection scheme. We have specifically re-fitted the proper
motion and parallax for our targets with a more stringent rejection of
outliers. We first reject any detection with flags indicating failures
in the photometry analysis as well as any detections with insufficient
coverage of unmasked pixels ($psf\_qf < 0.85$). We use 100 bootstrap
resamples of the dataset to measure the parallax and proper motion of
the object. For each of these samples we then measure the distance of
each point from the fitted path on the sky, scaled by the position
errors. Detections which are more than 5$\sigma$ from the path in more
than 50\% of the samples are marked as outliers and excluded from the
final fit. We also use 1000 bootstrap resample tests of the remaining
points to determine the errors on the fitted parameters.

\begin{figure*}
  \begin{center}
    \includegraphics[width=0.45\textwidth]{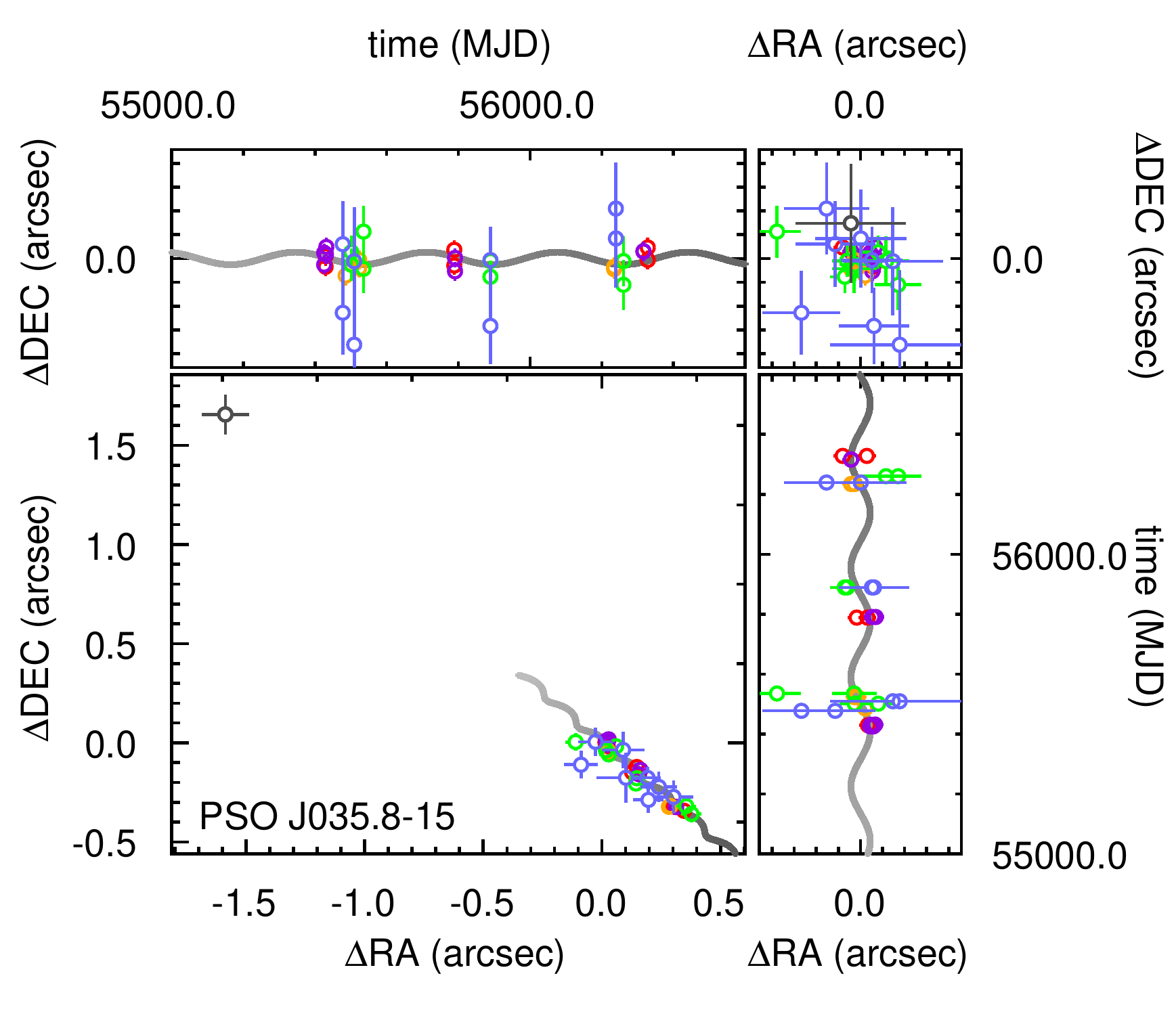}
    \includegraphics[width=0.45\textwidth]{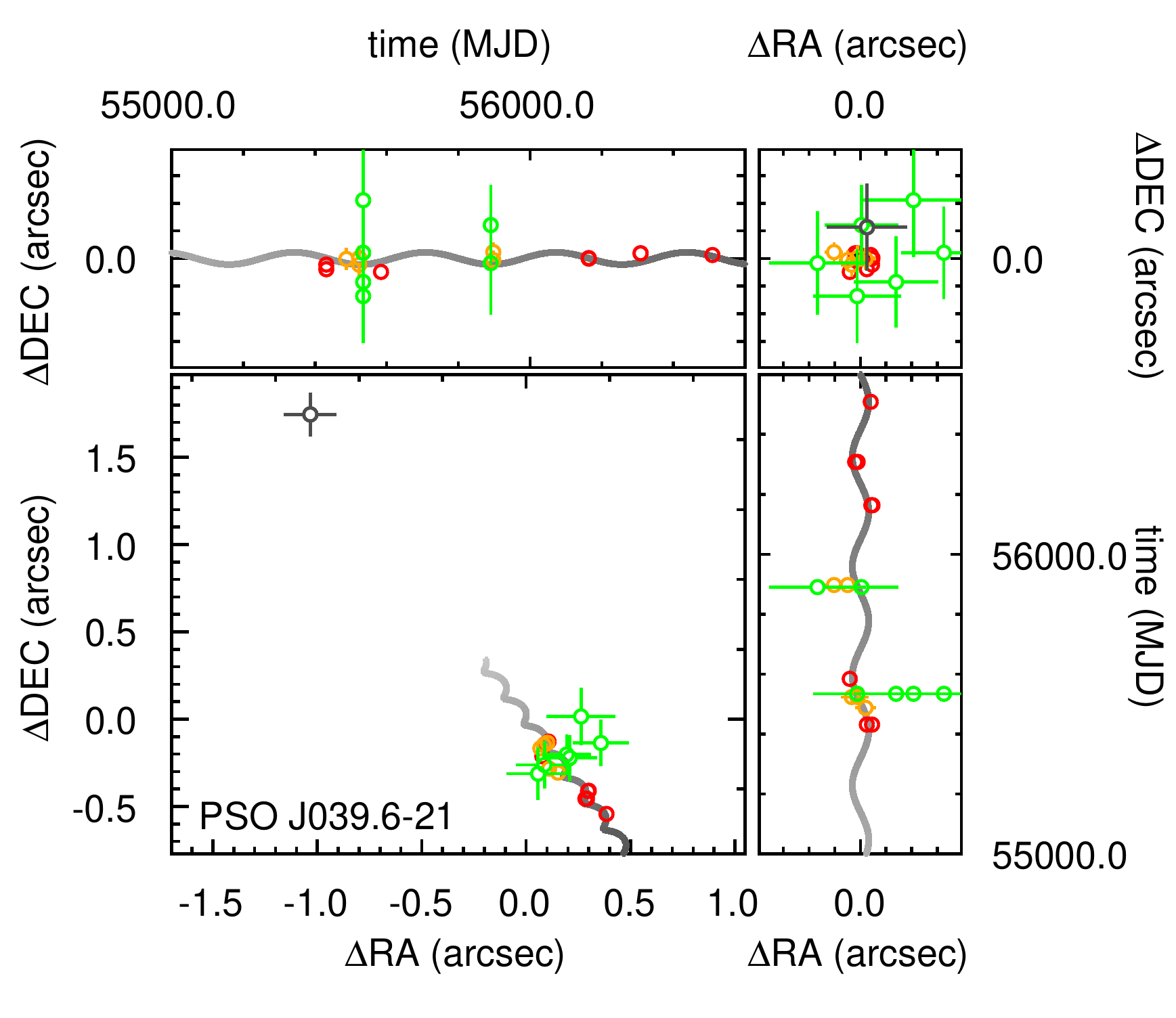}
    \includegraphics[width=0.45\textwidth]{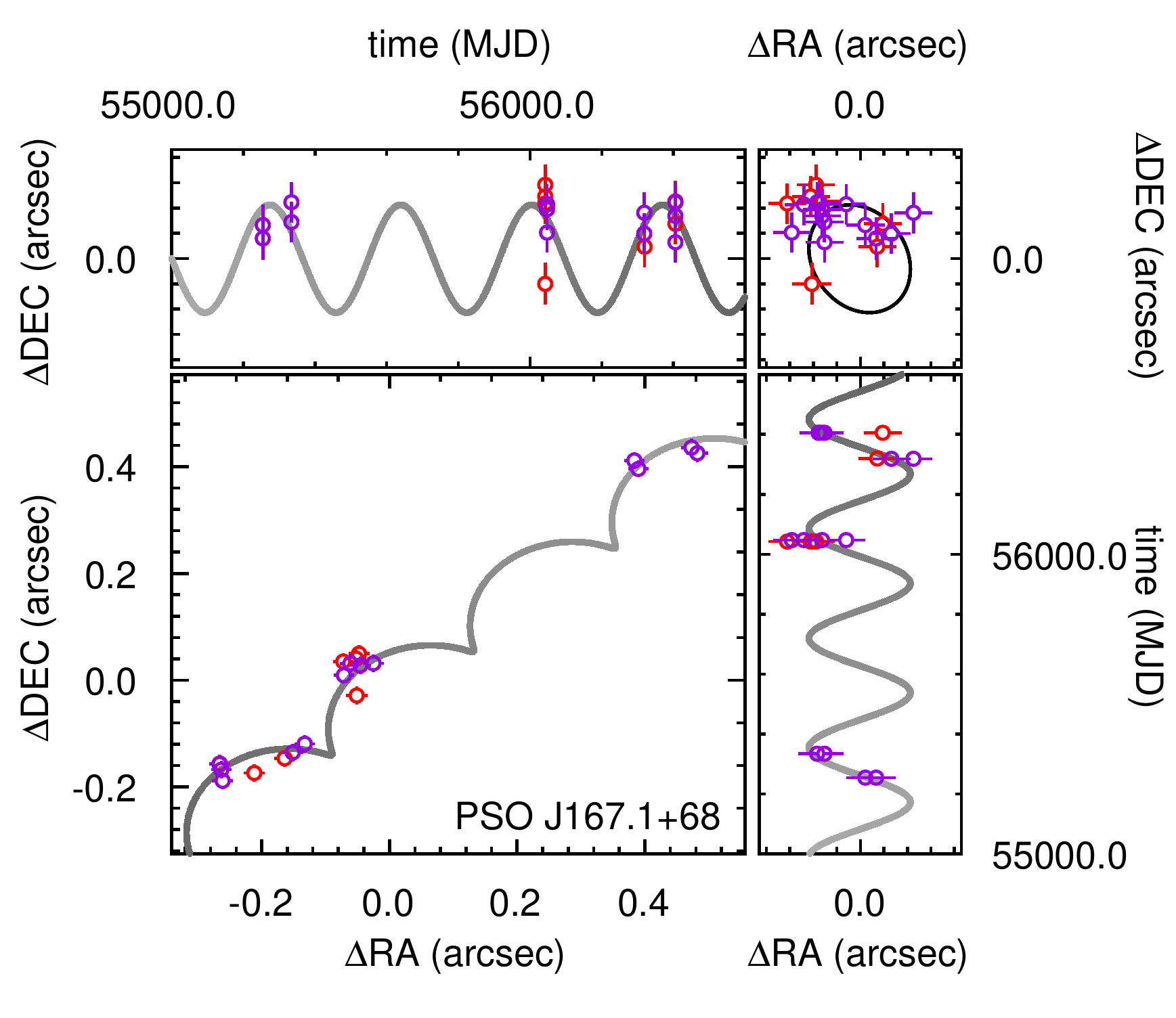}
    \includegraphics[width=0.45\textwidth]{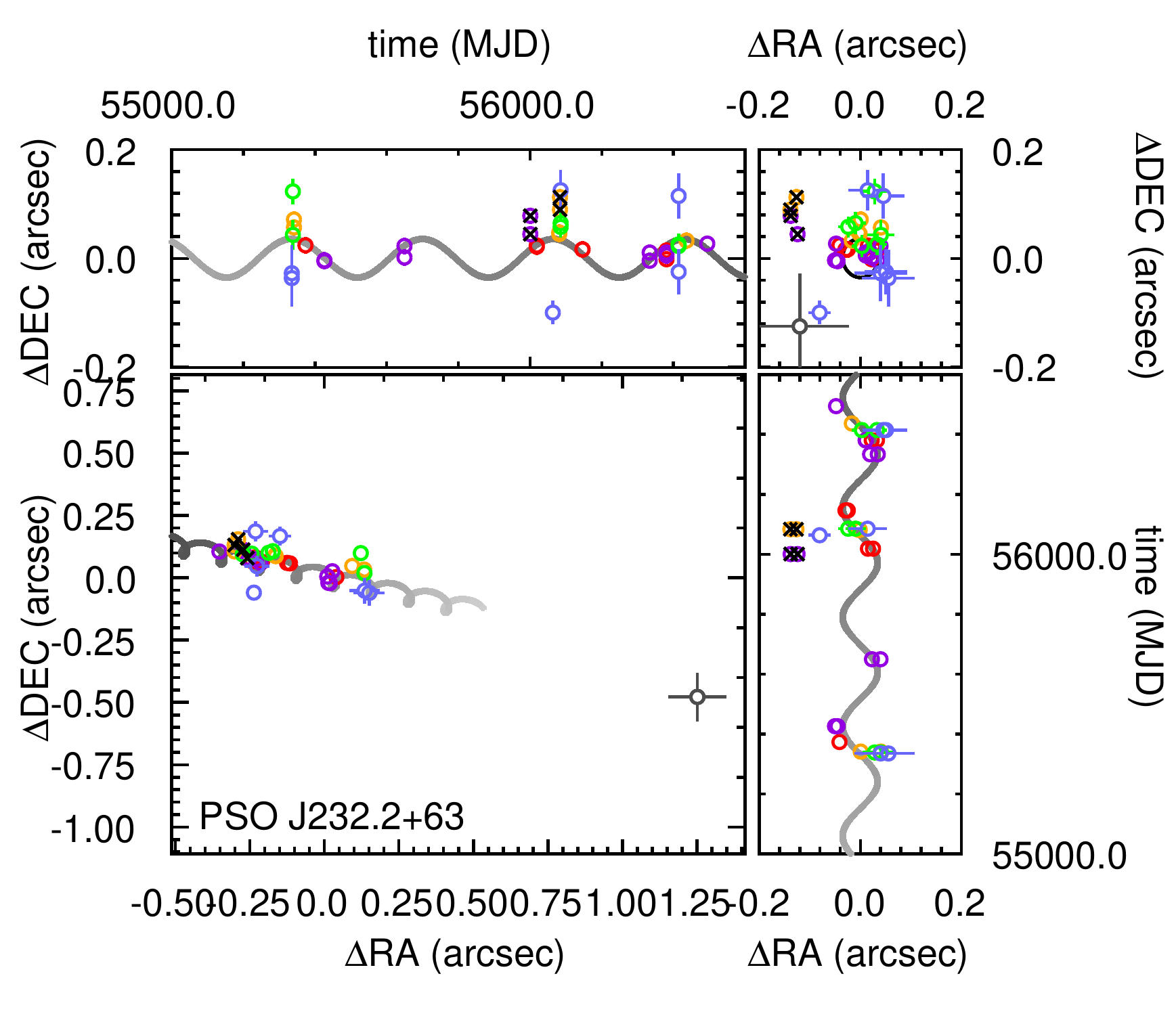}
    \caption{PS1 parallax motion and the motion in both R.A. and
      decl. for PSO~J035.8$-$15 (\emph{top left}), PSO~J039.6$-$21
      (\emph{top right}), PSO~J167.1+68 (\emph{bottom left}), and
      PSO~J232.3$-$63 (\emph{bottom right}). The different color
      symbols correspond to the filter used to determine the
      astrometric position for \gps~(\emph{blue}),
      \rps~(\emph{green}), \ips~(\emph{orange}), \zps~(\emph{purple}),
      \yps~(\emph{red}), and 2MASS (\emph{gray}). The \emph{x} marks
      denote points rejected as outliers during the astrometric
      fit. The \emph{thick gray line} denotes the best fit where the
      object is moving from the \emph{light--dark gray} over
      time. \label{fig:plxfit1}}
  \end{center}
\end{figure*}

\begin{figure*}
  \begin{center}
    \includegraphics[width=0.45\textwidth]{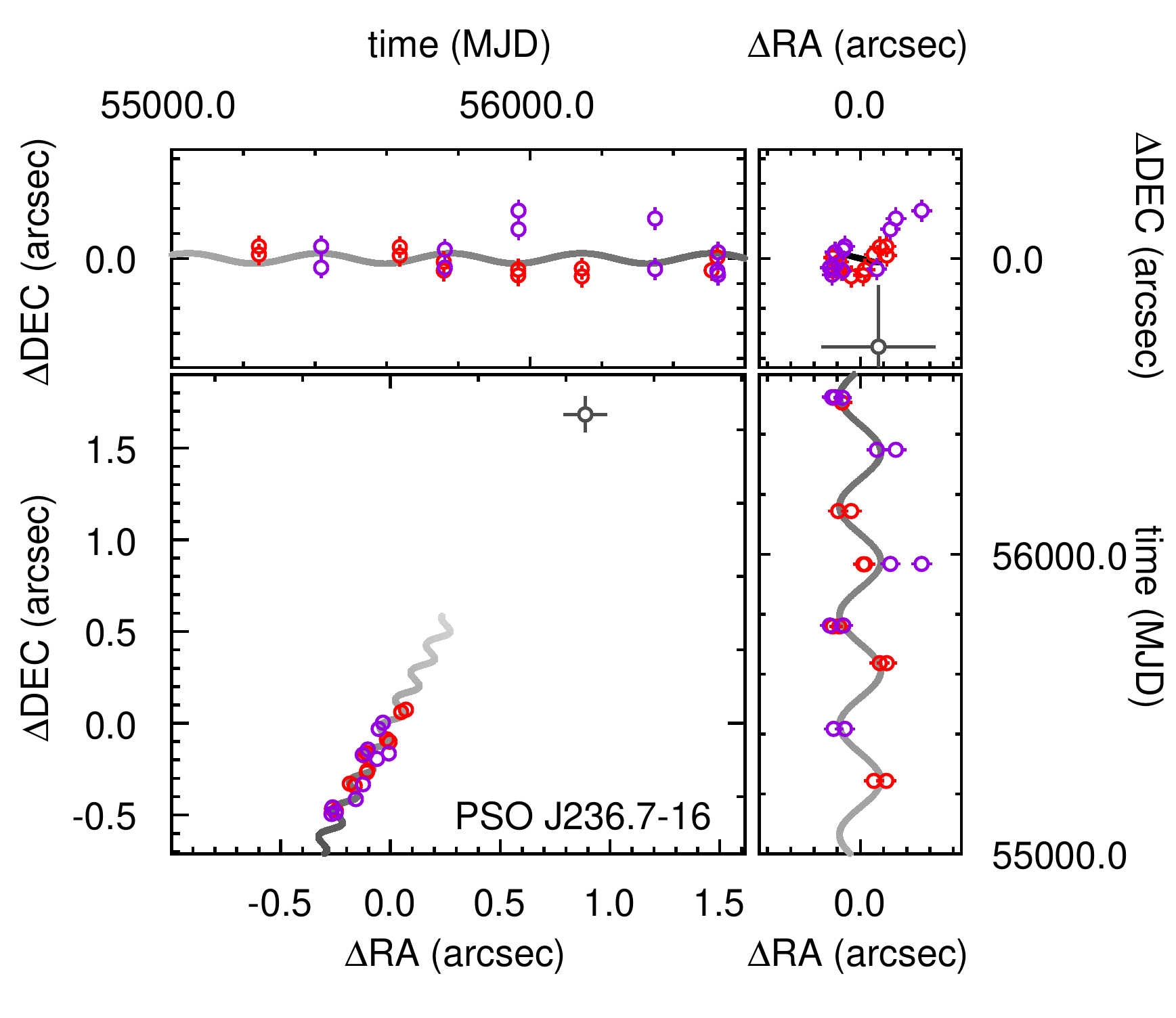}
    \includegraphics[width=0.45\textwidth]{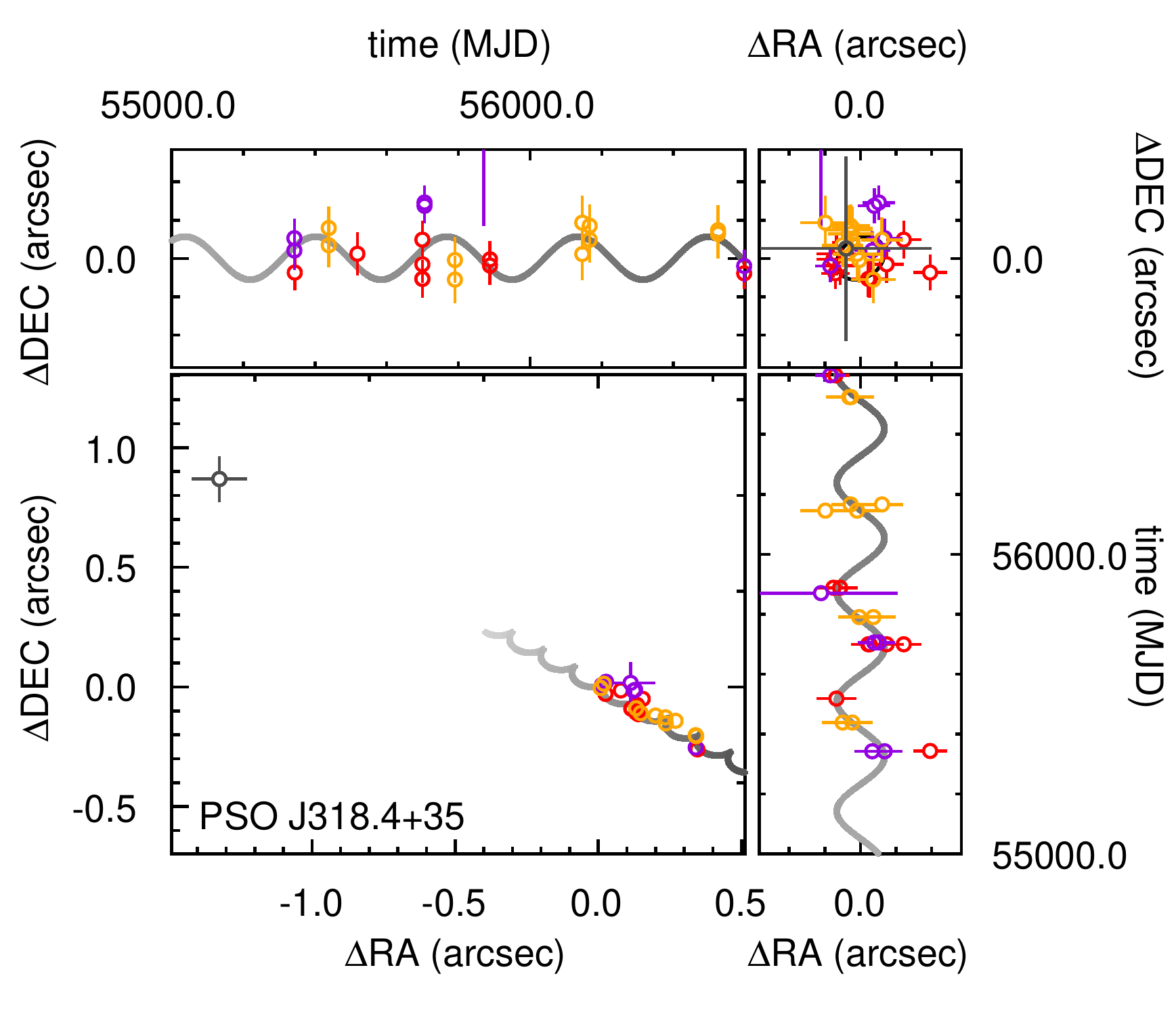}
    \includegraphics[width=0.45\textwidth]{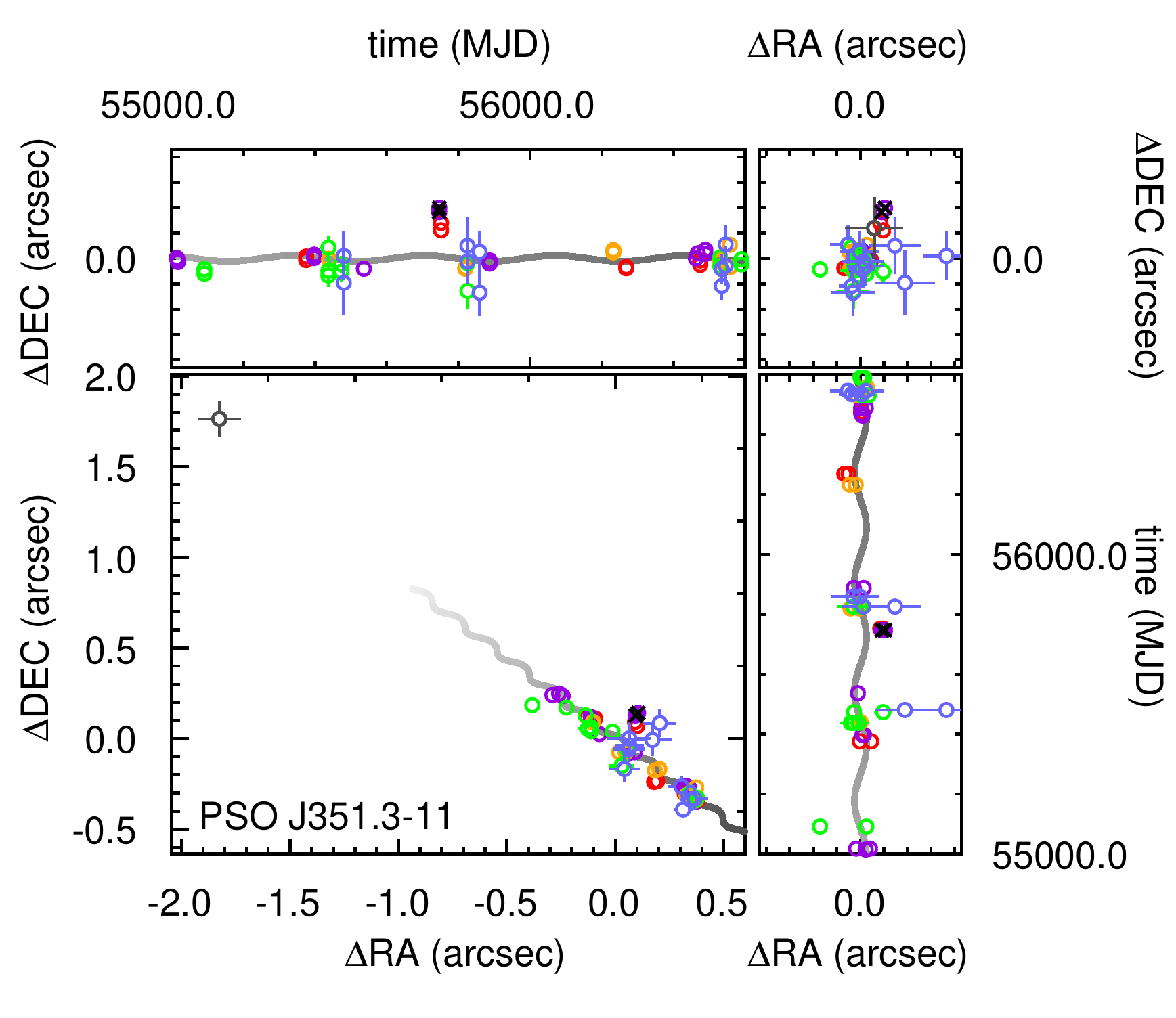}
    \includegraphics[width=0.45\textwidth]{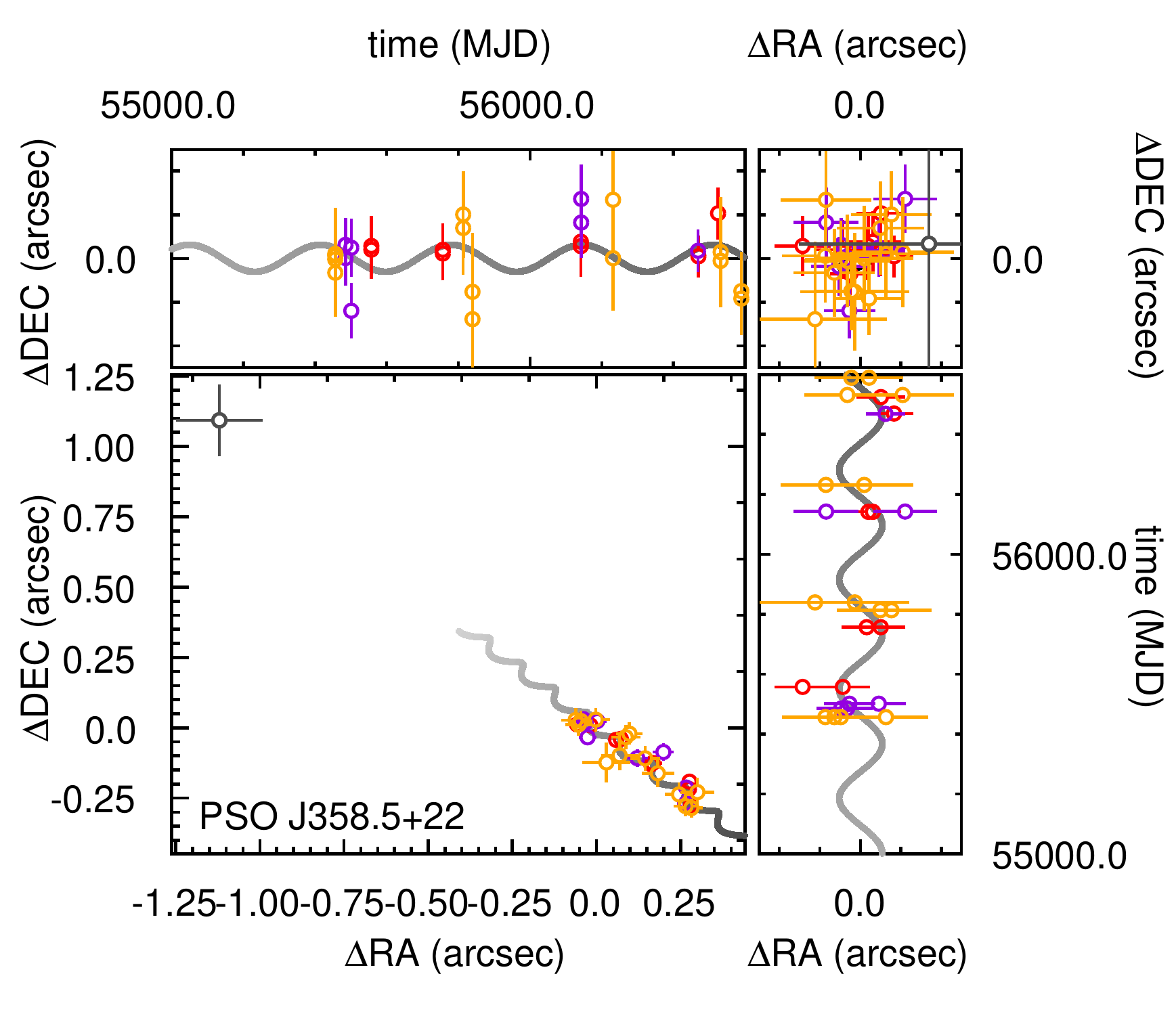}
    \includegraphics[width=0.45\textwidth]{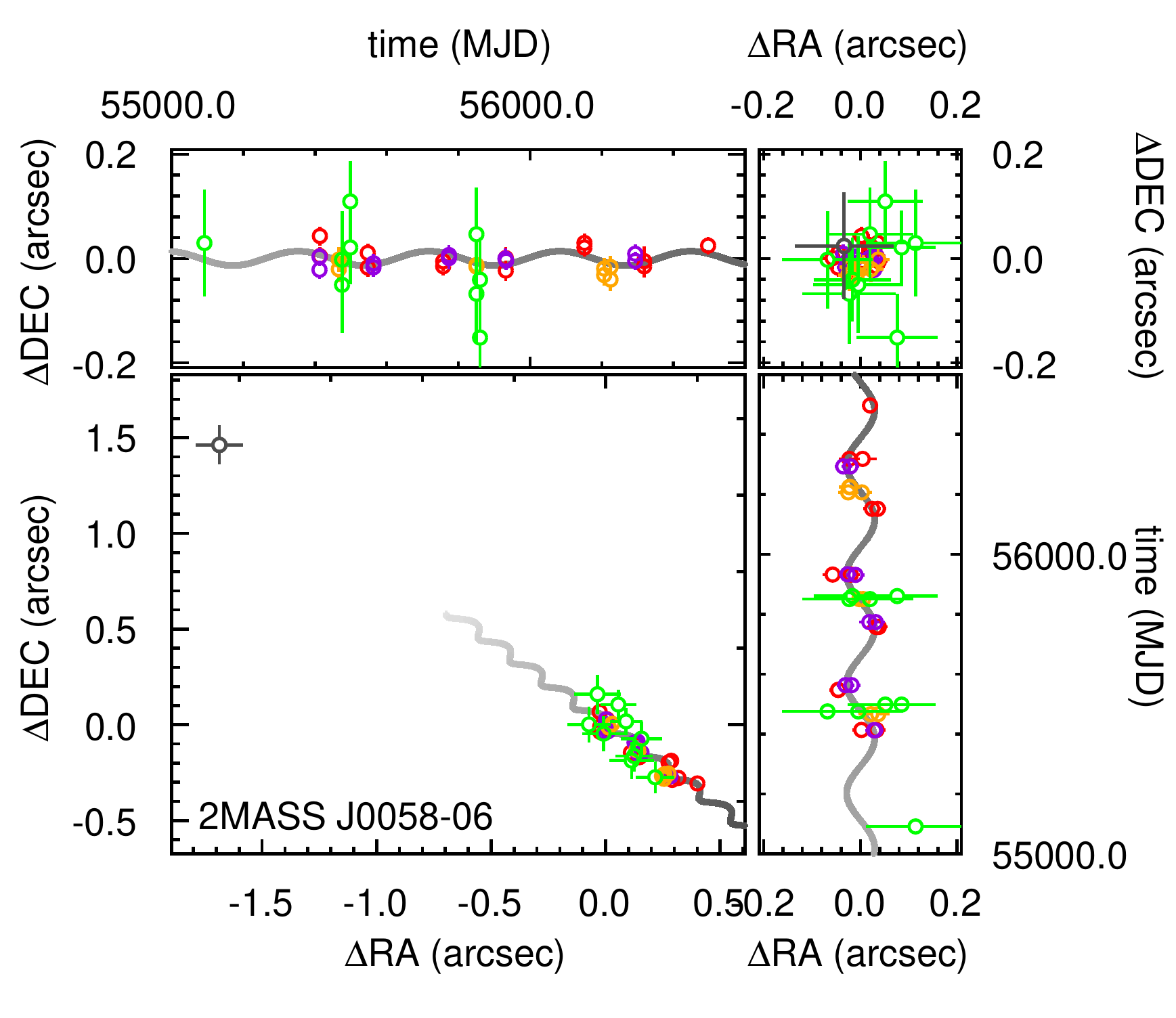}
    \caption{PS1 parallax motion and the motion in both R.A. and
      decl. for PSO~J236.8$-$16 (\emph{top left}), PSO~J318.4+35
      (\emph{top right}), PSO~J351.3$-$11 (\emph{middle left}),
      PSO~J358.5+22 (\emph{middle right}) and the
      previously-identified candidate member, 2MASS~J0058$-$06
      (\emph{bottom left}). The colors and symbols are the same as in
      Figure~\ref{fig:plxfit1} \label{fig:plxfit2}}
  \end{center}
\end{figure*}

\begin{figure*}  
  \begin{center}
    \includegraphics[width=0.3\textwidth]{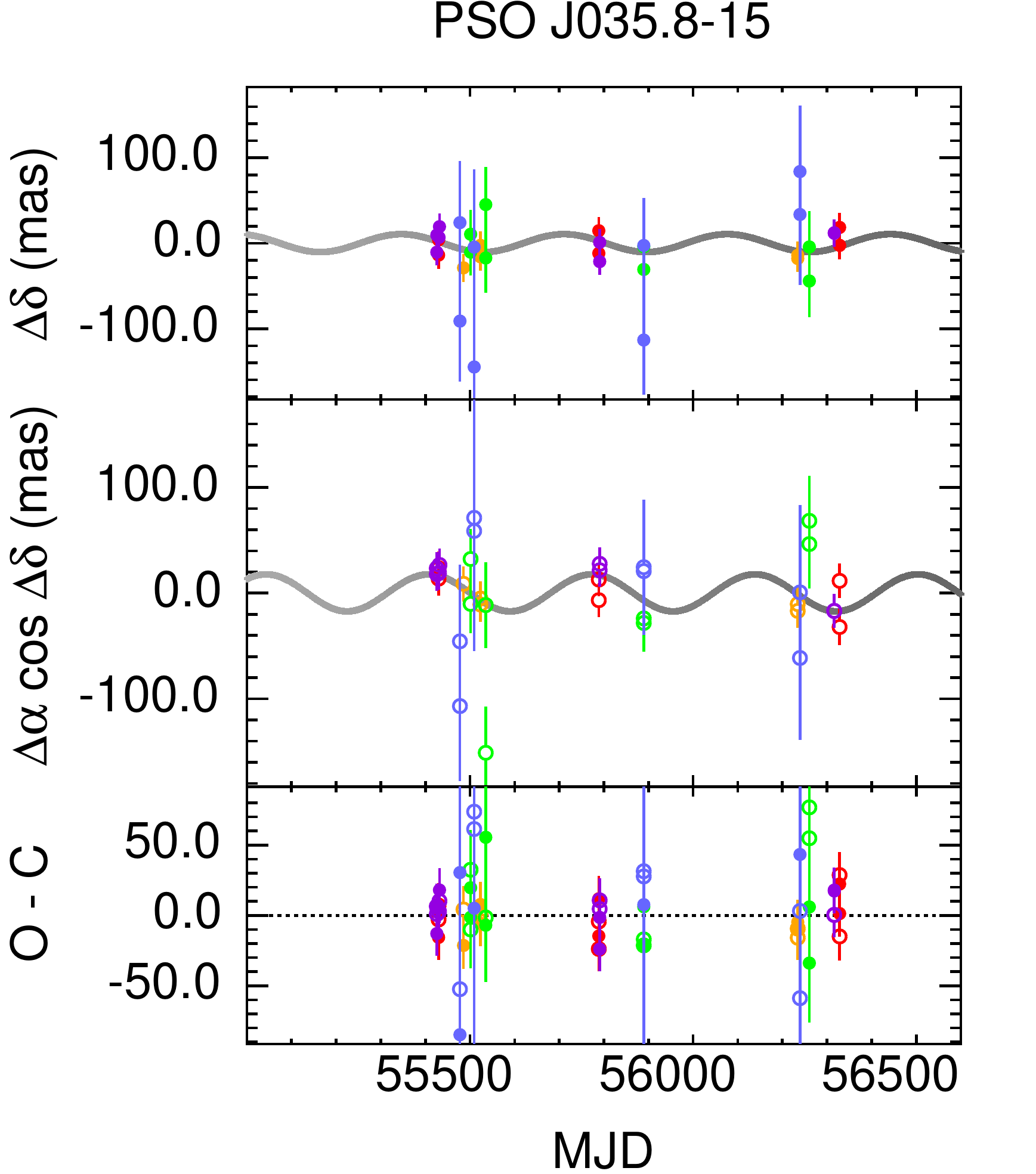}
    \includegraphics[width=0.3\textwidth]{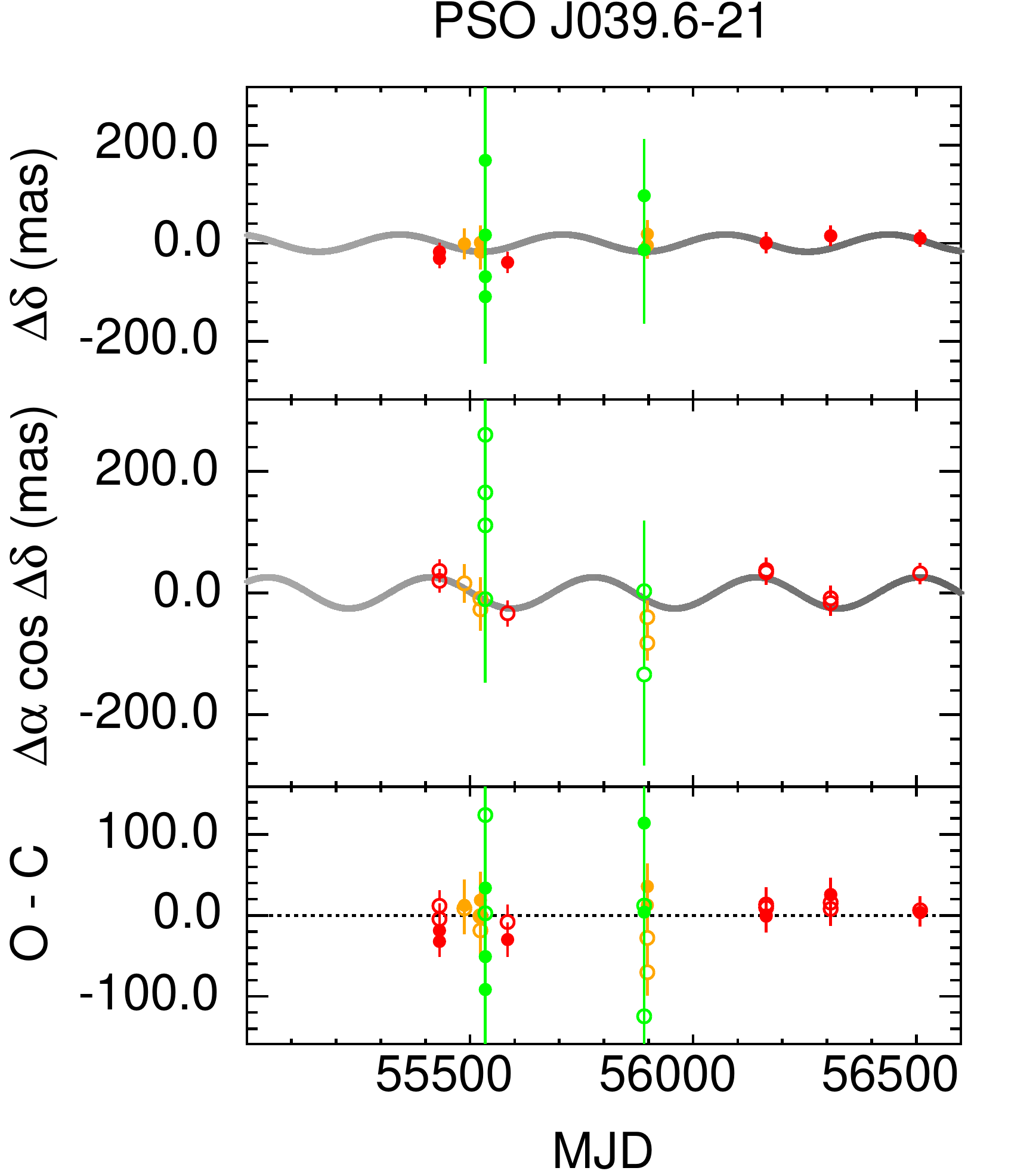}
    \includegraphics[width=0.3\textwidth]{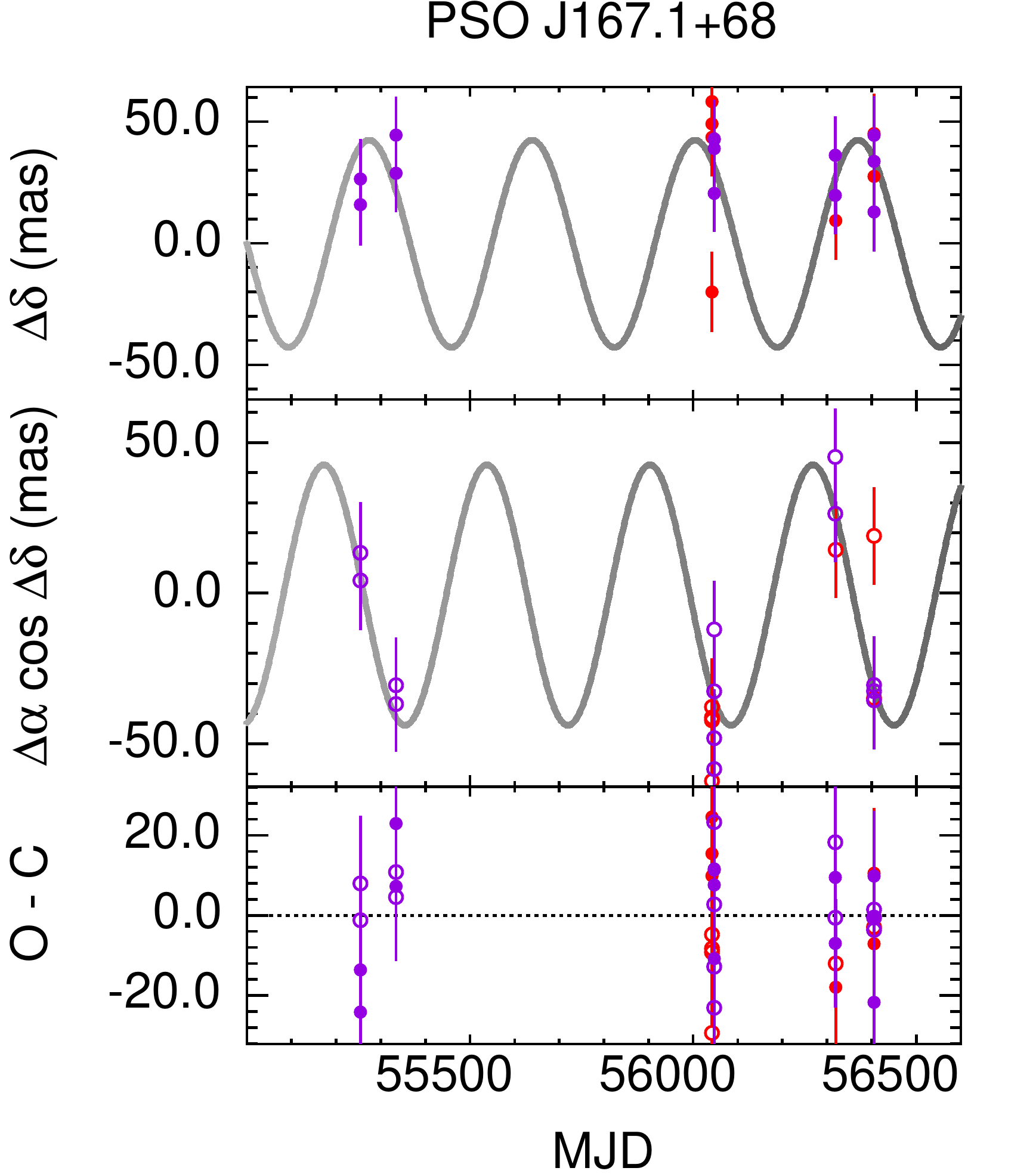}
    \includegraphics[width=0.3\textwidth]{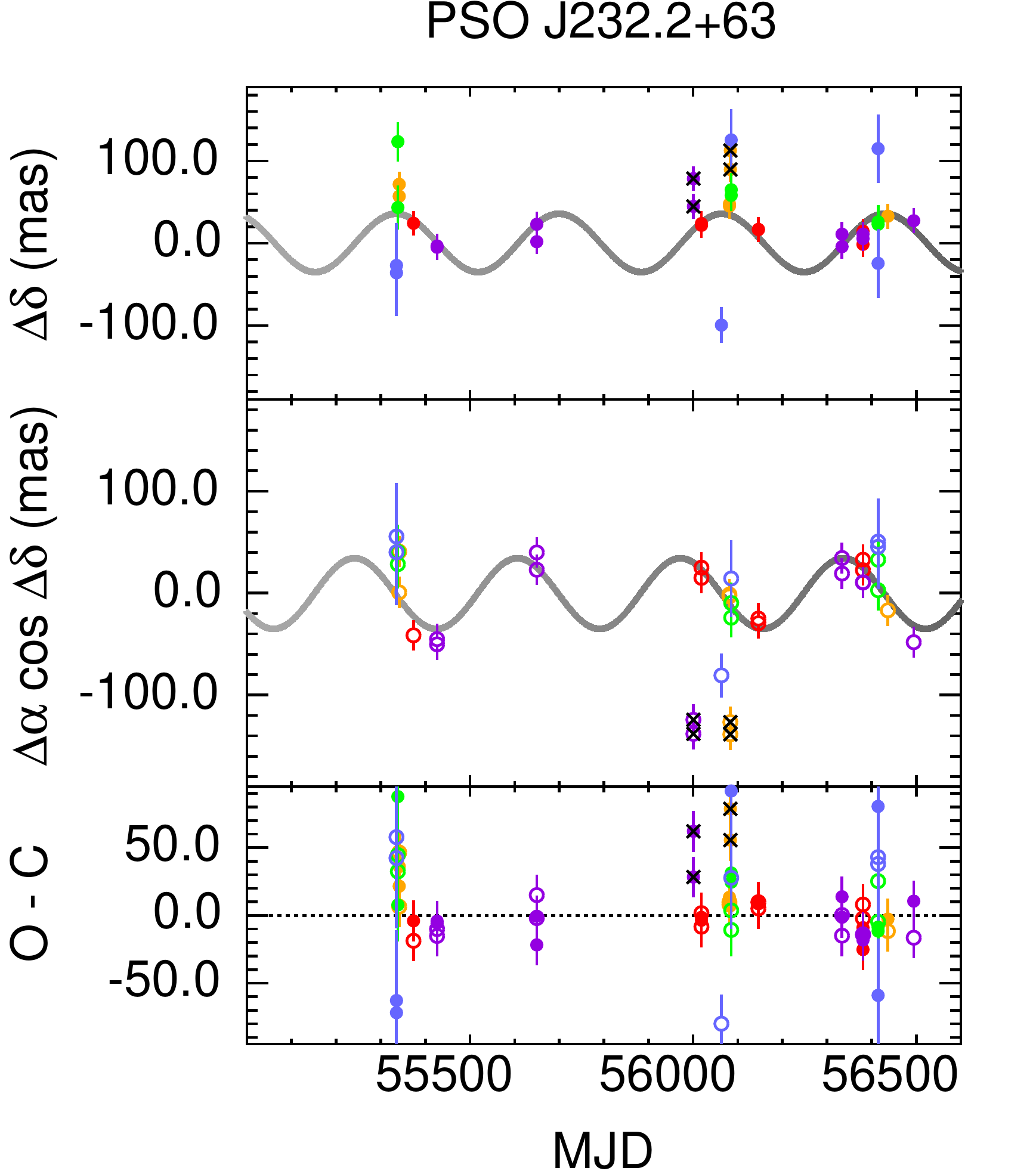}
    \includegraphics[width=0.3\textwidth]{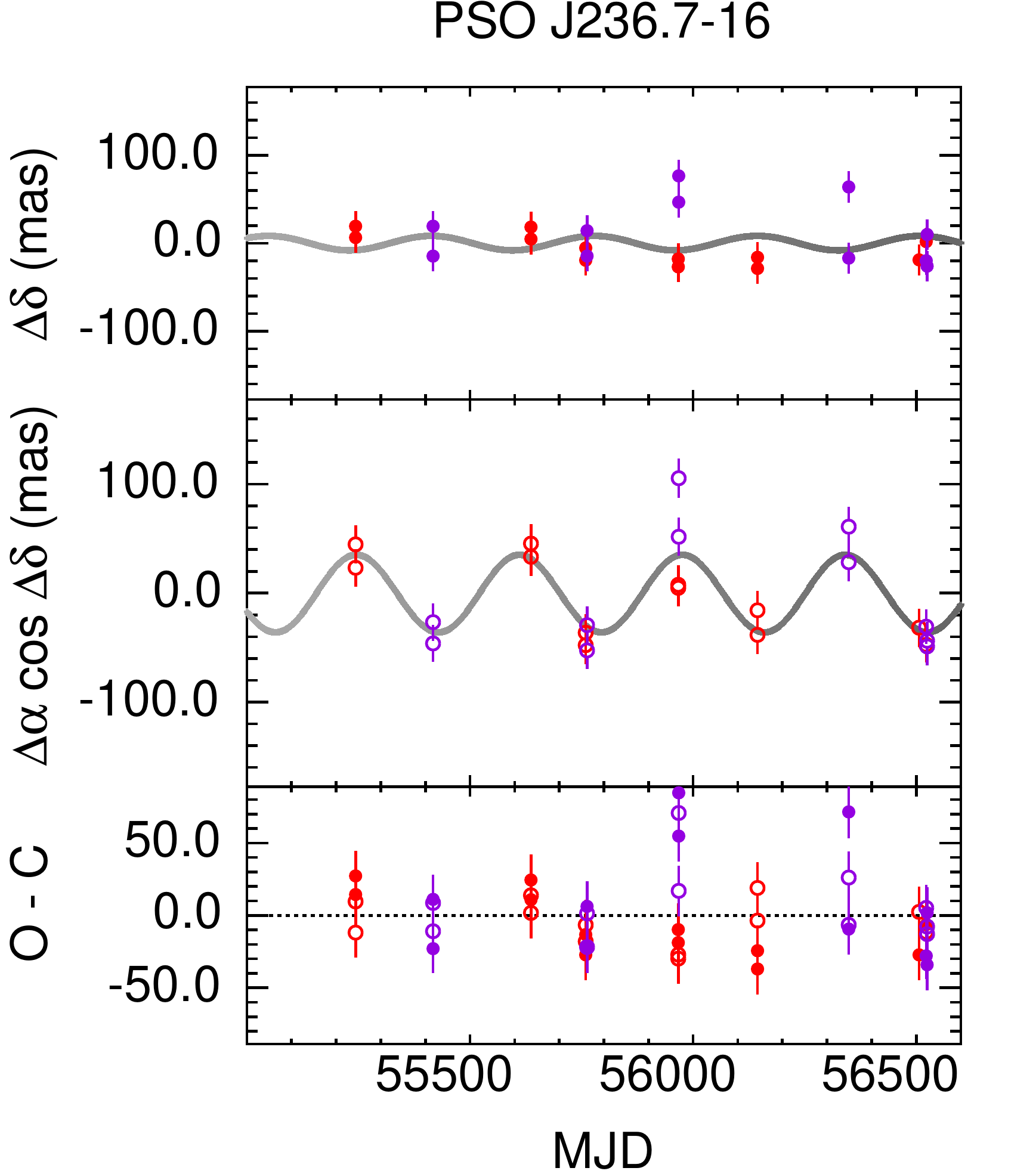}
    \includegraphics[width=0.3\textwidth]{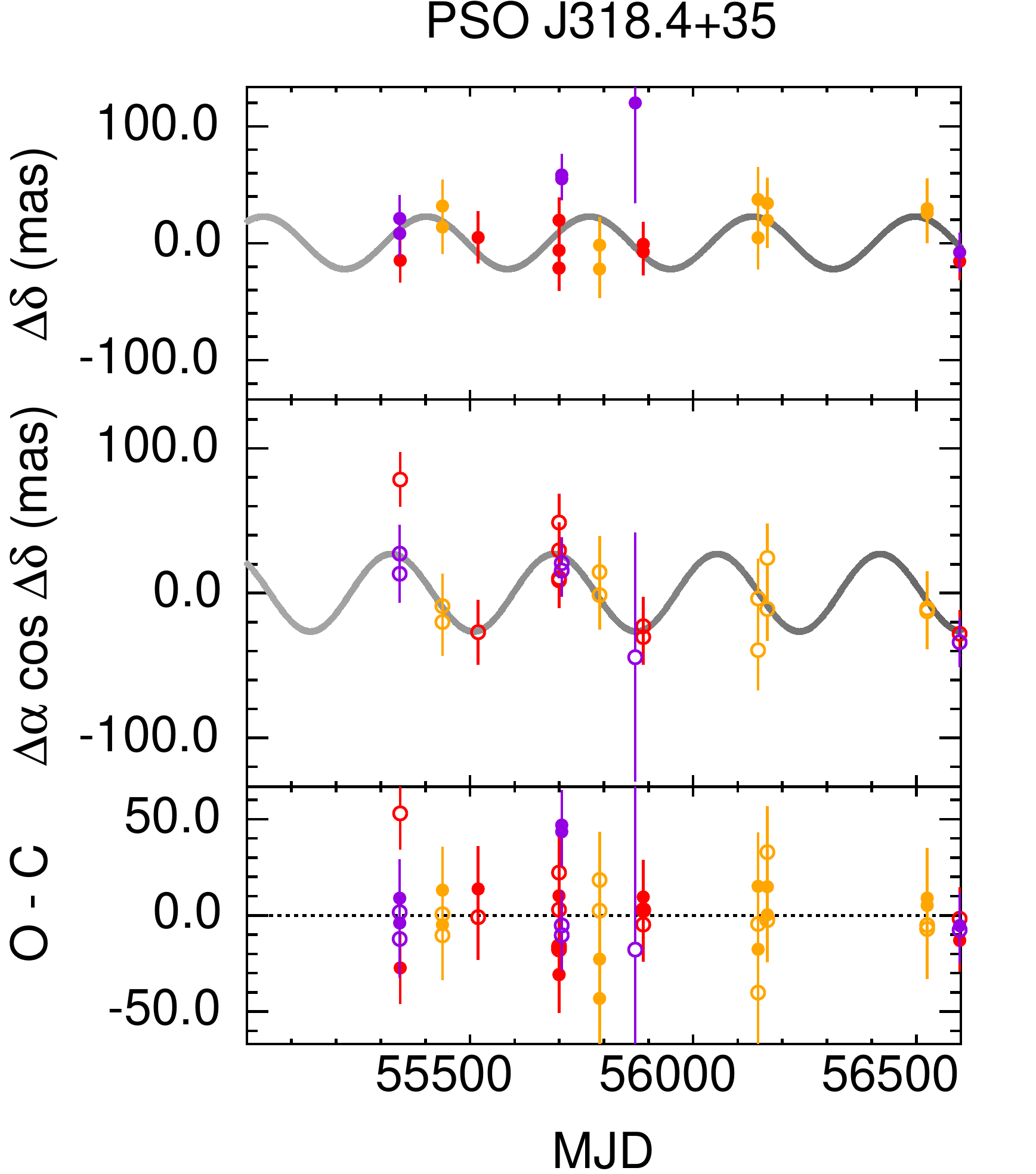}
    \includegraphics[width=0.3\textwidth]{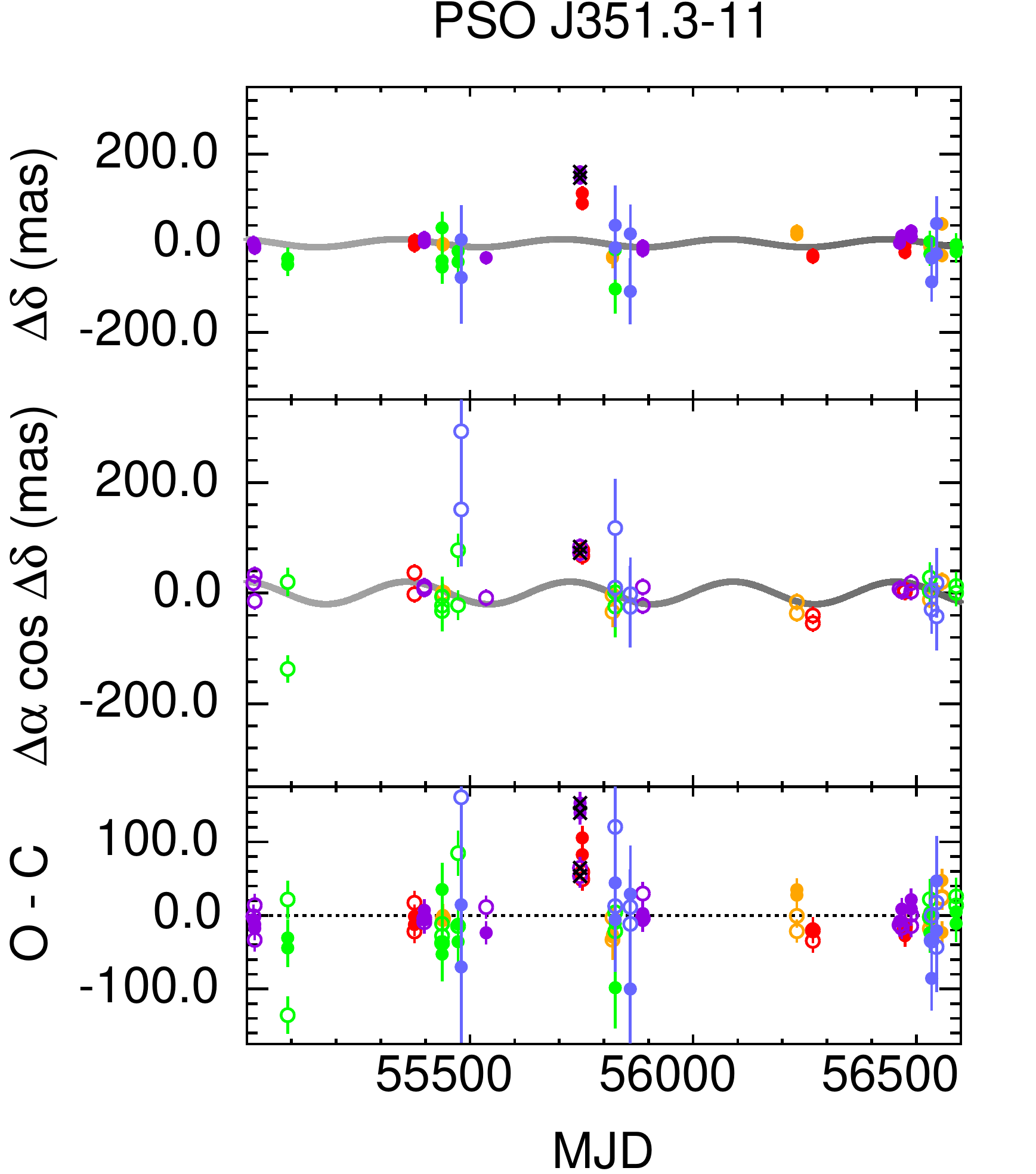}
    \includegraphics[width=0.3\textwidth]{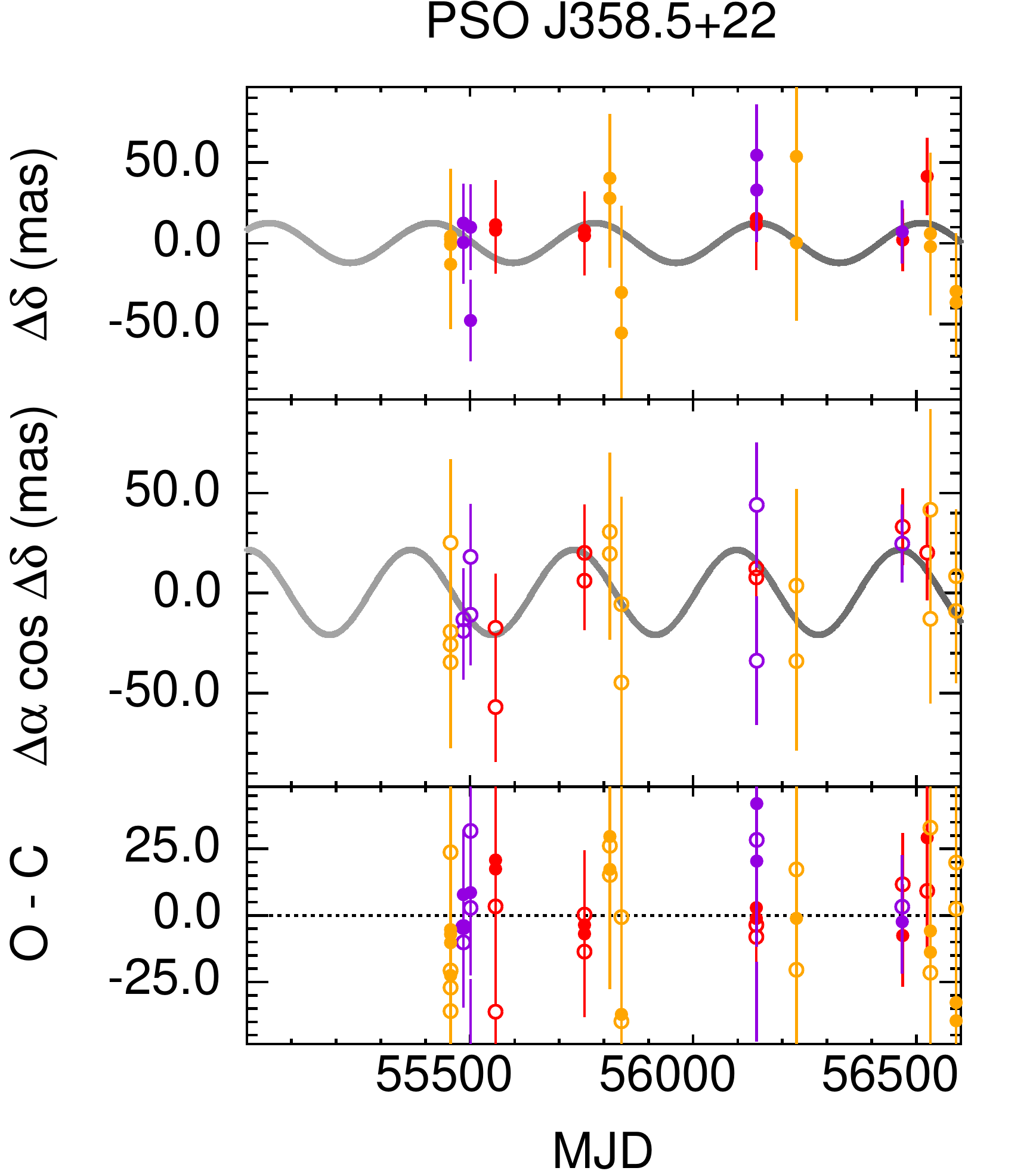}
    \includegraphics[width=0.3\textwidth]{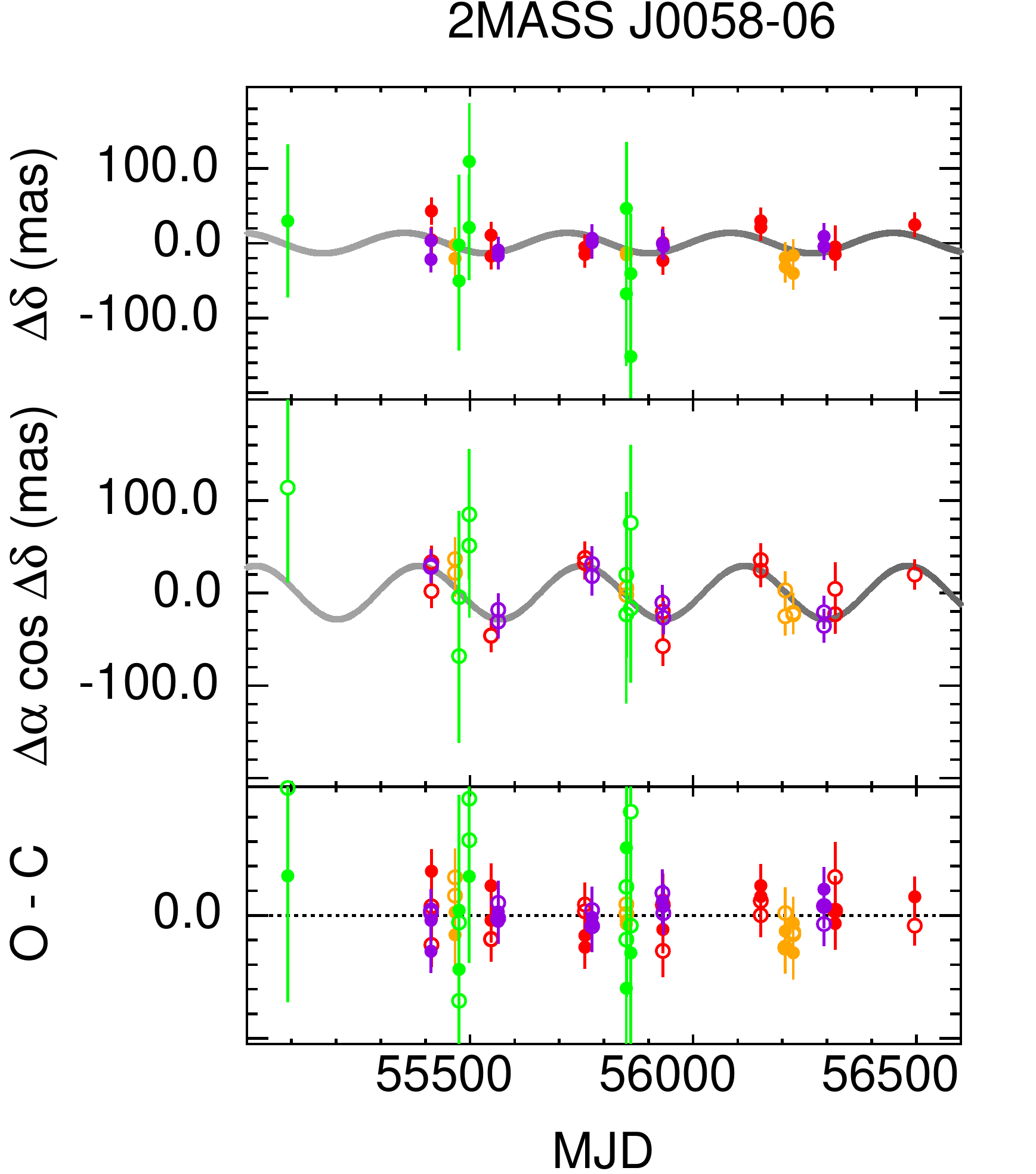}
    \caption{PS1 R.A. and decl. motion with the best fit overplotted in
      \emph{gray} in the top two panels of each plot with the
      residuals in the bottom panel. In order from top left to bottom
      right, our objects are: PSO~J035.8$-$15, PSO~J039.6$-$21,
      PSO~J167.1+68, PSO~J236.8$-$16, PSO~J318.4+35, PSO~J351.3$-$11,
      PSO~J358.5+22, and 2MASS~J0058$-$06. The colors and symbols are
      the same as in Figure~\ref{fig:plxfit1} except that \emph{open
        circles} and \emph{filled circles} correspond to R.A. and decl.
      positions, respectively. \label{fig:plxoc}}
  \end{center}
\end{figure*}

Using our custom astrometry analysis, we have determined parallaxes
for eight of our objects and one of the previously known substellar
candidate AB~Dor members, 2MASS~J0058425$-$0651239
\citep{2015ApJS..219...33G}. We show the final fits in
Figures~\ref{fig:plxfit1} and \ref{fig:plxfit2}. In
Figure~\ref{fig:plxoc} we plot the fitted parallax motion with the
data and the residuals. Our parallaxes are tabulated in
Table~\ref{table:obsprop} (for our new candidate members only) and in
Table~\ref{table:abdmembers}.

We also calculated photometric distances for all of our objects
(Table~\ref{table:obsprop}). We used the average 2MASS $J$-band
absolute magnitudes as a function of spectral type tabulated by
\citet{2012ApJS..201...19D} to determine photometric distances from
our NIR spectral types and $J$ apparent magnitude. While
\citet{2012ApJS..201...19D} also provide polynomial relations between
spectral type and absolute magnitude, these relations have a larger
dispersion compared to their tabulated averages because the spectral
type range (M6--T9) spanned by the polynomials is much larger than the
range for our objects (M6--L4). Therefore, using the average absolute
magnitudes as a function of spectral type is more accurate than using
the polynomial relation for our purposes. We assume that the
photometric distance uncertainty is 20$\%$, in accord with
\citet{2012ApJS..201...19D}.

Because young late-M~dwarfs can be overluminous compared with their
older counterparts \citep[e.g.][]{2013AN....334...85L}, we allow their
absolute magnitudes to be up to 1.5~mag brighter than the field value
since we do not know the ages of our objects. For our late-M~dwarfs,
we calculate two values for the photometric distance, one using the
field absolute magnitudes and one brighter by 1.5~mag. However, young
early L~dwarfs have absolute magnitudes consistent with those of their
older counterparts \citep[e.g.][]{2013AN....334...85L}. Therefore, for
our L~dwarfs we use the field value when converting from spectral type
to absolute magnitude. Thus the uncertainty in the photometric
distance for our M~dwarfs is significantly larger than for our
L~dwarfs. 

\begin{figure}
  \begin{center}
    \includegraphics[width=\textwidth]{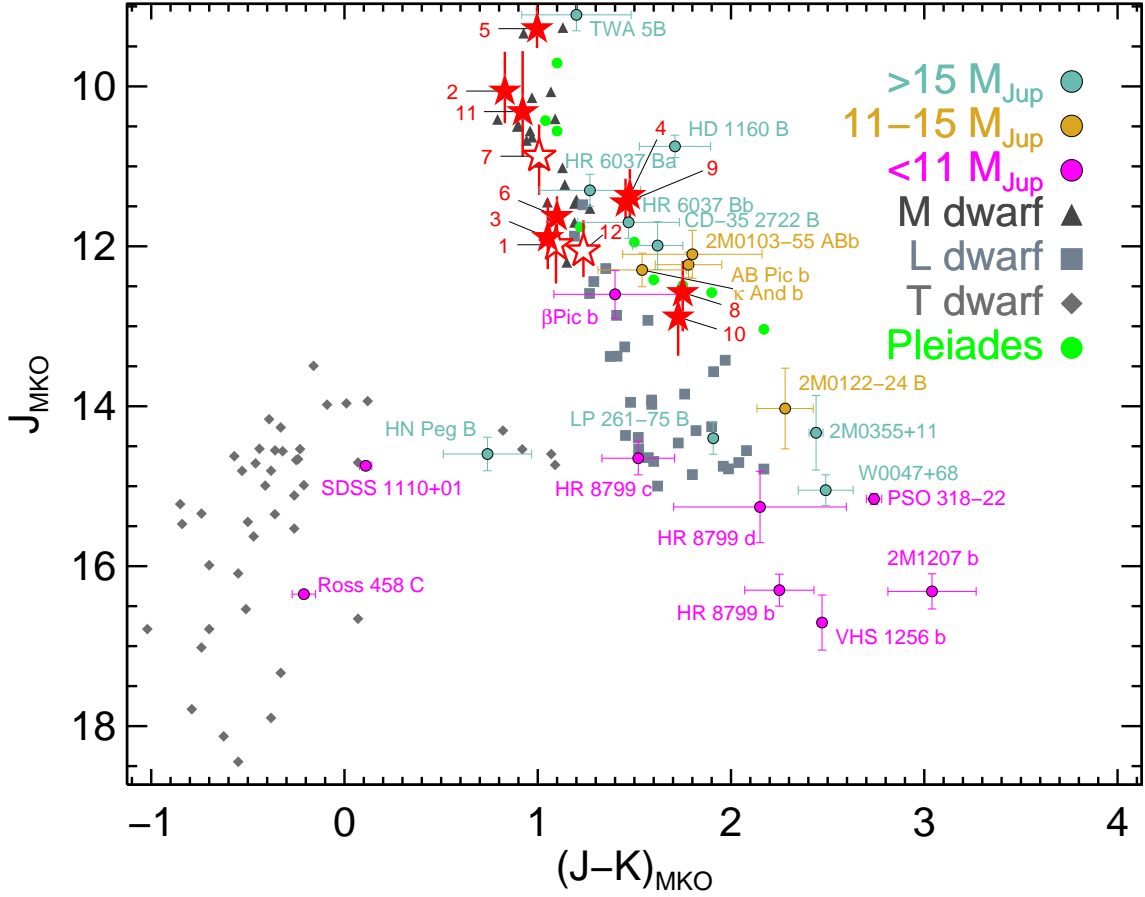}
    \caption{NIR $J-K$ color and absolute $J$ magnitude of our
      candidate AB~Dor Moving Group substellar members (\emph{red
        stars}), field dwarfs (\emph{gray symbols}), and known young
      brown dwarfs and planetary-mass objects (\emph{colored
        circles}). Our objects are numbered the same as in
      Figure~\ref{fig:prismind}: (1)~PSO~J004.7+41,
      (2)~PSO~J035.8$-$15, (3)~PSO~J039.6$-$21, (4)~PSO~J167.1+68,
      (5)~PSO~J232.2+63, (6)~PSO~J236.8$-$16, (7)~PSO~J292.9$-$06,
      (8)~PSO~J306.0+16, (9)~PSO~J318.4+35, (10)~PSO~J334.2+28,
      (11)~PSO~J351.3$-$11, (12)~PSO~J358.5+22. We use the compilation
      of known objects from \citet{2012ApJS..201...19D} for photometry
      and parallaxes of the field objects and for CD-35~2722~B,
      2MASS~J0355+11, and 2M1207~b. Photometric data is taken from the
      literature for TWA~5~B \citep{2013ApJ...762..118W}, HD~1060~B
      \citep{2012ApJ...750...53N}, HR~6037~Bab
      \citep{2013ApJ...776....4N}, 2MASS~J0103$-$55~ABb
      \citep{2013A&A...553L...5D}, AB~Pic~b
      \citep{2013ApJ...777..160B}, $\kappa$~And~b
      \citep{2013ApJ...763L..32C,2014A&A...562A.111B}, $\beta$~Pic~b
      \citep{2013A&A...555A.107B}, 2MASS~J0122$-$24~B LP~261-75~B
      \citep{2006PASP..118..671R,2013ApJ...774...55B},
      \citep{2013ApJ...774...55B}, WISEP~J0047+68
      \citep{2015ApJ...799..203G}, PSO~J318.5$-$22
      \citep{2013ApJ...777L..20L}, HN~Peg~B
      \citep{2007ApJ...654..570L}, SDSS~J1110+01
      \citep{2015ApJ...808L..20G}, Ross~458~C
      \citep{2011MNRAS.414.3590B}, VHS~1256~b
      \citep{2015ApJ...804...96G}, and the HR~8799 planets
      \citep{2008Sci...322.1348M}. All photometry for previously known
      objects are on the $MKO$ system. We synthesized $MKO$ photometry for
      our candidates from the NIR spectra. Known Pleiades members
      (\emph {green circles}) trace out the isochrones for an age of
      $\approx$125~Myr
      \citep{2007MNRAS.380..712L,2010A&A...519A..93B}. The AB~Dor
      Moving Group is $\approx$125~Myr old. For our candidate without a
      parallax, we use photometric distances calculated using the
      relations from \citet{2012ApJS..201...19D} and assume a
      photometric distance uncertainty of 20\%
      (Section~\ref{sec:dphot}). \label{fig:dphot}}
  \end{center}
\end{figure}

Figure~\ref{fig:dphot} shows the resulting color-magnitude diagram for
our objects compared to previously known young and field objects. For
our objects, we synthesized $MKO$ magnitudes in order to compare with
field dwarfs because of the larger amount of $MKO$ photometry available
for young companions and the Pleiades objects. When synthesizing
photometry we used the 2MASS \citep{2006AJ....131.1163S}\footnote{\url{http://www.ipac.caltech.edu/2mass/releases/second/doc/sec3\_1b1.html}}
and $MKO$ \citep{2002PASP..114..180T} filter profiles and used the 2MASS
photometry for flux calibration. We then compared our candidates
(Table~\ref{table:synthphot}) against known Pleiades members
\citep[e.g.][]{2007MNRAS.380..712L,2010A&A...519A..93B}, whose age
\citep[125~Myr; e.g.][]{1996ApJ...458..600B,1998ApJ...499L..61M}, is
similar to that of the AB~Dor Moving Group. In addition, we compared
our objects to field dwarfs and young substellar companions based on
the compilation by \citet{2012ApJS..201...19D}. Our candidate AB~Dor
Moving Group members have NIR absolute magnitudes and colors
consistent with the Pleiades sequence, as expected given the similar
ages of the two stellar associations. (Although by construction the
NIR absolute magnitudes, for objects without parallaxes, should be
consistent with their ages, the NIR colors would not necessarily be
the same for our candidates as for other objects with the same age and
absolute magnitude.)

\subsection{AB~Dor Moving Group Membership} \label{sec:membership}
We assessed our candidates' membership in the AB~Dor Moving Group
using the BANYAN~II webtool
\citep{2013ApJ...762...88M,2014ApJ...783..121G}, which calculates
membership probabilities for objects using Bayesian inference and the
proper motion, sky coordinates and parallactic (or photometric) distance
(Section~\ref{sec:dphot}). In addition, because our objects have
spectral signatures of youth, we could improve the accuracy of the
BANYAN-II membership probabilities by only using a young ($<$~1~Gyr)
field population to determine the field membership
probability. Although the BANYAN~II webtool analysis (which uses
kinematics only) is different from the full BANYAN~II analysis
\citep[][ incorporates kinematics and
  photometry]{2014ApJ...783..121G}, we have used their field
contamination rate curves to approximate the membership quality of our
candidates. These field contamination curves suggest that objects
(with distances but no radial velocities) with BANYAN~II membership
probabilities of $\gtrsim$15\%, $\approx$15--75\%, and $\gtrsim$75\%
would have field contamination rates of $\gtrsim$50$\%$,
$\approx$50-10\%, and $\gtrsim$10\%, respectively. Also, we note that
bona fide members (i.e. members with signatures of youth, proper
motions, RVs and parallaxes) can have BANYAN~II membership
probabilities from 10\% to 95\% \citep{2015ApJ...798...73G}. Therefore, we
have roughly divided our sample into three bins based on their
BANYAN~II webtool memberships, with a few exceptions
(Section~\ref{ssec:indivobj}): strong candidates ($\ge$75\%), possible
candidates (15--75\%), and probable young field interlopers ($<$15\%).

\begin{figure*}
  \begin{center}
    \includegraphics[width=0.9\textwidth]{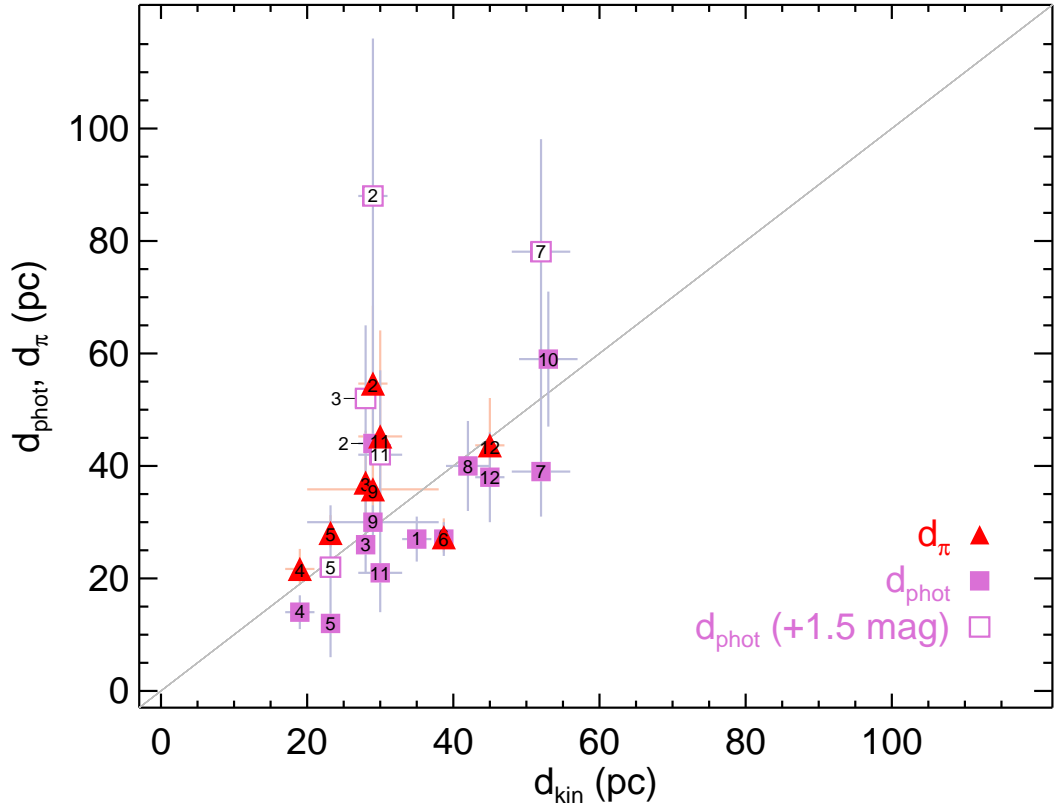}
    \caption{The kinematic distances (d$_{kin}$) of our
        candidate AB~Dor Moving Group members compared with their
        photometric distances (d$_{phot}$) in \emph{purple squares}
        and parallactic distances (d$_{\pi}$) as \emph{red
          triangles}. The \emph{open purple squares} are the
        d$_{phot}$ calculated assuming the object is overluminous
        compared with the field sequence by 1.5~mag which reflects the
        systematic uncertainty in the photometric distances for young
        late-M~dwarfs. Our objects are numbered as in
        Figure~\ref{fig:prismind}: (1)~PSO~J004.7+41,
        (2)~PSO~J035.8$-$15, (3)~PSO~J039.6$-$21, (4)~PSO~J167.1+68,
        (5)~PSO~J232.2+63, (6)~PSO~J236.8$-$16, (7)~PSO~J292.9$-$06,
        (8)~PSO~J306.0+16, (9)~PSO~J318.4+35, (10)~PSO~J334.2+28,
        (11)~PSO~J351.3$-$11, (12)~PSO~J358.5+22. Eight of our objects
        (except PSO~J004.7+41, PSO~J292.9$-$06, PSO~J306.0+16,
        PSO~J318.4+35, and PSO~J334.2+28) have parallactic distances
        from PS1. Seven of our objects have parallaxes which are
        consistent with their d$_{kin}$, the distance they would have
        if they were AB~Dor members. \label{fig:dpar}}
    \end{center}
\end{figure*}

The BANYAN~II webtool also computes statistical distances, i.e. the
most probable distance if an object were a member of a given moving
group (in this case, the AB~Dor Moving Group). We tabulate these
distances in Table~\ref{table:obsprop}. We compared both the
statistical distance and the d$_{kin}$ with the parallactic (or
photometric for objects without parallaxes) distances for all of our
objects. The statistical distances are consistent within 2$\sigma$
with the parallactic (or photometric) distances for all of our objects
except PSO~J035.8$-$15 where the large difference (3$\sigma$) between
the statistical and parallactic distances suggests it is a probable
young field interloper. For all of our objects, except
PSO~J236.8$-$16, the d$_{kin}$ is consistent within the 2$\sigma$ with
the parallactic (or photometric) distance (Figure~\ref{fig:dpar}). The
discrepancy between the parallactic distance and d$_{kin}$ for
PSO~J236.8$-$16 is consistent with it having a low AB~Dor Moving Group
membership probability and being a probable young field interloper. We
also note that the statistical distances are consistent within
2$\sigma$ of the d$_{kin}$ for all objects.

As another way of assessing membership, we compared the $\theta$ for
our objects with the $\theta$ for the known members (see
Section~\ref{sec:searchmethod}, Figure~\ref{fig:abdtheta}). All of our
objects have $\theta$ under 15$^{\circ}$ which is consistent with the
known members.

\begin{figure*}
  \centering
    \includegraphics[width=0.85\textwidth]{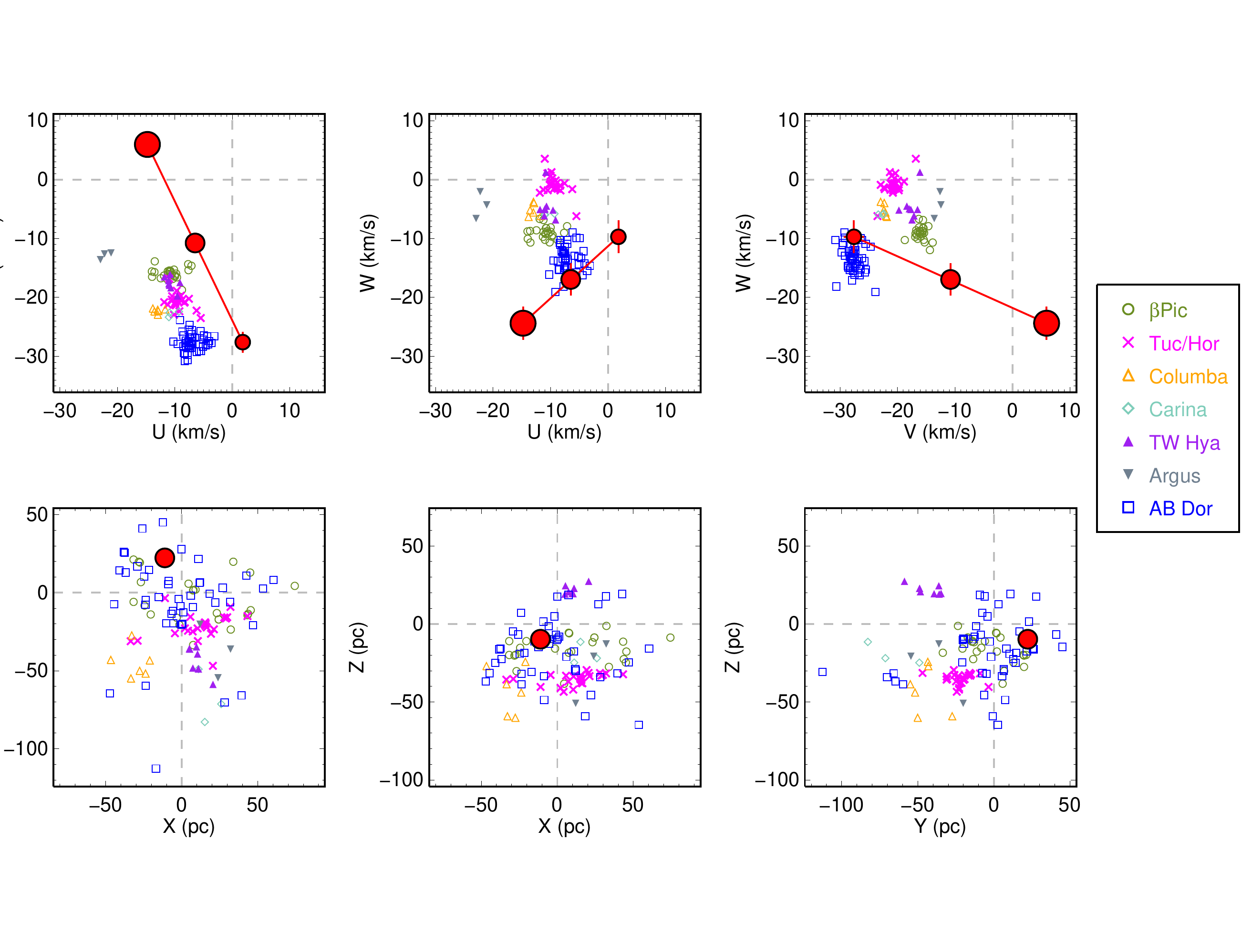}
    \vspace{-1 cm}
    \caption{The heliocentric space velocity ($UVW$) and positions ($XYZ$)
      of PSO~J004.7+41 compared with the kinematics and positions of
      YMG members from \citet{2008hsf2.book..757T} with membership
      probabilities of at least 75\%. We have used RVs and parallaxes
      from the literature for objects which had no measured values in
      \citet{2008hsf2.book..757T}. The \emph{red lines} show the $UVW$
      range for an RV between $-$20 and 20 km/s, reasonable for young
      objects (i.e. the RV range for young objects in
      \citealt{2008hsf2.book..757T}). The error bars are due to
      uncertainties in photometric distance and proper
      motion. \label{fig:ymg0019}}
\end{figure*}

\begin{figure*}
  \vspace{-2 cm}
  \centering
    \includegraphics[width=0.85\textwidth]{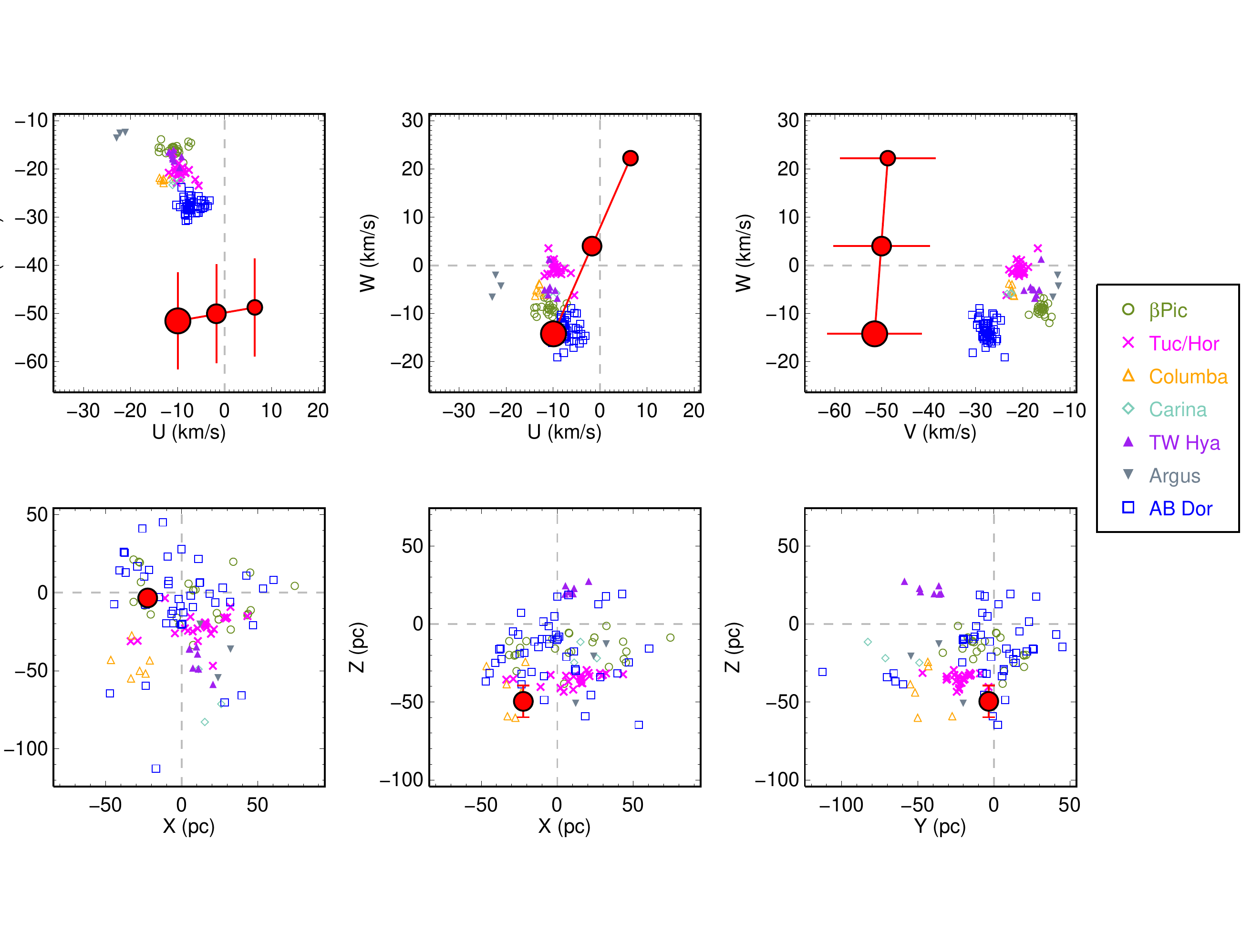}
    \vspace{-1 cm}
    \caption{The heliocentric space velocity ($UVW$) and positions ($XYZ$)
      of PSO~J035.8$-$15 compared with the kinematics and positions of
      known YMG members from \citet{2008hsf2.book..757T} as in
      Figure~\ref{fig:ymg0019}. The \emph{red lines} show the $UVW$
      range for an RV between $-$20 and 20 km/s, reasonable for young
      objects (i.e. the RV range for young objects in
      \citealt{2008hsf2.book..757T}). The error bars are due to
      uncertainties in photometric distance and proper
      motion. \label{fig:ymg0223}}
\end{figure*}

\begin{figure*}
  \centering
    \includegraphics[width=0.85\textwidth]{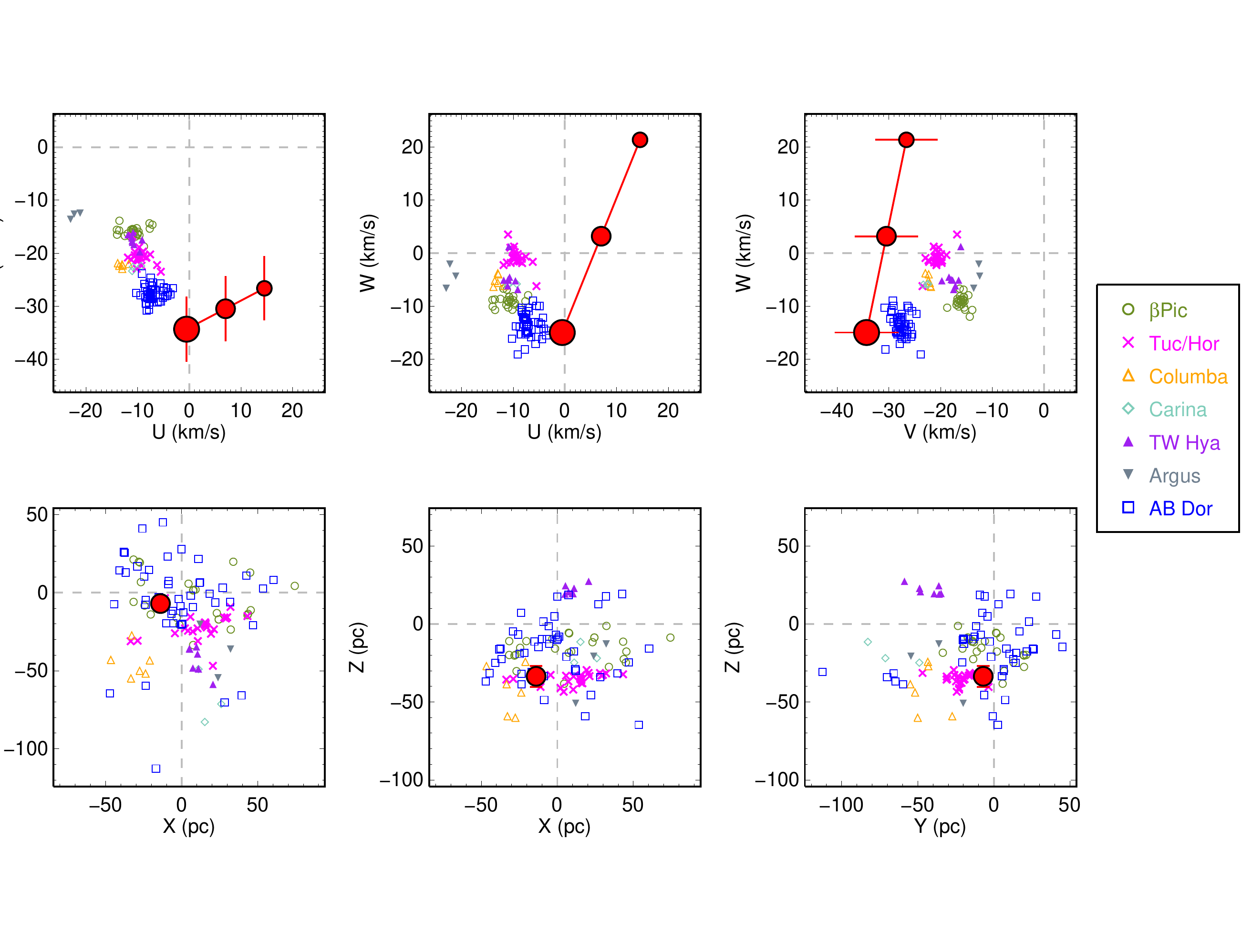}
    \vspace{-1 cm}
    \caption{The heliocentric space velocity ($UVW$) and positions ($XYZ$)
      of PSO~J039.6$-$21 compared with the kinematics and positions of
      known YMG members from \citet{2008hsf2.book..757T} as in
      Figure~\ref{fig:ymg0019}. The \emph{red lines} show the $UVW$
      range for an RV between $-$20 and 20 km/s, reasonable for young
      objects (i.e. the RV range for young objects in
      \citealt{2008hsf2.book..757T}). The error bars are due to
      uncertainties in parallax and proper
      motion. \label{fig:ymg0238}}
\end{figure*}

\begin{figure*}
  \vspace{-2 cm}
  \centering
    \includegraphics[width=0.85\textwidth]{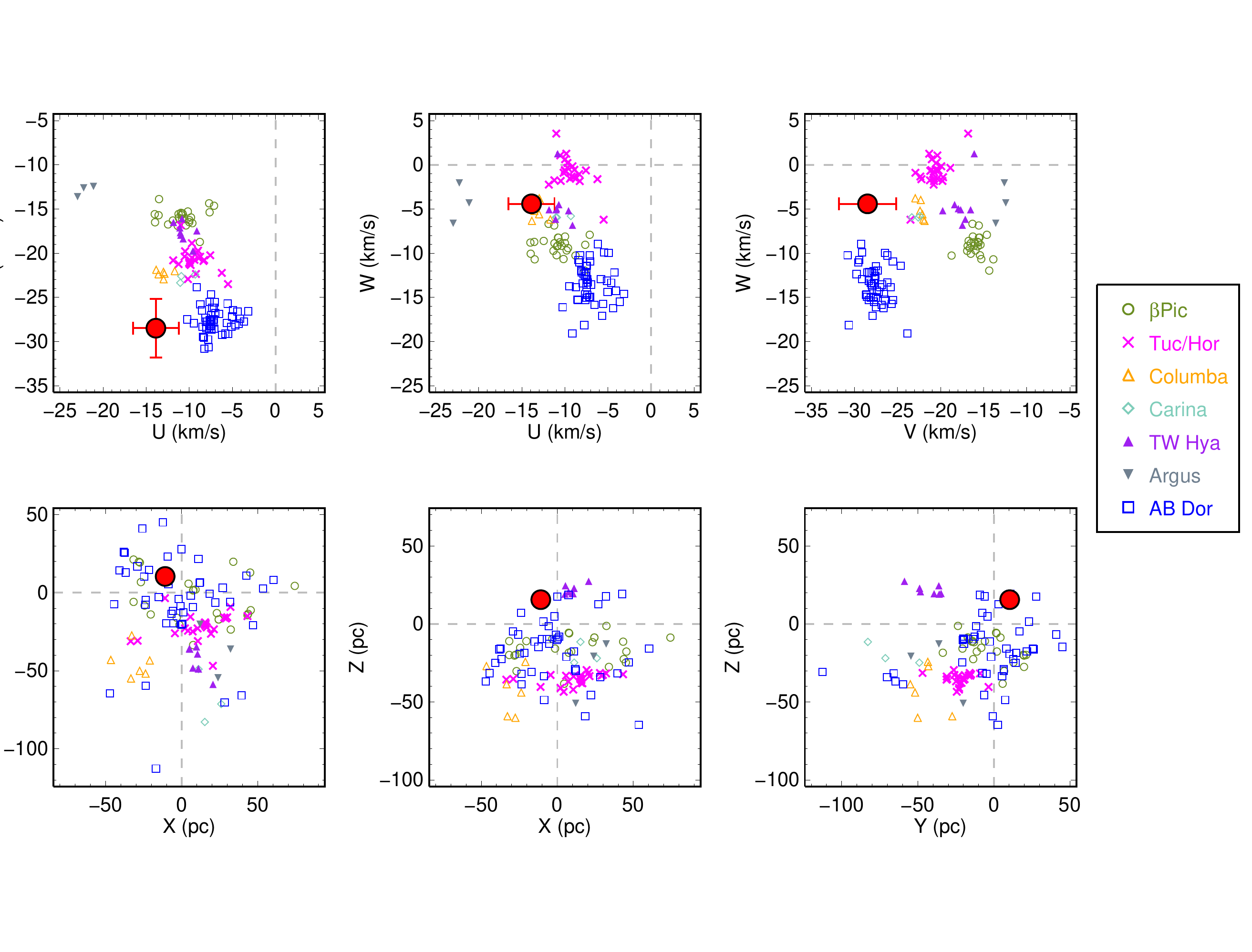}
    \vspace{-1 cm}
    \caption{The heliocentric space velocity ($UVW$) and positions ($XYZ$)
      of PSO~J167.1+68 compared with the kinematics and positions of
      known YMG members from \citet{2008hsf2.book..757T} as in
      Figure~\ref{fig:ymg0019}. The error bars are due to
      uncertainties in parallax, radial velocity, and proper
      motion. \label{fig:ymg1108}}
\end{figure*}

\begin{figure*}
  \centering
    \includegraphics[width=0.85\textwidth]{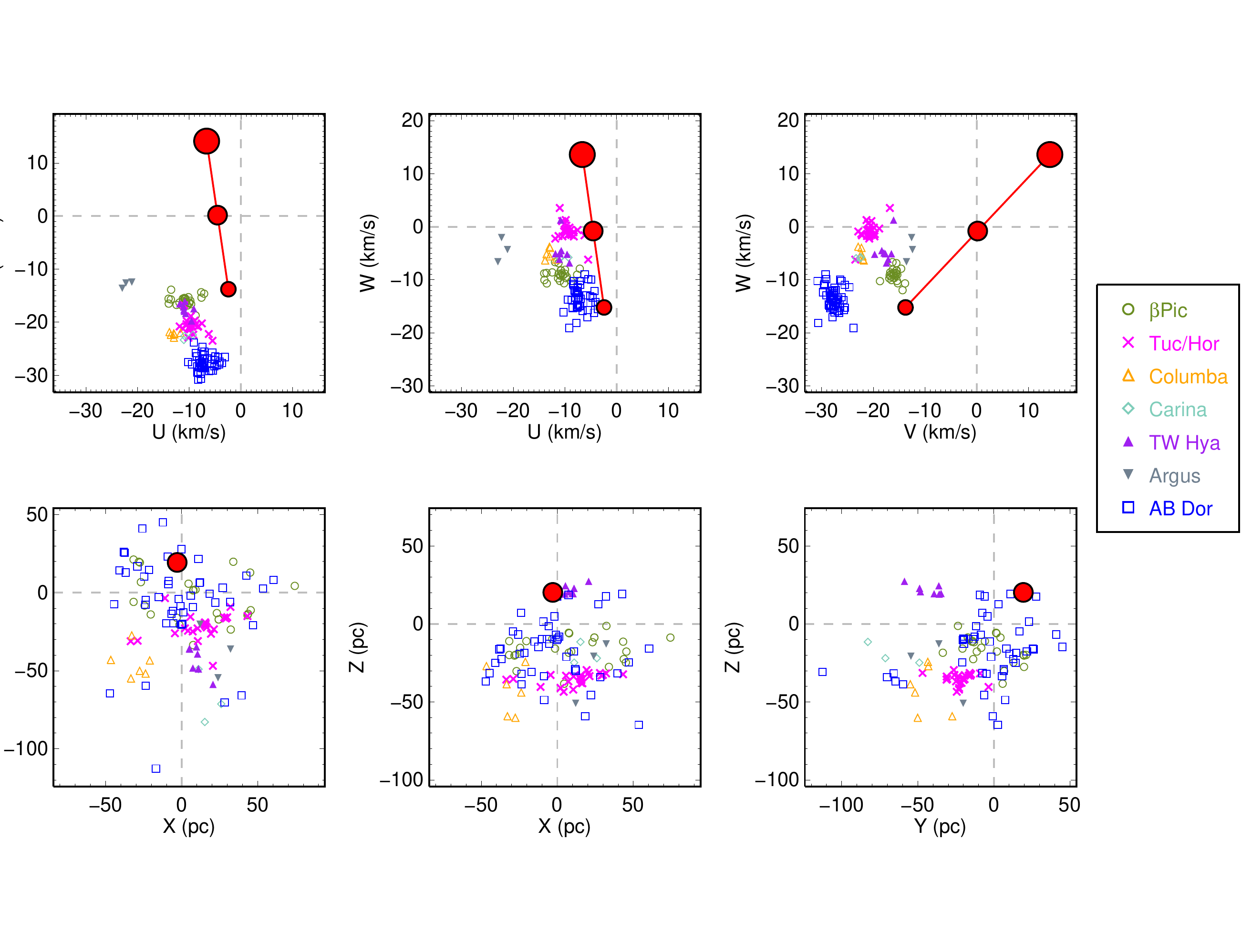}
    \vspace{-1 cm}
    \caption{The heliocentric space velocity ($UVW$) and positions ($XYZ$)
      of PSO~J232.2+63 compared with the kinematics and positions of
      known YMG members from \citet{2008hsf2.book..757T} as in
      Figure~\ref{fig:ymg0019}. The \emph{red lines} show the $UVW$
      range for an RV between $-$20 and 20 km/s, reasonable for young
      objects (i.e. the RV range for young objects in
      \citealt{2008hsf2.book..757T}). The error bars are due to
      uncertainties in parallax and proper motion. \label{fig:ymg232}}
\end{figure*}

\begin{figure*}
  \vspace{-2 cm}
  \centering
    \includegraphics[width=0.85\textwidth]{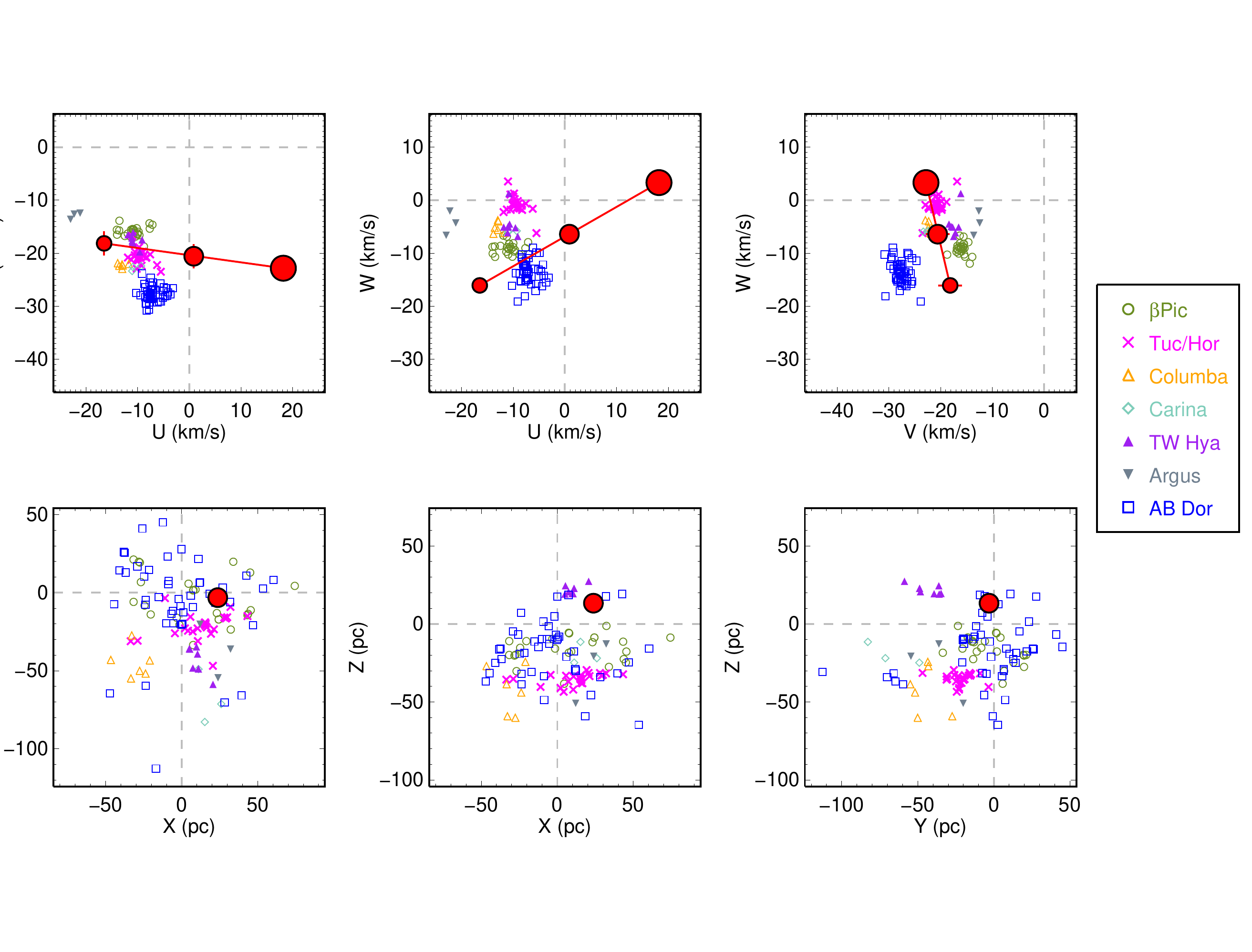}
    \vspace{-1 cm}
    \caption{The heliocentric space velocity ($UVW$) and positions ($XYZ$)
      of PSO~J236.8$-$16 compared with the kinematics and positions of
      known YMG members from \citet{2008hsf2.book..757T} as in
      Figure~\ref{fig:ymg0019}. The \emph{red lines} show the $UVW$
      range for an RV between $-$20 and 20 km/s, reasonable for young
      objects (i.e. the RV range for young objects in
      \citealt{2008hsf2.book..757T}). The error bars are due to
      uncertainties in parallax and proper
      motion. \label{fig:ymg1547}}
\end{figure*}

\begin{figure*}
  \centering
    \includegraphics[width=0.85\textwidth]{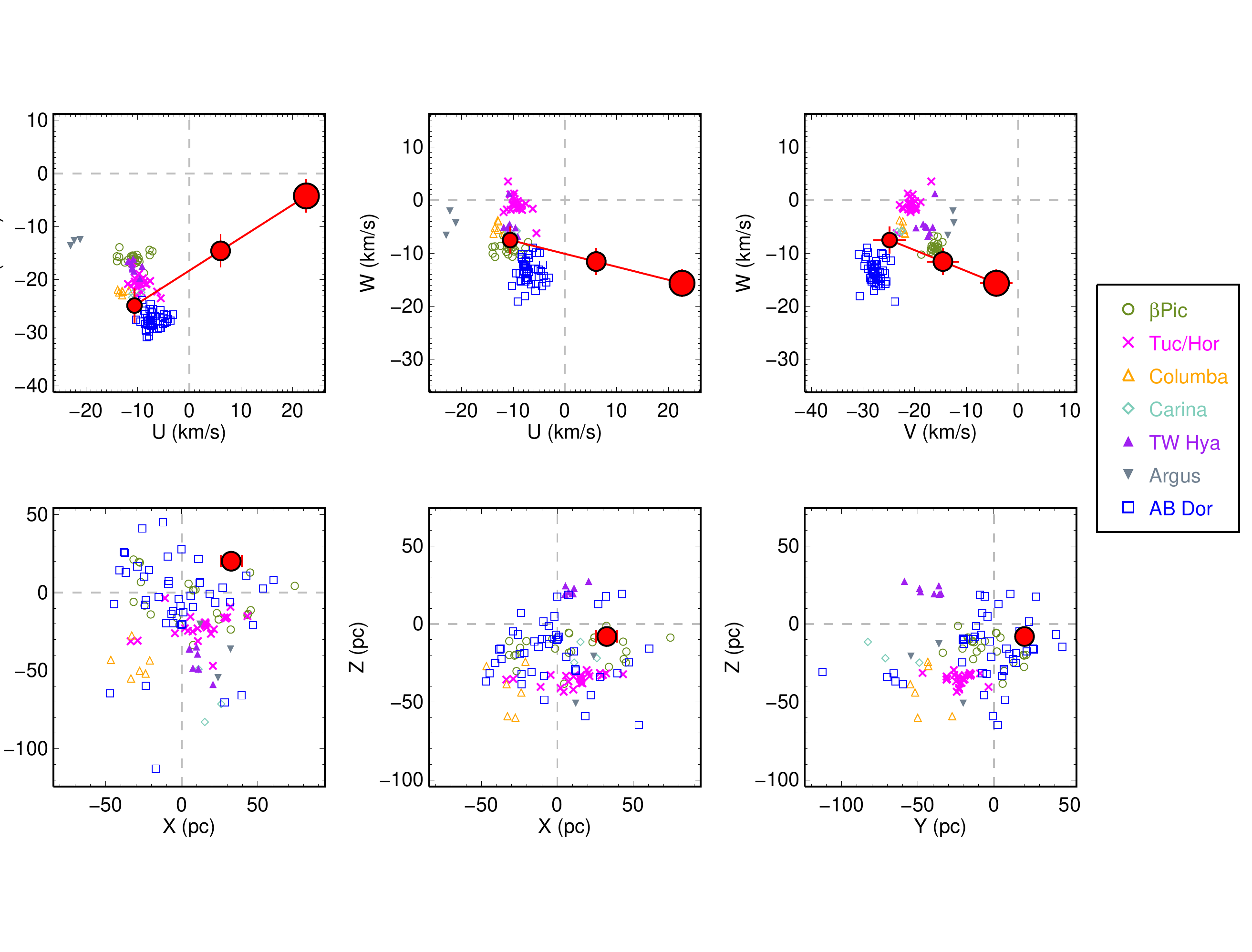}
    \vspace{-1 cm}
    \caption{The heliocentric space velocity ($UVW$) and positions ($XYZ$)
      of PSO~J292.9$-$06 compared with the kinematics and positions of
      known YMG members from \citet{2008hsf2.book..757T} as in
      Figure~\ref{fig:ymg0019}. The \emph{red lines} show the $UVW$
      range for an RV between $-$20 and 20 km/s, reasonable for young
      objects (i.e. the RV range for young objects in
      \citealt{2008hsf2.book..757T}). The error bars are due to
      uncertainties in photometric distance and proper
      motion. \label{fig:ymg292}}
\end{figure*}

\begin{figure*}
  \vspace{-2 cm}
  \centering
    \includegraphics[width=0.85\textwidth]{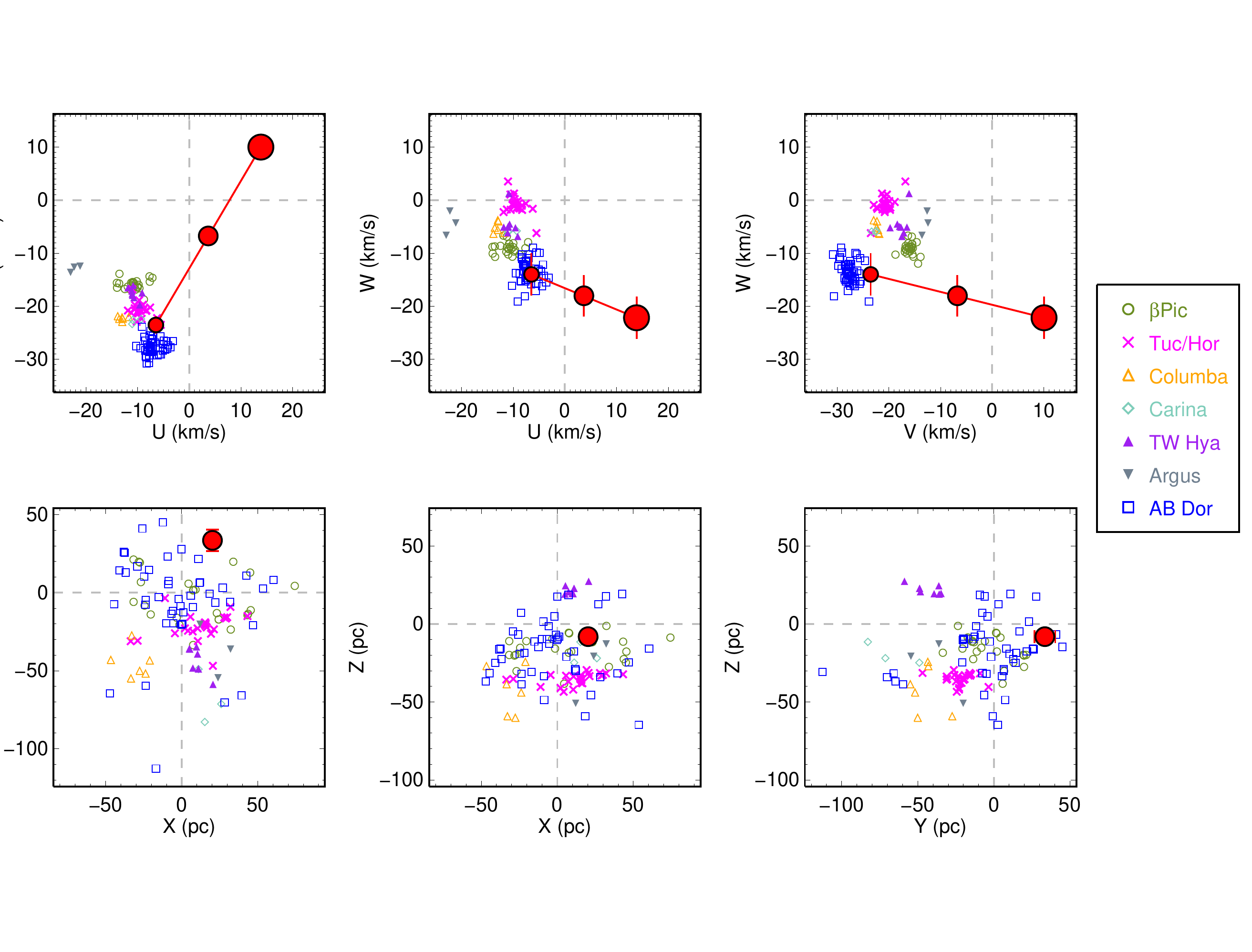}
    \vspace{-1 cm}
    \caption{The heliocentric space velocity ($UVW$) and positions ($XYZ$)
      of PSO~J306.0+16 compared with the kinematics and positions of
      known YMG members from \citet{2008hsf2.book..757T} as in
      Figure~\ref{fig:ymg0019}. The \emph{red lines} show the $UVW$
      range for an RV between $-$20 and 20 km/s, reasonable for young
      objects (i.e. the RV range for young objects in
      \citealt{2008hsf2.book..757T}). The error bars are due to
      uncertainties in photometric distance and proper
      motion. \label{fig:ymg306}}
\end{figure*}

\begin{figure*}
  \centering
    \includegraphics[width=0.85\textwidth]{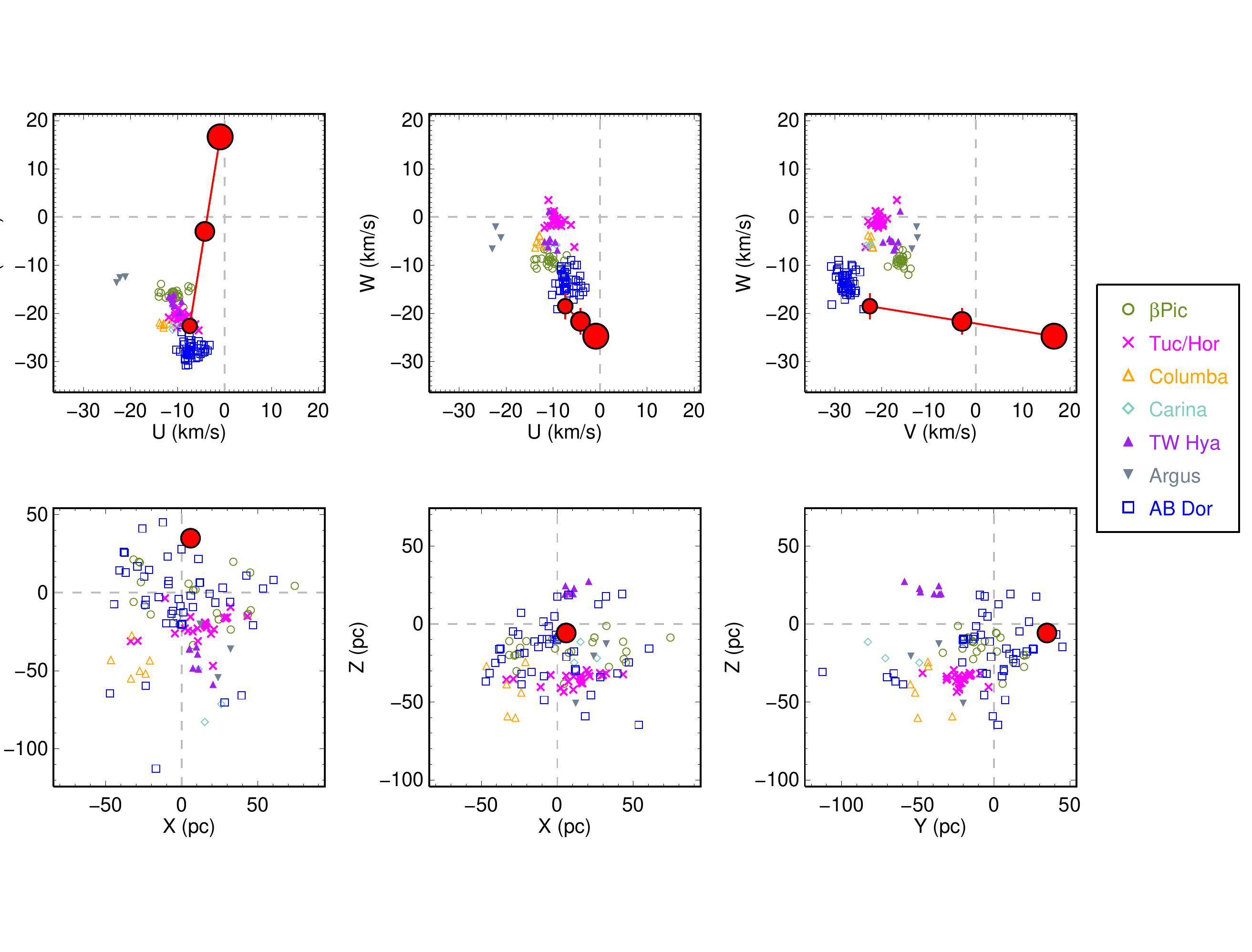}
    \vspace{-1 cm}
    \caption{The heliocentric space velocity ($UVW$) and positions ($XYZ$)
      of PSO~J318.4+35 compared with the kinematics and positions of
      known YMG members from \citet{2008hsf2.book..757T} as in
      Figure~\ref{fig:ymg0019}. The \emph{red lines} show the $UVW$
      range for an RV between $-$20 and 20 km/s, reasonable for young
      objects (i.e. the RV range for young objects in
      \citealt{2008hsf2.book..757T}). The error bars are due to
      uncertainties in parallax and proper
      motion. \label{fig:ymg2113}}
\end{figure*}

\begin{figure*}
  \vspace{-2 cm}
  \centering
    \includegraphics[width=0.85\textwidth]{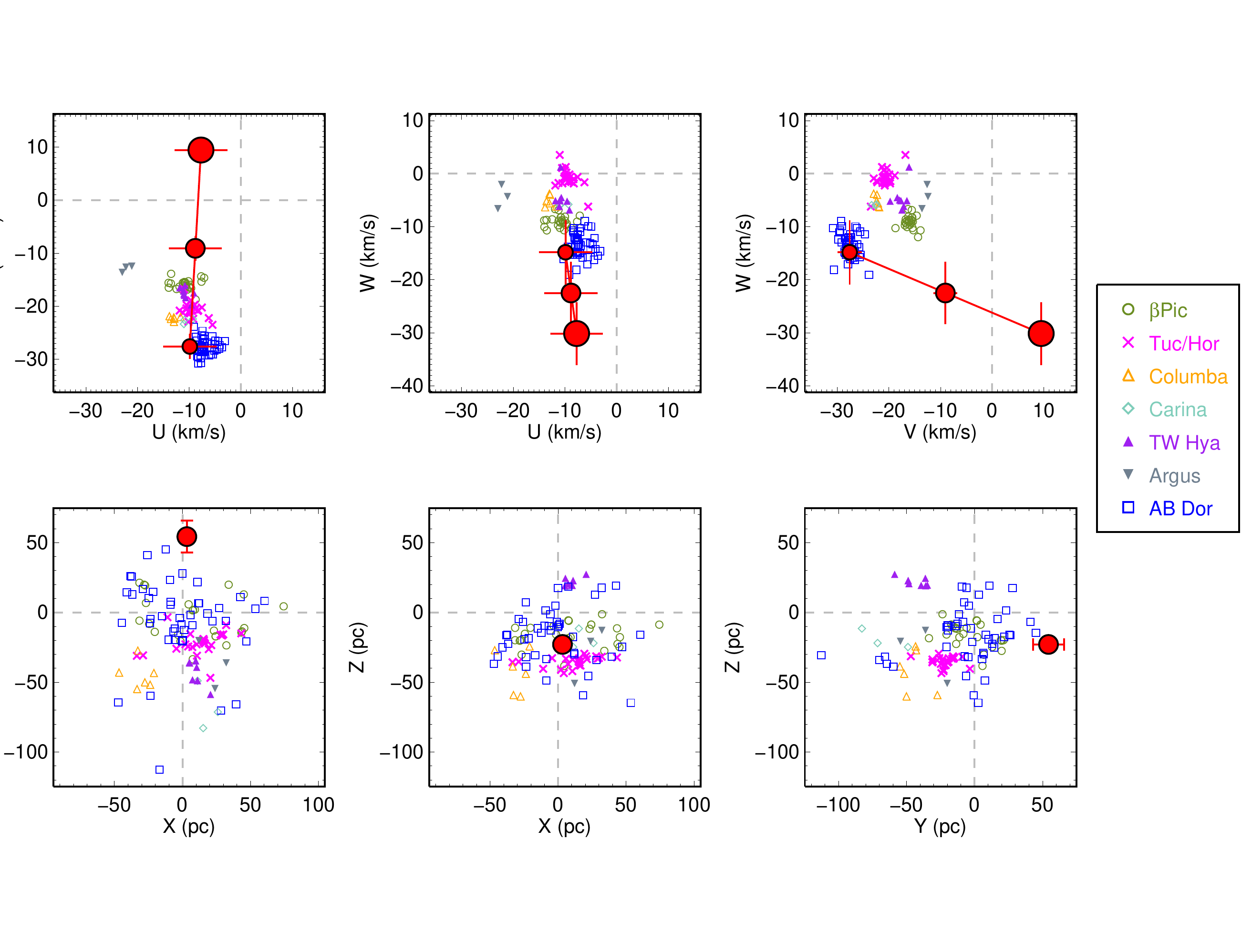}
    \vspace{-1 cm}
    \caption{The heliocentric space velocity ($UVW$) and positions ($XYZ$)
      of PSO~J334.2+28 compared with the kinematics and positions of
      known YMG members from \citet{2008hsf2.book..757T} as in
      Figure~\ref{fig:ymg0019}. The \emph{red lines} show the $UVW$
      range for an RV between $-$20 and 20 km/s, reasonable for young
      objects (i.e. the RV range for young objects in
      \citealt{2008hsf2.book..757T}). The error bars are due to
      uncertainties in photometric distance and proper
      motion. \label{fig:ymg334}}
\end{figure*}

\begin{figure*}
  \centering
    \includegraphics[width=0.85\textwidth]{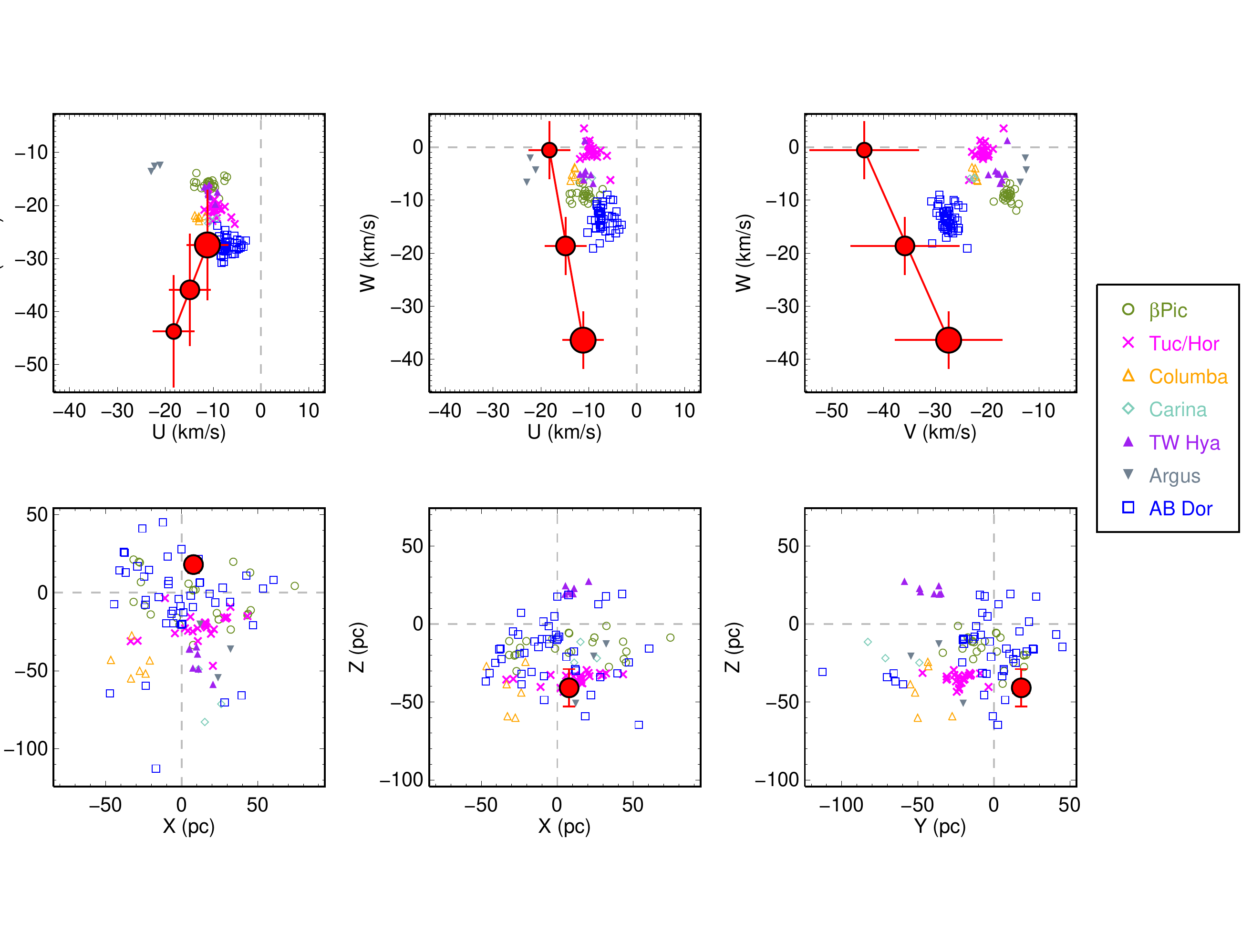}
    \vspace{-1 cm}
    \caption{The heliocentric space velocity ($UVW$) and positions ($XYZ$)
      of PSO~J351.3$-$11 compared with the kinematics and positions of
      known YMG members from \citet{2008hsf2.book..757T} as in
      Figure~\ref{fig:ymg0019}. The \emph{red lines} show the $UVW$
      range for an RV between $-$20 and 20 km/s, reasonable for young
      objects (i.e. the RV range for young objects in
      \citealt{2008hsf2.book..757T}). The error bars are due to
      uncertainties in parallax and proper
      motion. \label{fig:ymg2325}}
\end{figure*}

\begin{figure*}
  \vspace{-2 cm}
  \centering
    \includegraphics[width=0.85\textwidth]{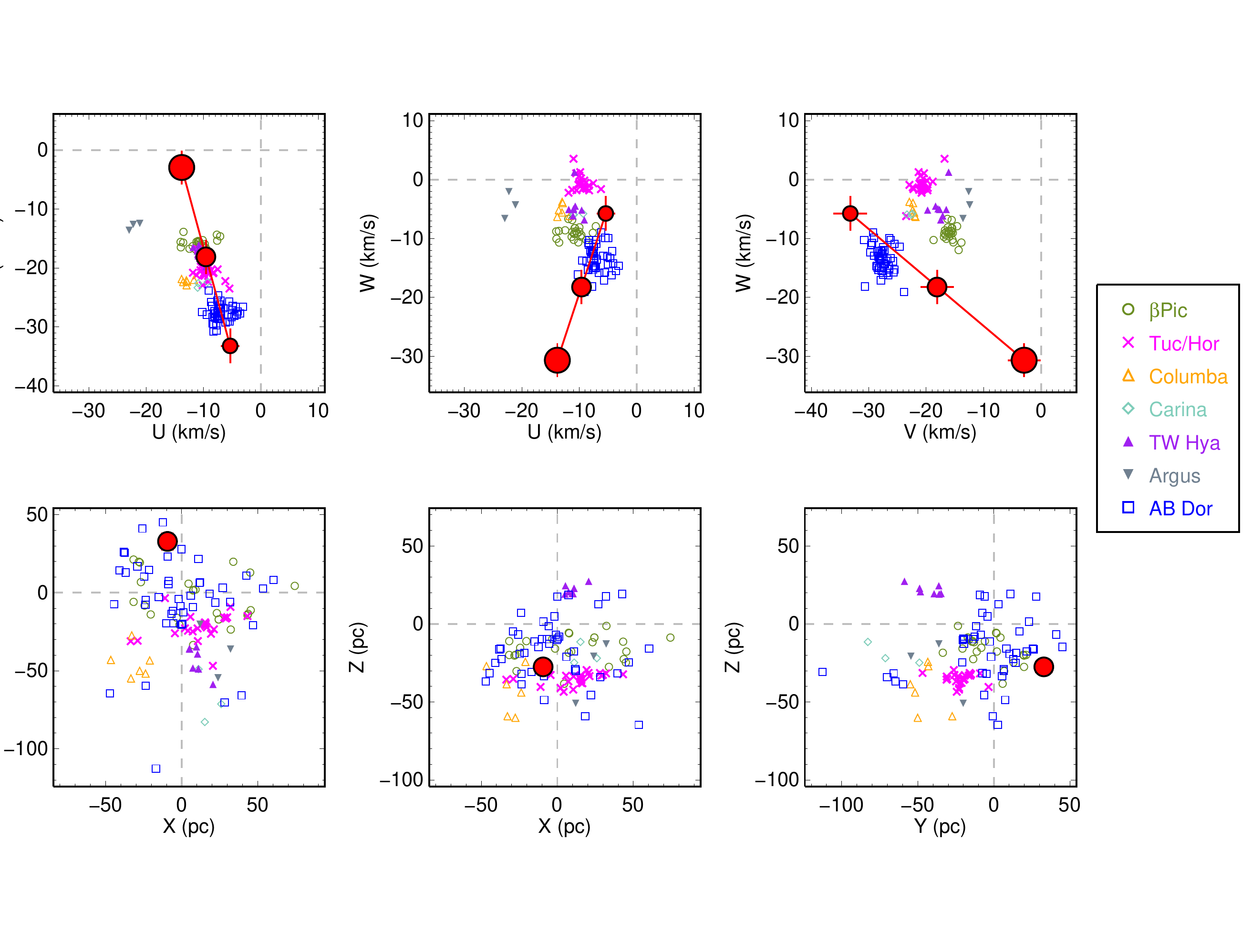}
    \vspace{-1 cm}
    \caption{The heliocentric space velocity ($UVW$) and positions
        ($XYZ$) of PSO~J358.5+22 compared with the kinematics and
        positions of known YMG members from
        \citet{2008hsf2.book..757T} as in
        Figure~\ref{fig:ymg0019}. The \emph{red lines} show the $UVW$
        range for an RV between $-$20 and 20 km/s, reasonable for
        young objects (i.e. the RV range for young objects in
        \citealt{2008hsf2.book..757T}). The error bars are due to
        uncertainties in parallax and proper
        motion. \label{fig:ymg2354}}
\end{figure*}

We also compared the heliocentric kinematics ($UVW$) and space
positions ($XYZ$) of our AB~Dor Moving Group candidates in
Figures~\ref{fig:ymg0019}--\ref{fig:ymg2354} with the YMG members from
\citet{2008hsf2.book..757T} with membership probabilities of at least
75\%. For the plotted YMG members, we used radial velocities and
parallaxes with values from the literature. For our candidates, we
assumed an probable RV range of $-$20 to +20~km/s, consistent with the
range of the bona fide YMG members from \citet{2008hsf2.book..757T}
plotted. Eight of our candidates' positions are consistent within
2$\sigma$ of the average positions of the known AB~Dor members within
their uncertainties. Note that 3~km/s is the uncertainty in the mean
$UVW$ position for the group given the uncertainties in the each
coordinate ($\approx$1-2~km/s) and that the spatial positions ($XYZ$)
of the known members are more spread out in comparison to the $UVW$
positions (10--50~pc encompasses the 1$\sigma$ distance
range). However, PSO~J004.7+41, PSO~J035.8$-$15, PSO~J167.1+68, and
PSO~J236.8$-$16 have $UVW$ positions inconsistent with AB~Dor
membership. We tabulate the distance between the mean group $UVW$ and
$XYZ$ positions and the closest possible positions of our candidates
(within the assumed RV range) in Table~\ref{table:obsprop}.

In total, we have three strong candidates, five possible candidates,
and four probable young field interlopers. We summarize the final
BANYAN~II webtool membership probabilities in
Table~\ref{table:obsprop}, and the BANYAN~II webtool and
\citet{2015ApJ...798...73G,2015ApJS..219...33G} membership
probabilities, when available, in
Table~\ref{table:abdmembers}. However, these probabilities are likely
a lower limit on the actual membership probabilities because our
candidates are $\approx$125~Myr, the age of the AB~Dor Moving Group,
much younger than the 1~Gyr old field population used in the BANYAN~II
web tool.

\subsection{Physical Properties \label{sec:physprop}}

We calculated the bolometric magnitudes, effective temperatures and
masses for our AB~Dor Moving Group candidate members, the previously
known bona fide substellar members, and candidate substellar members
with spectroscopically confirmed youth
(Table~\ref{table:physprop}). For all of these properties, we used a
Monte Carlo simulation to propagate measurement errors (in distance,
spectral type, and age) and determine the 68\% confidence limits for
each calculated parameter.

We use the \citet{2010ApJ...722..311L} $H$-band bolometric corrections
to determine the absolute bolometric magnitude because the $H$-band
corrections have the lowest dispersion and the bolometric correction
changes slowly with spectral type. To use these corrections, we used
our synthesized $MKO$ $H$-band magnitudes.

For our objects, we determined $T_{eff}$ and mass from our estimated
bolometric luminosities using the \citet{2000ApJ...542..464C}
evolutionary models. For our analysis, we assume the age of the AB~Dor
Moving Group is 125$\pm$20~Myr
\citep{2005ApJ...628L..69L,2007MNRAS.377..441O,2013ApJ...766....6B}
and propagate the uncertainties in age using a Monte Carlo
simulation. For our young field interlopers we adopt a more
conservative age of 150$\pm$100~Myr. Our late-M~dwarfs
PSO~J035.8$-$15, PSO~J236.8$-$16, and PSO~J351.3$-$11 have masses of
50--100~\Mjup. Our L~dwarfs, PSO~J004.7+41, PSO~J039.6$-$21,
PSO~J167.1+68, PSO~J318.4+35, and PSO~J358.5+22 have masses of
$\sim$35--45~\Mjup.

\subsection{Comparison with BASS} \label{sec:BASS}

Eight of our new young brown dwarfs are in the very large
($\sim$10$^{4}$ objects) input catalog of color-selected brown dwarfs
from the \citet{2015ApJ...798...73G} BASS program, an all-sky survey
constructed by combining 2MASS and {\it WISE}. Three of our new young
brown dwarf candidate members, PSO~J292.9$-$06, PSO~J306.0+16 and
PSO~J318.4+35, are missing from the BASS input catalog because their
low Galactic latitudes (b=$-$11$^{\circ}$.9, $-$11$^{\circ}$.7 and
$-$9$^{\circ}$.3, respectively) excluded them from the BASS search
($|$b$|$$\ge$15$^{\circ}$). PSO~J334.2+28 is also missing from the
BASS input catalog, possibly due to its faint 2MASS magnitudes.

\begin{figure*}
  \begin{center}
    \includegraphics[width=\textwidth]{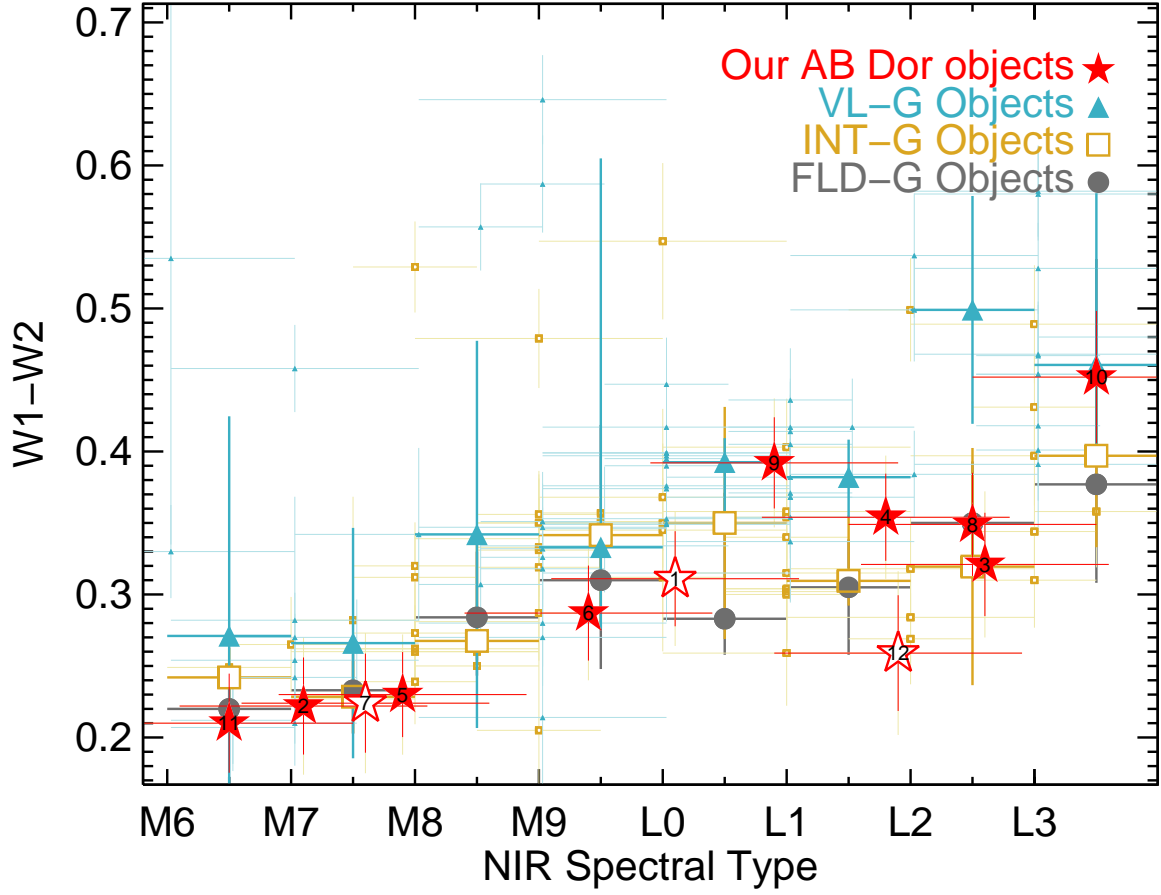}
    \caption{{\it WISE} $W1-W2$ color as a function of NIR spectral
      type for our candidate AB~Dor Moving Group members. Those
      classified as {\sc int-g} are the \emph{filled red stars} and
      those as {\sc int-g?} and {\sc fld-g} are the \emph{empty red
        stars}. Our objects are numbered (as in
      Figure~\ref{fig:prismind}): (1)~PSO~J004.7+41,
      (2)~PSO~J035.8$-$15, (3)~PSO~J039.6$-$21, (4)~PSO~J167.1+68,
      (5)~PSO~J232.2+63, (6)~PSO~J236.8$-$16, (7)~PSO~J292.9$-$06,
      (8)~PSO~J306.0+16, (9)~PSO~J318.4+35, (10)~PSO~J334.2+28,
      (11)~PSO~J351.3$-$11, (12)~PSO~J358.5+22.  The young field
      dwarfs with {\sc int-g} gravity classifications from
      \citet{2013ApJ...772...79A} are the \emph{small open yellow
        squares} and those with {\sc vl-g} are the \emph{small teal
        triangles}. We also include the \citet{2015ApJS..219...33G}
            {\sc int-g} (\emph{yellow}) and {\sc vl-g} (\emph{teal})
            substellar objects. The mean $W1-W2$ color for each
            spectral type bin and the 68$^{th}$ percentile confidence
            region are the \emph{filled gray circles} for old field
            objects, \emph{open yellow squares} for {\sc int-g}
            objects, and \emph{teal triangles} for {\sc vl-g}
            objects. The mean $W1-W2$ colors for each bin are plotted
            in the middle of the bin (e.g. for spectral type M6--M6.9,
            the mean color is plotted at M6.5) We note that the
            slightly blue $J-K$ colors of PSO~J236.7$-$16 are likely
            due to blending from the earlier-type closeby companion
            \citep[M5;][]{2015ApJS..219...33G}. We find that the {\sc
              vl-g} dwarfs tend to have redder $W1-W2$ colors than the
                  {\sc int-g} and then {\sc fld-g} dwarfs. However,
                  the $W1-W2$ colors for {\sc int-g} dwarfs are
                  consistent with their {\sc fld-g} counterparts
                  within the uncertainties. \label{fig:wisecolors}}
  \end{center}
\end{figure*}

\begin{figure*}
  \begin{center}
    \includegraphics[width=\textwidth]{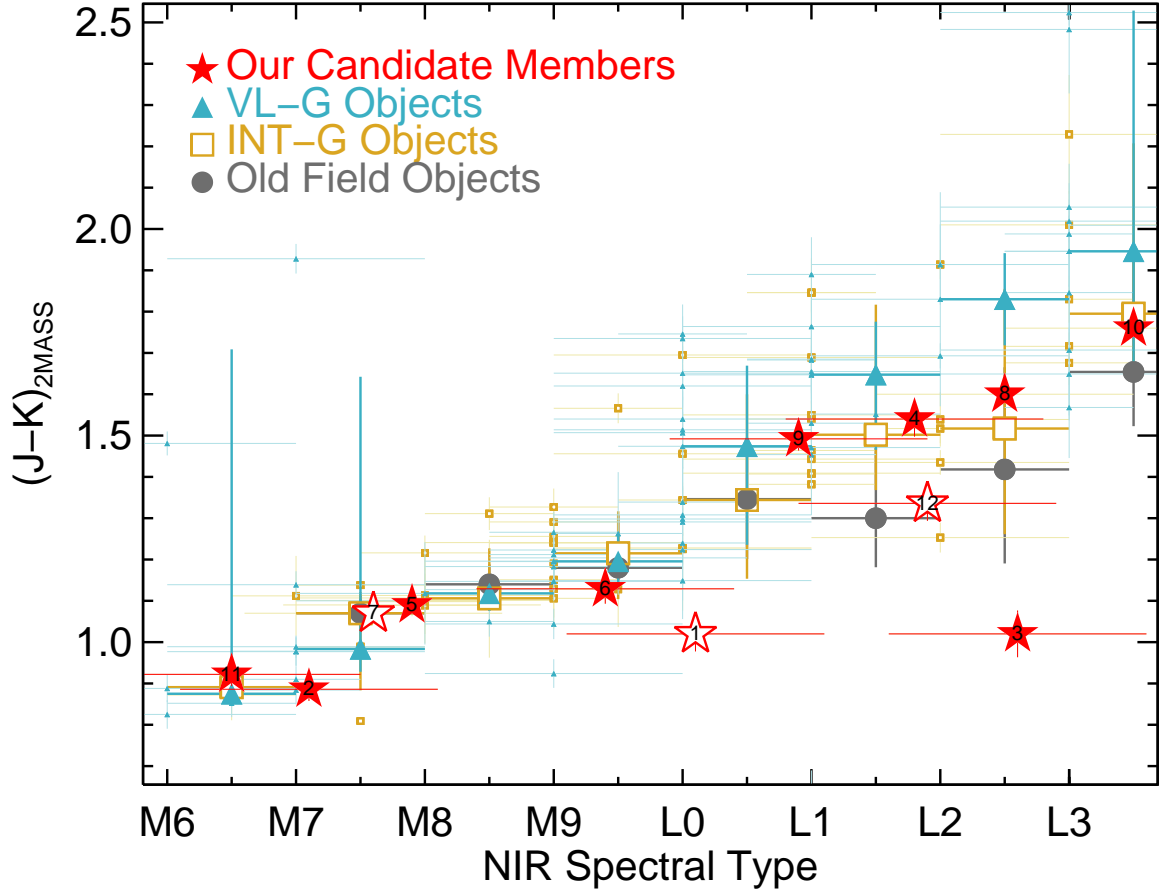}
    \caption{2MASS $J-K$ color as a function of NIR spectral type for
      our candidate AB~Dor Moving Group members. Those classified as
      {\sc int-g} are the \emph{filled red stars} and those as {\sc
        int-g?} or {\sc fld-g} are the \emph{empty red stars}. Our
      objects have the same numbers as in
      Figure~\ref{fig:wisecolors}. The young field dwarfs with {\sc
        int-g} gravity classifications from
      \citet{2013ApJ...772...79A} are the \emph{small open yellow
        squares} and those with {\sc vl-g} are the \emph{small teal
        triangles}. We also include the \citet{2015ApJS..219...33G}
            {\sc int-g} (\emph{yellow}) and {\sc vl-g} (\emph{teal})
            substellar objects. The mean $J-K$ color for each spectral
            type bin and the 68$^{th}$ percentile confidence region
            are the \emph{filled gray circles} for old field objects,
            \emph{open yellow squares} for {\sc int-g} objects, and
            \emph{teal triangles} for {\sc vl-g} objects. The mean
            $J-K$ colors for each bin are plotted in the middle of the
            bin (e.g. for spectral type M6--M6.9, the mean color is
            plotted at M6.5) We note that the slightly blue $J-K$
            colors of PSO~J236.7$-$16 are likely due to blending from
            the earlier-type closeby companion
            \citep[M5;][]{2015ApJS..219...33G}. There is no
            significant difference between the mean $J-K$ colors for
            M6--M9 dwarfs with different gravities ({\sc vl-g}, {\sc
              int-g} or {\sc fld-g}). However for L0--L3 dwarfs, the
            {\sc vl-g} objects are redder than the {\sc fld-g} objects
            while the {\sc int-g} objects have colors redder than
            their {\sc fld-g} counterparts but not as extreme as their
            {\sc vl-g} counterparts. \label{fig:jkcolors}}
  \end{center}
\end{figure*}

However, none of these eight candidate AB~Dor Moving Group members
included in the BASS input catalog, are in the final
\citet{2015ApJ...798...73G} catalog of $\sim$300 high priority
candidate YMG members. In creating this high priority catalog,
\citet{2015ApJ...798...73G} used 2MASS+{\it WISE} proper motions and a
color-magnitude diagram to select candidates that are redder than the
field sequence. However, based on our parallactic (or photometric)
distances to calculate absolute magnitudes, three of our {\sc int-g}
and one of our {\sc fld-g} objects in the BASS input catalog
(PSO~J035.8$-$15, PSO~J039.6$-$21, PSO~J236.8$-$16, and PSO~J004.7+41)
are slightly blue compared to other known young objects and are
consistent with the field given their spectral type
(Figures~\ref{fig:wisecolors} and \ref{fig:jkcolors}) and thus could have
been rejected from the high priority catalog.

Although our remaining four objects present in the BASS input catalog,
PSO~J167.1+68, PSO~J236.8$-$16, PSO~J232.2+63, and PSO~J358.5+22, may
have a red enough $J-K$ color compared with the model used in the BASS
survey, our absolute magnitudes may differ from theirs because we
calculate absolute magnitudes from photometric distance (from spectral
type) whereas they use the statistical distance from their Bayesian
analysis. Thus, the BASS survey may have placed these objects in
different location on a color-magnitude diagram and rejected them as
likely young objects in their analysis. Although the 2MASS+{\it WISE}
proper motion from \citet{2015ApJ...798...73G} could also have removed
these objects from their high priority catalog, our PS1+2MASS proper
motions are consistent within the uncertainties
(Table~\ref{table:obsprop}).

Three of our objects present in the BASS input catalog, PSO~J167.1+68
(2MASS~J11083081+6830169), PSO~J232.2+63 (2MASS~J15291017+6312539) and
PSO~J236.8$-$16 (2MASS~J15470557$-$1626303A), were independently found
as candidate moving group members in \citet{2015ApJS..219...33G} as
part of their less-restricted initial search (see
Section~\ref{ssec:indivobj} for details). Determining RVs for
PSO~J236.8$-$16 and PSO~J232.2+63 would be needed for any further
membership assessment. We conclude that PSO~J167.1+68 is a young field
interloper using the literature RV \citep{2010ApJ...723..684B} and our
PS1 parallax (Section~\ref{ssec:indivobj}).

\subsection{Summary of Properties of Individual Objects \label{ssec:indivobj}}

We have determined properties (i.e. spectral type and mass) and
assessed the group membership for our 12 objects. We summarize the
results in the following paragraphs (see
Section~\ref{sec:specanalysis}, Section~\ref{sec:membership},
Section~\ref{sec:physprop}, Section~\ref{sec:BASS}, and
Table~\ref{table:abdmembers} for details).

\emph{PSO~J004.7+41} is a {\sc fld-g}, L0.1 dwarf
(40$^{+11}_{-13}$~$\Mjup$), with a spectrum that shows hints of
youth. However, the $UVWXYZ$ positions are inconsistent with AB~Dor
Moving Group membership and the BANYAN~II membership is low, thus we
consider it to be a probable young field interloper.

\emph{PSO~J035.8$-$15} is an {\sc int-g} M7.1 dwarf
(80$^{+40}_{-30}$~$\Mjup$). Although the $UVWXYZ$ positions for
PSO~J035$-$15 are inconsistent with AB~Dor Moving Group membership it
has a moderate BANYAN~II webtool membership probability (49\%). We
speculate that the membership probability may be optimistic due to the
large distance uncertainty. Thus we consider this object to be a
probable young field interloper.

\emph{PSO~J039.6$-$21} is an {\sc int-g} L2 dwarf (37$\pm$5~$\Mjup$)
with a high membership probability. Its $UVWXYZ$ positions are also
consistent with AB~Dor Moving Group membership (except U, which is
consistent within 2.5$\sigma$). However, it appears to be
spectroscopically peculiar. When comparing to known field dwarfs, its
spectrum matches very well with the blue L2~dwarf 2MASS~J1431+14
\citep{2009AJ....137..304S}, a candidate subdwarf. We also note that
the overall continuum is more blue than both the {\sc fld-g} and {\sc
  vl-g} standards, also indicating spectral peculiarity. Thus, as the
\citet{2013ApJ...772...79A} gravity indices were not intended for use
on subdwarfs, the {\sc int-g} classification may be invalid. However,
the kinematics and position appear to be consistent with possible
membership with the AB~Dor Moving Group. Therefore we conclude that
although membership in the group is possible, it could also be a field
L-type subdwarf. Thus a radial velocity is still needed to conclude
its group membership, or lack thereof.

\emph{PSO~J167.1+68} is an {\sc int-g} L1.8 dwarf
(52$^{+14}_{-16}$~$\Mjup$) with H$\alpha$ emission which was first
discovered in \citet{2000AJ....120.1085G} as an L1 dwarf (optical
spectral type). It was also independently identified as a low
probability candidate Carina member in
\citet{2015ApJS..219...33G}. After combining the space
positions+kinematics ($UVWXYZ$; Figure~\ref{fig:ymg1108}), the
literature RV \citep{2010ApJ...723..684B} and our PS1 parallax, we
suggest that it is actually unlikely to be a Carina member (BANYAN~II
webtool probability of zero). As we can completely determine the
$UVWXYZ$ positions for this object we conclude that this object is a
young field member.

\emph{PSO~J232.2+63} is an {\sc int-g}, M7.8 dwarf
(130$^{+20}_{-40}$~$\Mjup$) with a membership probability of 37\% and
$UVWXYZ$ consistent with AB~Dor Moving Group membership. It was also
independently discovered as a candidate member by
\citet{2015ApJS..219...33G}. Although they note that it has a high
young field contamination probability, they did not have a parallactic
distance. Thus, because we have a parallactic distance and the RV is
still unknown, we still consider it a possible member. We note that
the parallactic distance is significantly closer than the photometric
distance, even if we assume that young M~dwarfs are more luminous by
1.5~mag than their field counterparts. The high estimated mass,
suggesting a stellar rather than substellar object, is likely due to
the overluminosity of this object.

\emph{PSO~J236.7$-$16} is an {\sc int-g} M9.4 dwarf
(44$^{+12}_{-15}$~$\Mjup$) with a very low membership probability and
$UVWXYZ$ positions inconsistent with AB~Dor Moving Group membership. It
was also independently discovered in \citet{2015ApJS..219...33G} as a
low probability member with a high field contamination
probability. With the addition of our PS1 parallax to the membership
analysis, we conclude that PSO~J236.7$-$16 is a likely young field
interloper. We note that \citet{2015ApJS..219...33G} also propose that
the nearby object, 2MASS~J15470557$-$1626303B is a low-gravity stellar
companion with a spectral type of M5$\pm$2.

\emph{PSO~J292.9$-$06} is an {\sc int-g} M7.6 dwarf
($\approx$55--110~$\Mjup$). Although it has a low membership
probability, the $UVWXYZ$ positions are consistent with AB~Dor Moving
Group membership. Thus we consider it to be a possible member.

\emph{PSO~J306.0+16} is an {\sc int-g?} L2.3 dwarf
(34$^{+5}_{-6}$~$\Mjup$) with a membership probability of 36\% and
$UVWXYZ$ consistent with AB~Dor Moving Group membership. The spectrum
shows hints of youth but may still be an older field object. We
consider this object as a possible candidate member, requiring
parallax and RV to further assess membership.

\emph{PSO~J334.2+28} is an {\sc int-g} L3.5 dwarf
(31$^{+6}_{5}$~$\Mjup$) with a low BANYAN~II webtool membership
probability (0.63\%) but with heliocentric kinematics ($UVW$) and space
positions ($XYZ$) which are consistent with AB~Dor Moving Group
membership. One possible reason for the low BANYAN~II membership
probability is that the distance is 59$\pm$12~pc, further than the
1$\sigma$ distance range of the bona fide members which were used to
develop the BANYAN~II model. Thus we still consider this object to be
a possible candidate.

\emph{PSO~J351.3$-$11} is an {\sc int-g} M6.5 dwarf
(70$^{+50}_{-30}$~$\Mjup$) with a membership probability of 50\%. As
the $UVWXYZ$ positions are also consistent with AB~Dor Moving Group
membership, we consider it as a possible member.

\emph{PSO~J358.5+22} is an {\sc int-g?} L1.9 dwarf
(36$^{+5}_{-6}$~$\Mjup$) with a high membership probability (79\%),
$UVWXYZ$ positions consistent with group membership. Thus, we consider
it to be a strong candidate member.

\section{Discussion}\label{sec:discussion}

Low gravity (i.e. young) and dusty field L~dwarfs are known to have
redder {\it WISE} colors ($W1-W2$) and NIR colors ($J-K$) compared
with their old field counterparts of the same spectral type
\citep[e.g.][]{2012AJ....144...94G}. Our work sheds light on the
colors of these objects at intermediate ages ($\approx$125~Myr) and
intermediate gravities ({\sc int-g}). We compared the $W1-W2$ and
$J-K$ colors of our candidate AB~Dor Moving Group members with the
mean $W1-W2$ and mean $J-K$ colors both young ({\sc vl-g} and {\sc
  int-g}) and old ({\sc fld-g}) dwarfs (Figures~\ref{fig:wisecolors}
and \ref{fig:jkcolors}).

The average $W1-W2$ colors for {\sc vl-g} dwarfs tend to be redder
than both the {\sc fld-g} and the {\sc int-g} dwarfs from
\citet{2013ApJ...772...79A}, while the {\sc int-g} dwarfs have colors
more consistent with the {\sc fld-g} dwarfs. Three of our candidates
appear slightly redder than the mean $W1-W2$ color of old field
objects, given their spectral type. But the other nine candidates have
$W1-W2$ colors consistent with the field values. Interestingly, there
is also no obvious segregation in the $W1-W2$ colors between our
objects classified as {\sc int-g} and those classified as {\sc int-g?}
or {\sc fld-g}.

The average $J-K$ colors for M6--M9 dwarfs with gravity
classifications of {\sc vl-g}, {\sc int-g} and {\sc fld-g} are
consistent within the uncertainties. However, the $J-K$ colors appear
to have a stronger dependency with gravity for dwarfs with spectral
types of L0 and later. The average $J-K$ colors of {\sc vl-g} L0--L3
dwarfs are redder than for the {\sc fld-g} L0--L3 dwarfs, with the
{\sc int-g} L0--L3 dwarf average colors intermediate between {\sc
  vl-g} and {\sc fld-g}. Yet, there is significant scatter about these
trends. Two of our objects, PSO~J004.7+41 ({\sc fld-g}) and
PSO~J039.6$-$21 ({\sc int-g}) have slightly blue $J-K$ colors compared
with the field values. Another four objects, PSO~J167.1+68,
PSO~J306.0+16, PSO~J318.4+35 and PSO~J334.2+28, (all {\sc int-g}) have
$J-K$ colors slightly redder than the field sequence. The remaining
objects have $J-K$ colors consistent within the uncertainties in the
average field values.

One possible reason that our candidate AB~Dor moving group members do
not have such distinct IR colors from field ultracool dwarfs is that
they have older ($\approx$125~Myr) than the young brown dwarf members of
the TW Hydrae or $\beta$~Pic moving groups ($\sim$10--20~Myr), which
are classified as {\sc vl-g} in \citet{2013ApJ...772...79A} as
compared to the {\sc fld-g}, {\sc int-g?}, and {\sc int-g}
classifications of our objects. Our AB~Dor Moving Group candidates
have gravity classifications of {\sc int-g} or {\sc fld-g}, which
suggests that at $\approx$125~Myr, ultracool dwarfs tend to have higher
gravities than their younger (i.e. $\sim$10--50~Myr)
counterparts. With these intermediate gravities, our $\approx$125~Myr-old
dwarfs also have less extreme IR colors compared with the lower
gravity, younger, dwarfs.

Table~\ref{table:abdmembers} summarizes the full list of AB~Dor Moving
Group substellar members from our work and the literature. It includes
the properties (kinematics, spectral type, gravity classification,
parallax, d$_{phot}$ and RV) of the bona fide substellar members, the
previously published substellar candidate AB~Dor Moving Group members,
and our new candidates. We also include our new parallax for
2MASS~J0058$-$06 from PS1 data. We have calculated the BANYAN~II
webtool membership probabilities for the candidate members using our
proper motions when available, or otherwise those from
\citet{2015ApJ...798...73G}, in addition to our (parallactic or
photometric) distances. Our PS1 proper motions agree with the
\citet{2015ApJ...798...73G} proper motions within the uncertainties so
do not likely contribute significantly to any membership probability
differences. For objects also in
\citet{2015ApJ...798...73G,2015ApJS..219...33G}, we include their
tabulated BANYAN~II probabilities, which also use SEDs to determine
membership probabilities and thus will not agree with the webtool
probabilities. Our probabilities may also disagree due to differences
in the adopted distances.

\section{Conclusions}\label{sec:conclusions}

We have used PS1, 2MASS and {\it WISE} photometry coupled with proper
motions from PS1 + 2MASS to search for AB~Dor Moving Group substellar
candidates. Our search method combines color selection with SED
fitting and proper motion analysis, thereby significantly decreasing
the number of field ultracool dwarf interlopers compared with a solely
color-selected sample. We have obtained low and moderate-resolution
NIR spectroscopy of our candidates and confirmed the youth of nine
objects, six of which we conclude are likely AB~Dor Moving Group
members. We have also determined the \citet{2013ApJ...772...79A}
gravity classification of our objects and assessed their AB~Dor Moving
Group membership with the BANYAN~II web tool probabilities
\citep{2013ApJ...762...88M,2014ApJ...783..121G} and their $UVWXYZ$
positions.

We report the discovery of eight AB~Dor Moving Group candidate members
with spectral types of M6--L4 and masses down to
$\approx$30~$\Mjup$. Six of these have {\sc int-g} gravity
classifications (thought one of these {\sc int-g} objects is
spectroscopically peculiar and may be an ultracool subdwarf). The
remaining two objects have uncertain gravity classifications of {\sc
  int-g?} but have kinematics and spatial positions consistent with
group membership. In order to distinguish any of our candidate members
from being either an AB~Dor Moving Group member or unassociated with
any known YMG \citep[e.g.][]{2009ApJ...699..649S,2012ApJ...758...56S},
we still need to determine parallaxes (for three more objects) and
radial velocities, which are currently underway.

Finally, we find four objects that we conclude to be probable field
interlopers. Three objects are {\sc int-g} ultracool dwarfs and one
object is a {\sc fld-g} ultracool dwarf with a spectrum showing
possible hints of youth.

Although many known low gravity (i.e. young) substellar objects have
redder IR colors than their old field analogs, our intermediate
gravity ($\approx$125~Myr) brown dwarfs do not seem to follow this
trend. The $W1-W2$ color for the majority of our objects are
consistent with {\sc fld-g} dwarfs (i.e. older dwarfs) with the same
spectral type. Also, the average $J-K$ colors for mid--late M~dwarfs
with both {\sc vl-g} and {\sc int-g} gravities are consistent with
those of their {\sc fld-g} counterparts. However, the average $J-K$
colors for early L~dwarfs are redder for lower gravity objects
compared to those of the {\sc fld-g} objects, such that {\sc vl-g}
dwarfs are very red and {\sc int-g} dwarfs are slightly red. This
suggests that some young brown dwarfs with ages $\gtrsim$100~Myr may
have IR colors consistent with field objects, unlike the case for
younger brown dwarfs ($\lesssim$100~Myr).

Our updated census of the AB~Dor Moving Group brown dwarfs provides a
snapshot of brown dwarf evolution. In this era of ample large-area
surveys which are sensitive to red, faint substellar objects, refined
methods such as ours and those of others
\citep[e.g.][]{2014ApJ...783..121G} will be able to efficiently
uncover the substellar members of the known YMGs. An ensemble of young
brown dwarfs age benchmarks will allow us to characterize the brown
dwarf spectral evolution. Because young brown dwarfs can be analogs of
directly imaged planets \citep[e.g.][]{2013ApJ...777L..20L}, this
evolutionary sequence will also be valuable to compare with future
spectra of young exoplanets.

\acknowledgments 

KMA's research was supported by the National Science Foundation
Graduate Research Fellowship under Grant No. DGE-1329626. KMA's
research was also supported by NSF grant AST-1518339. Any opinion,
findings, and conclusions or recommendations expressed in this
material are those of the authors' and do not necessarily reflect the
views of the National Science Foundation. The Pan-STARRS1 Surveys
(PS1) have been made possible through contributions of the Institute
for Astronomy, the University of Hawaii, the Pan-STARRS Project
Office, the Max-Planck Society and its participating institutes, the
Max Planck Institute for Astronomy, Heidelberg and the Max Planck
Institute for Extraterrestrial Physics, Garching, The Johns Hopkins
University, Durham University, the University of Edinburgh, Queen's
University Belfast, the Harvard-Smithsonian Center for Astrophysics,
the Las Cumbres Observatory Global Telescope Network Incorporated, the
National Central University of Taiwan, the Space Telescope Science
Institute, the National Aeronautics and Space Administration under
Grant No. NNX08AR22G issued through the Planetary Science Division of
the NASA Science Mission Directorate, the National Science Foundation
under Grant No. AST-1238877, the University of Maryland, and Eotvos
Lorand University (ELTE). We also use data products from the Two
Micron All Sky Survey, which is a joint project of the University of
Massachusetts and the Infrared Processing and Analysis
Center/California Institute of Technology, funded by the National
Aeronautics and Space Administration and the National Science
Foundation. In addition, we use of data products from the {\it
  Wide-field Infrared Survey Explorer}, which is a joint project of
the University of California, Los Angeles, and the Jet Propulsion
Laboratory/California Institute of Technology, funded by the National
Aeronautics and Space Administration. This research has also benefited
from the M, L, and T dwarf compendium housed at DwarfArchives.org and
maintained by Chris Gelino, Davy Kirkpatrick, and Adam Burgasser. We
also use the IRTF Spectral Library housed at
\url{http://irtfweb.ifa.hawaii.edu/spex/IRTF\_Spectral\_Library/index.html}
and maintained by Michael Cushing. We have also made use of TOPCAT
\citep{2005ASPC..347...29T}, an interactive tool for manipulating and
merging tabular data. We would like to thank Katelyn Allers for
providing us with her custom version of SpeXtool for reducing
{\it Gemini}/GNIRS data. Finally, mahalo nui loa to the kama'\={a}ina of
Hawai'i for allowing us to operate telescopes on Mauna Kea. We wish to
acknowledge the very significant cultural role Mauna Kea has within
the indigenous Hawaiian community and that we are very fortunate to be
able to conduct observations.

Facilities: \facility{IRTF (SpeX)}, \facility{Gemini:North (GNIRS)}, \facility{Pan-STARRS1}, \facility{2MASS}, \facility{{\it WISE}}.


\begin{appendix}
\section{Comparison of PS1 and SDSS Proper Motions in Stripe 82 \label{appendix}}

In order to evaluate our proper motion quality, we compared the proper
motions of stars in the SDSS Stripe 82 with those measured by our
methods using PS1+2MASS data. Stripe 82 \citep{2008MNRAS.386..887B}
has a total sky area of $\approx$250~$\deg^{2}$ and spans 99$^{\circ}$
in R.A. ($\alpha$~=~20.7$^{h}$--3.3$^{h}$) and 2.52$^{\circ}$ in
decl. ($\delta$~=~$-$1.26$^{\circ}$ to +1.26$^{\circ}$). The 7~year
time baseline of SDSS yielded proper motions to an accuracy of
$\approx$5~mas~yr$^{-1}$. We matched the SDSS Stripe 82 stars with
their counterparts in the PS1 catalog using a 2\arcsec~ search
radius. We then chose the subset of matches which had at least 20
epochs of PS1+2MASS data and had proper motions below
1\arcsec~yr$^{-1}$. Also, since the PS1 and SDSS $z$-band filters are
similar \citep[$\Delta$z$\le$0.2~mag for stellar
  objects;][]{2012ApJ...750...99T}, we excluded stars with
significantly deviant PS1 and SDSS $z$ magnitudes
($|\Delta$z$|$~$\ge$~0.2~mag) in order to remove potential
mismatches. Finally, we also required all PS1 objects to have good
quality $z$-band photometry as defined in the same fashion as for our
search (Section~\ref{sec:data}).

\begin{figure*}
  \begin{center}
    \includegraphics[width=0.7\textwidth]{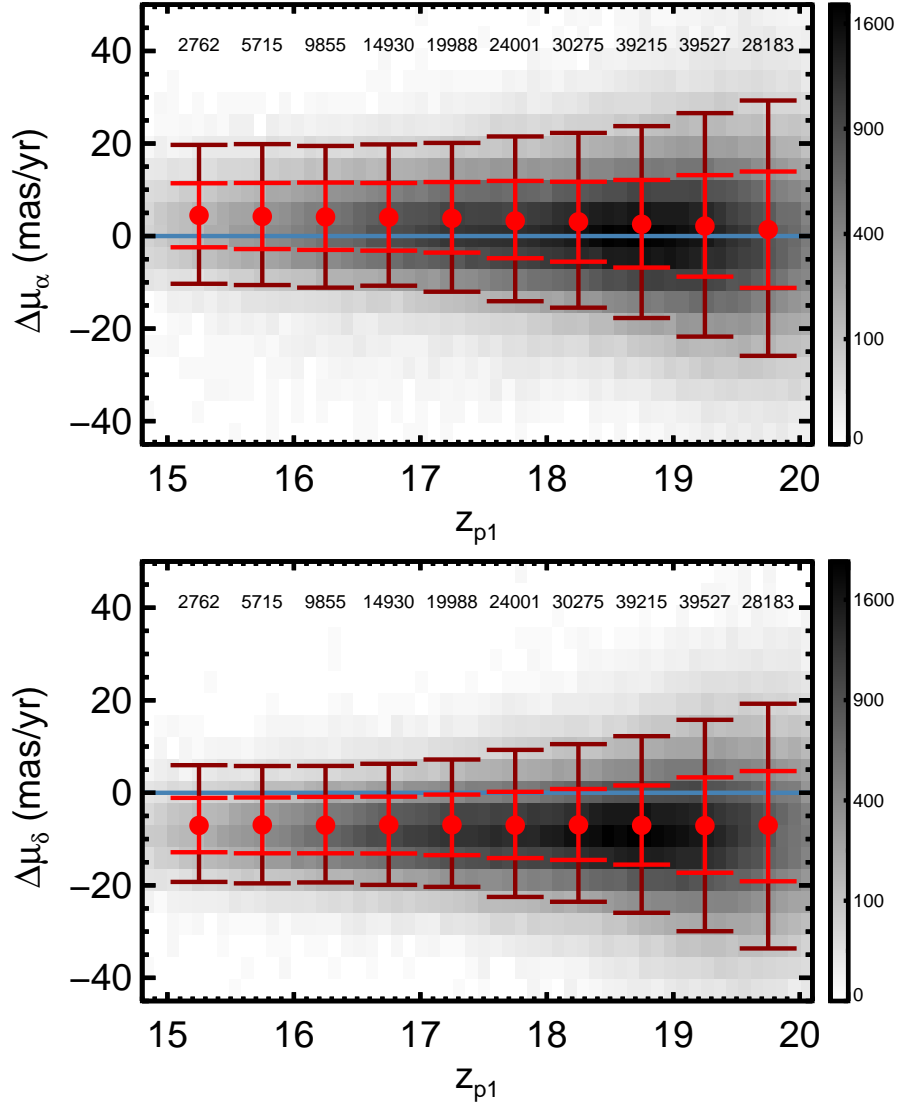}
    \caption{The deviation between the SDSS Stripe 82 and our
      PS1+2MASS proper motions in $\mu_{\alpha}$ and $\mu_{\delta}$ as
      a function of \zps~ for all objects with PS1+2MASS proper motion
      uncertainties less than 20~mas~yr$^{-1}$ and at least 20 epochs
      (\emph{gray-scale image}). In order to better show the
      distribution of our objects, we constructed a gray-scale image
      of the scatter plot of our data by binning by 0.2~mag in \zps~
      and 5~mas/yr in proper motion uncertainties. We scale the image
      by its squareroot for clarity. The median proper motion
      difference for each bin in \zps~ magnitude are the \emph{red
        dots} and the 68.5$^{th}$ and 95.4$^{th}$ confidence intervals
      are enclosed within the \emph{red} and \emph{brown} bars. The
      numbers at the top represent the number of objects in each
      magnitude bin. The proper motion difference between SDSS and
      PS1+2MASS does not change significantly as a function of
      \zps. The median deviation in $\mu_{\alpha}$ is
      3$\pm$11~mas~yr$^{-1}$ and in $\mu_{\delta}$ is
      $-$7$\pm$10~mas~yr$^{-1}$. \label{fig:dpm}}
  \end{center}
\end{figure*}

\begin{figure*}
  \begin{center}
    \includegraphics[width=0.8\textwidth]{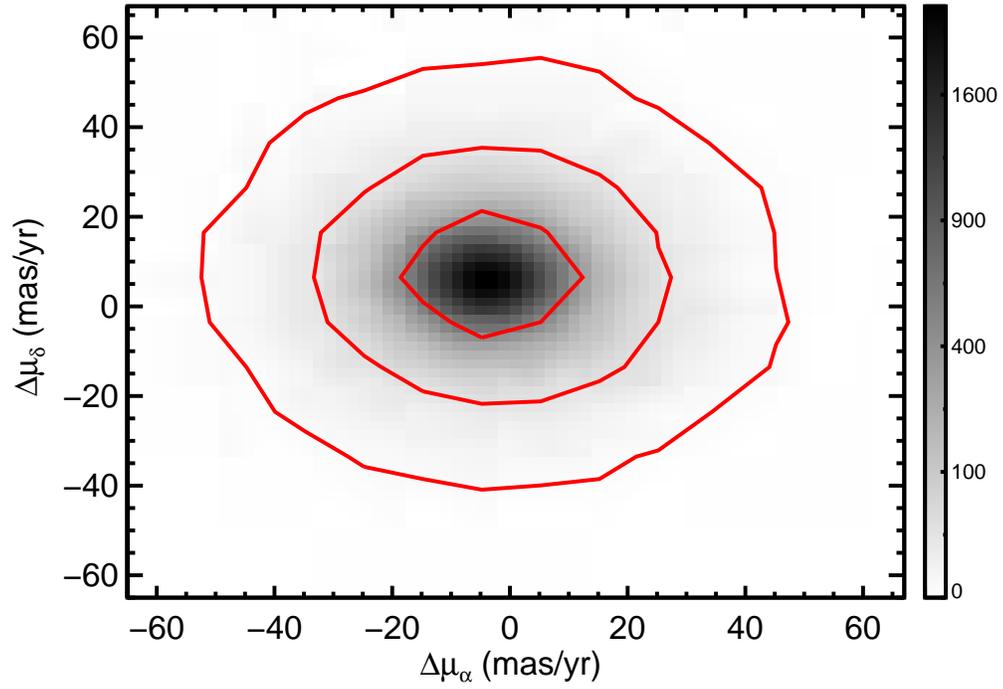}
    \caption{A gray-scale image of a scatter plot of the devation
      between the SDSS Stripe 82 and our PS1+2MASS proper motions for
      all objects with PS1+2MASS proper motion uncertainties less than
      20~mas~yr$^{-1}$ and at least 20 epochs. We constructed the
      image by binning the data into 2~mas/yr wide bins. Again, we
      scale the image by its square root for clarity. The x-axis is
      $\Delta\mu_{\alpha}$
      ($\mu_{\alpha,PS1+2MASS}-\mu_{\alpha,SDSS}$) and the y-axis is
      $\Delta\mu_{\delta}$
      ($\mu_{\delta,PS1+2MASS}-\mu_{\delta,SDSS}$). The red contour
      lines mark the 1$\sigma$, 2$\sigma$, and 3$\sigma$ confidence
      intervals. \label{fig:scatter}}
  \end{center}
\end{figure*}

The total matched sample consisted of 216,902 stars with PS1 proper
motion uncertainties below 20~mas~yr$^{-1}$ in both proper motion in
R.A. ($\mu_{\alpha}$) and in decl. ($\mu_{\delta}$). We separated our
analysis into three PS1 proper motion error bins:
$\le$~10~mas~yr$^{-1}$, 10--20~mas~yr$^{-1}$, and
$\le$~20~mas~yr$^{-1}$. The median and the 68.5\% range of the
differences between the PS1 and SDSS proper motions was
3$\pm$8~mas~yr$^{-1}$ in $\mu_{\alpha}$ and $-$7$\pm$7~mas~yr$^{-1}$
in $\mu_{\delta}$ for objects with PS1 proper motion uncertainties
below 10~mas~yr$^{-1}$. For objects with PS1 proper motion
uncertainties between 10 and 20~mas~yr$^{-1}$, the differences were
3$\pm$18~mas~yr$^{-1}$ in $\mu_{\alpha}$ and $-$8$\pm$21~mas~yr$^{-1}$
in $\mu_{\delta}$. For all the objects with PS1 proper motion
uncertainties below 20~mas~yr$^{-1}$, differences were
3$\pm$11~mas~yr$^{-1}$ in $\mu_{\alpha}$ and $-$7$\pm$10~mas~yr$^{-1}$
in $\mu_{\delta}$ (Figures~\ref{fig:dpm} and \ref{fig:scatter}). Thus, we
conclude that the PS1 and SDSS proper motions are consistent within
the uncertainties and that the PS1 proper motions are reliable.
\end{appendix}
\\
\\

\bibliographystyle{apj}

\onecolumngrid
\newpage
\clearpage

\newpage
\clearpage

\clearpage


\newpage
\onecolumngrid
\clearpage
\begin{landscape}
  \LongTables
  \tabletypesize{\footnotesize}
  \renewcommand{\tabcolsep}{2pt}
  \begin{deluxetable}{lccccccccccccc}
    \tablecolumns{13}
    \tablecaption{Observed Properties and Membership \label{table:obsprop}}
    \tablewidth{0pt}
    \tablehead{
      \colhead{Property} &
      \colhead{PSO~J004.7+41} &
      \colhead{PSO~J035.8$-$15} &
      \colhead{PSO~J039.6$-$21} &
      \colhead{PSO~J167.1+68} &
      \colhead{PSO~J232.2+63} & 
      \colhead{PSO~J236.8$-$16} &
      \colhead{PSO~J292.9$-$06} & 
      \colhead{PSO~J306.0+16} & 
      \colhead{PSO~J318.4+35} &
      \colhead{PSO~J334.2+28} & 
      \colhead{PSO~J351.3$-$11} &
      \colhead{PSO~J358.5+22}
    }
    \startdata
    $\mu_{\alpha}cos\delta$ (mas/yr) & 100.5$\pm$4.4    & 135.1$\pm$1.9    & 95.5$\pm$1.5     & $-$221.9$\pm$3.1 & $-$125.6$\pm$3.4 & $-$70.1$\pm$1.5  & 21.0$\pm$2.9         & 63.5$\pm$4.3     & 109.0$\pm$1.9   &  76.8$\pm$26.7  & 148.8$\pm$2.3     & 97.0$\pm$2.1 \\
    $\mu_{\delta}$ (mas/yr)          & $-$130.3$\pm$1.1 & $-$137.7$\pm$2.4 & $-$150.4$\pm$4.3 & $-$193.7$\pm$3.9 & 32.5$\pm$3.4     & $-$148.9$\pm$3.7 & $-$105.7$\pm$2.6     & $-$83.0$\pm$6.2  & $-$71.0$\pm$1.6 & $-$52.6$\pm$11.0 & $-$132.3$\pm$1.9  & $-$88.3$\pm$1.9 \\
    $\mu_{\alpha}cos\delta_{lit}$\tablenotemark{a} (mas/yr) & 94.2$\pm$5.6    & 147.8$\pm$8.8    & 102$\pm$7.3   & $-$238$\pm$5.8 & $-$119.7$\pm$3.6 & -64.1$\pm$5.9 & \nodata  & \nodata & \nodata & \nodata & 146.1$\pm$6.9 & 93.5$\pm$7.3 \\
    $\mu_{\delta,lit}$\tablenotemark{a} (mas/yr)     & $-$138$\pm$8.4 & $-$148.8$\pm$9 & $-$158.4$\pm$9.4 & $-$197.7$\pm$8.4 & 44.5$\pm$6.9 & $-$129.6$\pm$6.6 & \nodata  & \nodata  & \nodata & \nodata & $-$144.0$\pm$6.9 & $-$98.9$\pm$8.9 \\
    \ips (mag) & 18.09$\pm$0.01 & 16.98$\pm$0.04 & 19.12$\pm$0.01 & 17.733$\pm$0.005 & 15.40$\pm$0.01 & 18.42$\pm$0.015  & 17.48$\pm$0.01  & 20.258$\pm$0.015   & 18.91$\pm$0.01 & 21.56$\pm$0.08 & 16.816$\pm$0.006 & 19.74$\pm$0.02 \\
    \zps (mag) & 16.60$\pm$0.02 & 16.00$\pm$0.06   & 17.47$\pm$0.06 & 16.214$\pm$0.002 & 14.06$\pm$0.01 & 16.905$\pm$0.002 & 16.15$\pm$0.01 & 18.843$\pm$0.011  & 17.36$\pm$0.01 & 20.21$\pm$0.03 & 15.7$\pm$0.004 & 18.2$\pm$0.02 \\
    \yps (mag) & 15.80$\pm$0.01 & 15.24$\pm$0.06   & 16.55$\pm$0.01 & 15.211$\pm$0.002 & 13.30$\pm$0.01 & 15.931$\pm$0.003 & 15.43$\pm$0.01 & 17.872$\pm$0.0124  & 16.30$\pm$0.01 & 19.20$\pm$0.03 & 15.1$\pm$0.005 & 17.2$\pm$0.02 \\
    J\tablenotemark{b} (mag) & 14.10$\pm$0.03 & 14.0$\pm$0.02 & 14.8$\pm$0.04  & 13.12$\pm$0.02 & 11.64$\pm$0.02 & 13.86$\pm$0.03 & 13.86$\pm$0.03  & 15.58$\pm$0.06  & 14.3$\pm$0.03 & 16.84$\pm$0.17 & 13.6$\pm$0.02 & 15.4$\pm$0.05 \\
    H\tablenotemark{b} (mag) & 13.50$\pm$0.03 & 13.3$\pm$0.03 & 14.2$\pm$0.04 & 12.24$\pm$0.02 & 10.94$\pm$0.03  & 13.24$\pm$0.03 & 13.20$\pm$0.02  & 14.56$\pm$0.06  & 13.4$\pm$0.03 & 15.9$\pm$0.2 & 13.0$\pm$0.03 & 14.6$\pm$0.04 \\
    K\tablenotemark{b} (mag) & 13.10$\pm$0.03 & 13.0$\pm$0.02 & 13.8$\pm$0.05 & 11.58$\pm$0.02 & 10.55$\pm$0.02  & 12.74$\pm$0.03 & 12.79$\pm$0.02  & 13.98$\pm$0.05  & 12.8$\pm$0.03 & 15.08$\pm$0.16 & 12.7$\pm$0.03 & 14.0$\pm$0.05 \\
    W1 (mag)        & 12.76$\pm$0.02 & 12.63$\pm$0.02 & 13.43$\pm$0.02 & 11.12$\pm$0.02 & 10.30$\pm$0.02 & 12.43$\pm$0.02 & 12.52$\pm$0.02  & 13.32$\pm$0.03  & 12.23$\pm$0.02 & 14.34$\pm$0.03 & 12.45$\pm$0.03 & 13.63$\pm$0.03 \\
    W2 (mag)        & 12.45$\pm$0.03 & 12.41$\pm$0.02 & 13.11$\pm$0.03 & 10.76$\pm$0.02 & 10.06$\pm$0.02 & 12.14$\pm$0.02 & 12.29$\pm$0.03  & 12.98$\pm$0.03  & 11.85$\pm$0.02 & 13.89$\pm$0.04 & 12.24$\pm$0.03 & 13.37$\pm$0.04 \\
    d$_{phot}$\tablenotemark{c} (pc) & 27$\pm$4 & 44$\pm$14 (90$\pm$30) & 26$\pm$5 & 14$\pm$3 & 12$\pm$6 (22$\pm$11) & 26$\pm$6 (52$\pm$13) & 39$\pm$8 (80$\pm$20) & 40$\pm$8 & 29$\pm$9 & 59$\pm$12 & 21$\pm$7 (42$\pm$15) & 38$\pm$8 \\
    d$_{\pi}$ (pc) & \nodata & 54.6$^{+14}_{-9.2}$ & 37$^{+9}_{-6}$ & 22$^{+4}_{-3}$ & 28$^{+3}_{-3}$ & 27$^{+3}_{-3}$ & \nodata  & \nodata & 36$^{+5}_{-4}$ &  \nodata & 45$^{+19}_{-10}$ & 43$^{+8}_{-6}$ \\
    d$_{stat}$\tablenotemark{d} (pc) & 34.6$\pm$2.4 & 28.9$\pm$1.4 & 26.1$\pm$1.6 & 16.9$\pm$1.2 & 22.5$\pm$2.8 & 37.4$\pm$1.8 & 49.4$\pm$2.8 & 40.2$\pm$4.0 & 28.9$\pm$3.0 & 45.8$\pm$5.4 & 33.3$\pm$1.6 & 46.6$\pm$2.8 \\
    d$_{kin}$ (pc) & 35$^{+2}_{-2}$ & 29$^{+2}_{-2}$ & 28.$^{+0.7}_{-0.7}$ & 19$^{+2}_{-2}$ & 23.2$^{+0.6}_{-0.6}$ & 38.7$^{+0.9}_{-0.9}$ & 52.4$^{+3}_{-2}$ & 42$^{+3}_{-3}$ & 30$^{+3}_{-3}$ & 53$^{+4}_{-3}$ & 32.9$^{+2}_{-1.5}$ & 45$^{+2}_{-2}$ \\
    $\theta$ (degrees) & 12$^{+3}_{-3}$ & 2.0$^{+3}_{-3}$ & 14$^{+3}_{-3}$ & 4$^{+4}_{-4}$ & 1.0$^{+1.1}_{-0.7}$ & 1.2$^{+1.3}_{-0.9}$ & 3.7$^{+2}_{-2}$ & 4$^{+4}_{-3}$ & 3$^{+3}_{-2}$ & 6$^{+3}_{-3}$ & 6$^{+2}_{-2}$ & 3$^{+3}_{-2}$ \\
    $U_{best}$\tablenotemark{e} (km/s) & [$-$0.3$\pm$1.3]  & [$-$8.3$\pm$1.0]   & [0.2$\pm$1.6]     & $-$13.8$\pm$2.7\tablenotemark{f} & [$-$13.3$\pm$5.2] & [$-$7.8$\pm$1.3]  & [$-$7.8$\pm$2.3]  & [$-$6.5$\pm$5.0]  & [$-$7.4$\pm$2.4]   & [$-$9.3$\pm$6.7]  & [$-$14.2$\pm$4.4]  & [$-$7.3$\pm$1.6] \\
    $V_{best}$\tablenotemark{e} (km/s) & [$-$23.2$\pm$2.1] & [$-$50.9$\pm$10.0] & [$-$34.0$\pm$6.2] & $-$28.5$\pm$3.4\tablenotemark{f} & [$-$24.5$\pm$5.5] & [$-$19.4$\pm$2.3] & [$-$23.9$\pm$3.4] & [$-$23.9$\pm$8.2] & [$-$22.0$\pm$13.8] & [$-$24.7$\pm$4.2] & [$-$34.4$\pm$10.6] & [$-$25.8$\pm$3.1] \\
    $W_{best}$\tablenotemark{e} (km/s) & [$-$11.6$\pm$2.8] & [$-$10.6$\pm$1.9] & [$-$13.2$\pm$1.3] & $-$4.4$\pm$0.5\tablenotemark{f}  & [$-$2.6$\pm$5.0]  & [$-$11.3$\pm$1.0] & [$-$8.5$\pm$2.6]  & [$-$14.3$\pm$4.4] & [$-$18.5$\pm$3.5]  & [$-$16.0$\pm$7.0] & [$-$21.8$\pm$6.0]  & [$-$11.5$\pm$3.0] \\
    $\Delta(v)_{UVW}$\tablenotemark{e} (km/s) & [8.6$\pm$1.7] & [23.6$\pm$9.9] & [10.0$\pm$4.1]    & 11.2$\pm$1.6\tablenotemark{f}    & [12.4$\pm$5.1]    & [8.7$\pm$2.3]     & [6.6$\pm$3.0]     & [4.5$\pm$8.3]     & [7.4$\pm$9.2]      & [4.4$\pm$5.5]     & [12.1$\pm$8.3]     & [2.5$\pm$3.0] \\
    d$_{XYZ}$\tablenotemark{e} (pc) & [28.5$\pm$3.5]  & 37.9$\pm$9.2    & 20.1$\pm$6.1      & 35.4$\pm$2.0\tablenotemark{f}      & 50.9$\pm$8.5      & 41.2$\pm$2.1      & [46.9$\pm$6.2]    & [46.8$\pm$6.2]    & 42.4$\pm$4.2       & [60.0$\pm$11.2]   & 13.8$\pm$2.2       & 39.5$\pm$5.2 \\
    $P_{AB Dor}$ & 19.19\% & 49.06\% & 93.25\% & 0.23\% & 36.92 & 0.25\% & 6.77\% (0\%) & 42.76\% & 88.1\% & 0.63\% & 50.37\% & 79.04\% \\
    \enddata

    \tablenotetext{a}{The literature proper motion is taken from
      \citet{2015ApJ...798...73G}.}  

    \tablenotetext{b}{$J$, $H$, and $K$ are 2MASS magnitudes.}

    \tablenotetext{c}{Properties in parenthesis are calculated
      assuming the object is overluminous compared with the field
      sequence by 1.5~mag. This reflects the systematic uncertainty in
      the photometric distances for our late-M~dwarf candidates}

    \tablenotetext{d}{The AB~Dor Moving Group statistical distances
      are determined using the BANYAN~II webtool.}

    \tablenotetext{e}{The U$_{best}$, V$_{best}$ and W$_{best}$
      positions are determined by the RV (between $-$20 to +20 km/s)
      that minimizes the distance between the $UVW$ positions of our
      candidates and the mean $UVW$ positions of the known AB~Dor Moving
      Group Members with membership probabilities of at least 75\%
      from \citet{2008hsf2.book..757T}. The distances,
      $\Delta(v)_{UVW}$ and d$_{XYZ}$, are the distances between the
      $XYZ$ and the best fit UVW position of our candidates and the mean
      positions of the known group members. As for all objects, except
      PSO~J167.1+68\tablenotemark{f}, the values for $U_{best}$,
      $V_{best}$, $W_{best}$, $\Delta{v}_{UVW}$, and d$_{XYZ}$ (for
      objects without parallaxes) are not true measurements, we have
      enclosed them in brackets.}

    \tablenotetext{f}{As PSO~J167.1+68 has an RV from
      \citet{2010ApJ...723..684B}, we use this RV measurement to
      determine the $UVW$ positions and the $\Delta(v)_{UVW}$. Thus, the
      values are not enclosed in brackets.}

  \end{deluxetable}
\renewcommand{\tabcolsep}{6pt}
\clearpage
\end{landscape}
\newpage


\LongTables
\begin{deluxetable}{lccccccccc}
  \tablecaption{Spectroscopic Observations \label{table:obs}}
  \tabletypesize{\small}
  \tablewidth{0pt}
  \setlength{\tabcolsep}{0.07in}
  \tablehead{
    \colhead{Name} &
    \colhead{R.A. (PS1)} &
    \colhead{Decl. (PS1)} &
    \colhead{Date} &
    \colhead{$T_{exp}$} &
    \colhead{A0V Standard} &
    \colhead{Setup} &
    \colhead{Telescope} &
    \colhead{Mean S/N}
    \\ %
    \colhead{ } &
    \colhead{(J2000)} &
    \colhead{(J2000)} &
    \colhead{(UT)} &
    \colhead{(s) } &
    \colhead{ } &
    \colhead{ } &
    \colhead{ } &
    \colhead{($J$,$H$,$K$)}
  }
  \startdata
  PSO~J004.7+41 & 00:19:07.65 & +41:01:23.30 & 2014 Jan 17 & 1680 & HD~23594 & prism & IRTF & 57,61,68 \\
  PSO~J035.8$-$15 & 02:23:28.40 & $-$15:11:37.64 & 2013 Nov 23 & 960 & HD~20911 & SXD & IRTF & 48,56,53 \\
  PSO~J039.6$-$21 & 02:38:32.47 & $-$21:46:28.78 & 2013 Sep 22 & 600 & HD~31506 & prism & IRTF & 33,30,30 \\
  PSO~J167.1+68 & 11:08:30.25 & +68:30:14.39 & 2015 Jun 24 & 956 & HD~89239 & SXD & IRTF & 65,86,102 \\
  PSO~J232.2+63 & 15:29:09.96 & +63:12:54.50 & 2013 Aug 09 & 180 & HD~172728 & SXD & IRTF & 167,171,184 \\
  PSO~J236.8$-$16 & 15:47:05.52 & $-$16:26:32.20 & 2015 Jan 28 & 87.6 & HD~133569 & prism & IRTF & 59,53,40 \\
  PSO~J292.9$-$06 & 19:31:44.93 & $-$06:20:48.91 & 2015 Sep 25 & 59.8 & HD~190454 & prism & IRTF & 160,134,124 \\
  PSO~J306.0+16 & 20:24:03.05 & +16:47:49.09 & 2015 Jul 15 & 717.3 & HD~192538 & prism & IRTF & 75,75,80 \\
  PSO~J318.4+35 & 21:13:41.83 & +35:07:39.95 & 2013 Sep 22 & 180 & HD~209932 & prism & IRTF & 100,95,85 \\
  PSO~J334.2+28 & 22:17:02.98 & +28:56:37.98 & 2015 Jul 01 & 3000 & HIP~111538 & SXD & {\it Gemini} & 30,44,41 \\
  PSO~J351.3$-$11 & 23:25:22.42 & $-$11:21:05.18 & 2013 Dec 11 & 1440 & HD~3604 & SXD & IRTF & 35,32,29 \\
  PSO~J358.5+22 & 23:54:12.66 & +22:08:21.70 & 2013 Nov 23 & 720 & HD~1561 & prism & IRTF & 46,33,28
  \enddata
\end{deluxetable}

\LongTables
\begin{deluxetable}{lccccccc}
  \tablecaption{Spectral Type \label{table:spt}}
  \tablecolumns{8}
  \tablewidth{0pt}
  \tablehead{
    \colhead{Name} &
    \colhead{$J$ SpT\tablenotemark{a}} &
    \colhead{$K$ SpT\tablenotemark{a}} &
    \colhead{H$_{2}$O} &
    \colhead{H$_{2}$O--D} &
    \colhead{H$_{2}$O--1} &
    \colhead{H$_{2}$O--2} &
    \colhead{Final SpT}
  }
  \startdata
  PSO~J004.7+41 & M9$\pm$1 & M9$\pm$1 & L0.1$_{-0.4}^{+0.4}$ & \nodata \tablenotemark{b} & L1.3$_{-1.1}^{+1.1}$ & L0.4$_{-0.5}^{+0.5}$ & L0.1$\pm$1.0 \\
  PSO~J035.8$-$15 & M7$\pm$1 & M7$\pm$1 & M6.9$_{-0.4}^{+0.4}$ & \nodata \tablenotemark{b} & M8.6$_{-1.0}^{+1.1}$ & M7.2$_{-0.5}^{+0.5}$ & M7.1$\pm$1.0 \\
  PSO~J039.6$-$21 & L1$\pm$1 & L3$\pm$1 & L3.9$_{-1.0}^{+0.9}$ & L2.0$_{1.0}^{+0.9}$ & L4.8$_{-1.1}^{+1.0}$ & L0.9$_{-0.9}^{+1.0}$ & L2.6$\pm$1.0 \\
  PSO~J167.1+68 & L2$\pm$1 & L2$\pm$1 & L2.1$_{-0.3}^{+0.4}$ & L1.1$_{-0.7}^{+0.8}$ & L2.5$_{-1.0}^{+1.1}$ & L1.6$_{0.5}^{+0.4}$ & L1.8$\pm$1.0 \\
  PSO~J232.2+63 & M7$\pm$1 & M7$\pm$1 & M7.8$^{+0.3}_{-0.4}$ & \nodata & M8.6$^{+1.1}_{-1.1}$ & M7.8$^{+0.4}_{-0.5}$ & M7.8$\pm$1.0 \\
  PSO~J236.8$-$16 & M9$\pm$1 & L0$\pm$1 & L0.1$_{-0.6}^{+0.6}$ & L1.1$_{-0.9}^{+0.9}$ & M9.8$_{-1.0}^{+1.1}$ & M7.8$_{-0.6}^{+0.6}$ & M9.4$\pm$1.0 \\
  PSO~J292.9$-$06 & M8$\pm$1 & M8$\pm$1 & M7.2$^{+0.4}_{-0.4}$ & \nodata & M8.7$^{+1.0}_{-1.1}$ & M7.7$^{+0.5}_{-0.5}$ & M7.6$\pm$1.0 \\
  PSO~J306.0+16 & L3.5$\pm$1 & L2.5$\pm$1 & L2.0$^{+0.4}_{-0.4}$ & L3.0$^{+0.8}_{-0.8}$ & L2.1$^{+1.1}_{1.1}$ & L2.1$^{+0.6}_{0.5}$ & L2.3$\pm$1 \\
  PSO~J318.4+35 & L2$\pm$1 & L1$\pm$1 & L0.8$_{-0.4}^{+0.5}$ & L2.0$_{-0.7}^{+0.7}$ & L2.1$_{-1.0}^{+1.1}$ & L0.1$_{-0.5}^{+0.5}$ & L0.9$\pm$1.0 \\
  PSO~J334.2+28 & L4.5$\pm$1 & L3$\pm$1 & L3.8$_{-0.4}^{+0.5}$ & L2.6$^{+0.7}_{-0.7}$ & L3.4$^{+1.0}_{-1.1}$ & \nodata & L3.5$\pm$1 \\
  PSO~J351.3$-$11 & M6$\pm$1 & M7$\pm$1 & M6.8$_{-0.4}^{+0.4}$ & \nodata \tablenotemark{b} & M6.6$_{-1.1}^{+1.1}$ & M6.2$_{-0.5}^{+0.5}$ & M6.5$\pm$1.0 \\
  PSO~J358.5+22 & L1$\pm$1 & L3$\pm$1 & L1.2$_{-1.0}^{+0.9}$ & L1.8$_{-1.0}^{+1.0}$ & L2.6$_{-1.2}^{+1.3}$ & L2.1$_{-0.8}^{+0.6}$ & L1.9$\pm$1.0 \\
  \enddata
  \tablenotetext{a}{This spectral type is determined by visual classification.}
  \tablenotetext{b}{The H$_{2}$O--D index is undefined for this spectral type.}
\end{deluxetable}
\clearpage

\newpage
\clearpage
\begin{landscape}
  \begin{deluxetable}{lcccccccccccccc}
  \tablecaption{Gravity Indices and Classification \label{table:gravity}}
  \tabletypesize{\small}
  \setlength{\tabcolsep}{0.05in}
  \tablewidth{0pt}
  \tablehead{
    \colhead{Name} &
    \colhead{FeH$_{z}$} &
    \colhead{FeH$_{J}$} &
    \colhead{VO$_{z}$} &
    \colhead{KI$_{J}$} & 
    \colhead{H-cont} &
    \colhead{Index} &
    \colhead{NaI} &
    \colhead{KI} &
    \colhead{KI} &
    \colhead{KI} &
    \colhead{EW} & 
    \colhead{Final} & 
    \colhead{Overall} &
    \colhead{Overall} 
    \\
    \colhead{ } &
    \colhead{ } &
    \colhead{ } &
    \colhead{ } &
    \colhead{ } &
    \colhead{ } &
    \colhead{Scores\tablenotemark{a}} &
    \colhead{ } &
    \colhead{[1.169]} &
    \colhead{[1.177]} &
    \colhead{[1.253]} &
    \colhead{Index} &
    \colhead{Gravity} &
    \colhead{Gravity} &
    \colhead{Gravity} 
    \\
    \colhead{ } &
    \colhead{ } &
    \colhead{ } &
    \colhead{ } &
    \colhead{ } &
    \colhead{ } &
    \colhead{ } &
    \colhead{ } &
    \colhead{ } &
    \colhead{ } &
    \colhead{ } &
    \colhead{Scores\tablenotemark{a}} &
    \colhead{Scores\tablenotemark{a}} &
    \colhead{Value\tablenotemark{b}} &
    \colhead{Class} 
  }
  \startdata
PSO~J004.7+41 & 1.26$_{-0.01}^{+0.01}$ & 1.17$_{-0.01}^{+0.01}$ & 1.082$_{-0.006}^{+0.006}$ & 1.099$_{-0.004}^{+0.003}$ & 0.943$_{-0.003}^{+0.003}$ & 01001 (01001) & 14$_{-1}^{+1}$ & 7.2$_{-0.5}^{+0.4}$ & 9.7$_{-0.4}^{+0.4}$ & 7.6$_{-0.3}^{+0.4}$ & 0000 (0000) & 1001 (100?) & 0.5$_{-0.0}^{+0.0}$ & {\sc fld-g}\tablenotemark{c} \\
PSO~J035.8$-$15 & 1.079$_{-0.008}^{+0.008}$ & 1.06$_{-0.01}^{+0.01}$ & 1.020$_{-0.006}^{+0.005}$ & 1.040$_{-0.003}^{+0.003}$ & 0.991$_{-0.004}^{+0.003}$ & 11n12 (11n12) & 12$_{-1}^{+1}$ & 2.2$_{-0.6}^{+0.6}$ & 5.1$_{-0.5}^{+0.4}$ & 2.2$_{-0.4}^{+0.3}$ & 0101 (0101) & 1n12 (1n12) & 1.0$_{-0.0}^{+1.0}$ & {\sc int-g} \\
PSO~J039.6$-$21 & 1.17$_{-0.02}^{+0.02}$ & 1.05$_{-0.02}^{+0.02}$ & 1.06$_{-0.01}^{+0.01}$ & 1.068$_{-0.009}^{+0.008}$ & 0.91$_{-0.02}^{+0.01}$ & 1n021 (1n02?) & \nodata & \nodata & \nodata & \nodata & \nodata & 1021 (102?) & 1.0$_{-0.5}^{+0.0}$ & {\sc int-g} \\
PSO~J167.1+68 & 1.214$_{-0.009}^{+0.008}$ & 1.120$_{-0.008}^{+0.009}$ & 1.245$_{-0.005}^{+0.005}$ & 1.100$_{-0.003}^{+0.003}$ & 0.942$_{-0.002}^{+0.002}$ & 11211 (11211) & 8.9$_{-0.5}^{+0.5}$ & 5.6$_{-0.3}^{+0.3}$ & 7.0$_{-0.3}^{+0.3}$ & 4.3$_{-0.2}^{+0.2}$ & 1111 (1111) & 1211 (1211) & 1.0$_{-0.0}^{+0.0}$ & {\sc int-g} \\
PSO~J232.2+63 & 1.129$_{-0.002}^{+0.002}$ & 1.114$^{+0.003}_{-0.003}$ & 1.057$^{+0.0018}_{-0.0016}$ & 0.960$^{+0.001}_{-0.001}$ & 1.065$^{+0.001}_{-0.001}$ &  11n10 (11n10) & 9.8$_{-0.3}^{+0.3}$ & 2.8$^{+0.15}_{-0.15}$ & 5.0$^{+0.13}_{-0.13}$ & 2.6$^{+0.12}_{-0.12}$ & 1112 (1112) & 1n10 (1n10) & 1.0$_{-0.0}^{+0.0}$ & {\sc int-g} \\
PSO~J236.8$-$16 & 1.156$_{-0.016}^{+0.017}$ & \nodata & 1.078$_{-0.013}^{+0.013}$ & 1.077$_{-0.011}^{+0.010}$ & 0.943$_{-0.008}^{+0.008}$ & 1nn11 (1nn11) & \nodata & \nodata & \nodata & \nodata & \nodata & 1n11 (1n??) & 1.0$_{-0.0}^{+0.0}$ & {\sc int-g} \\
PSO~J292.9$-$06 & 1.104$^{+0.005}_{-0.004}$ & 1.013$_{-0.008}^{+0.008}$ & 1.042$^{+0.004}_{-0.004}$ & 0.984$^{+0.004}_{-0.004}$ & 1.068$^{+0.004}_{-0.004}$ & 1nn01 (1nn0?) & \nodata & \nodata & \nodata & \nodata & \nodata & 1n01 (1n0?) & 1.0$_{-1.0}^{+0.0}$ & {\sc int-g} \\
PSO~J306.0+16 & 1.324$^{+0.02}_{-0.019}$ & \nodata & 1.116$^{+0.011}_{-0.010}$ & 0.900$^{+0.006}_{-0.006}$ & 1.125$^{+0.009}_{-0.008}$ & 0n110 (0n110) & \nodata & \nodata & \nodata & \nodata & \nodata & 0101 (010?) & 0.5$_{-0.0}^{+0.5}$ & {\sc int-g?} \\
PSO~J318.4+35 & 1.16$_{-0.02}^{+0.02}$ & \nodata & 1.18$_{-0.01}^{+0.01}$ & 1.10$_{-0.006}^{+0.006}$ & 0.934$_{-0.005}^{+0.006}$ & 1n111 (1n111) & \nodata & \nodata & \nodata & \nodata & \nodata & 1111 (1111) & 1.0$_{-0.0}^{+0.0}$ & {\sc int-g} \\
PSO~J334.2+28 & \nodata & 1.21$^{+0.07}_{-0.06}$ & 1.10$^{+0.02}_{-0.02}$ & 0.926$^{+0.008}_{-0.007}$ &  1.106$_{-0.009}^{+0.009}$ & n1111 (n1111) & 10.9$\pm$0.9 & $-$1.5$\pm$4.5 & 7.5$\pm$2.5 & 6.6$\pm$0.7 & 1211 (021?) & 1111 (111?) & 1.0$_{-0.0}^{+0.0}$ & {\sc int-g} \\
PSO~J351.3$-$11 & 1.05$_{-0.01}^{+0.01}$ & 1.06$_{-0.01}^{+0.02}$ & 1.033$_{-0.008}^{+0.009}$ & 1.043$_{-0.004}^{+0.004}$ & 0.968$_{-0.006}^{+0.006}$ & 21n10 (21n10) & 8$_{-2}^{+2}$ & 0.6$_{-0.8}^{+0.8}$ & 3.0$_{-0.7}^{+0.7}$ & 2.0$_{-0.6}^{+0.6}$ & 1212 (1212)  & 2n20 (2n20) & 1.0$_{-0.0}^{+1.0}$ & {\sc int-g} \\
PSO~J358.5+22 & 1.36$_{-0.05}^{+0.07}$ & \nodata & 1.13$_{-0.02}^{+0.03}$ & 1.13$_{-0.014}^{+0.010}$ & 0.920$_{-0.015}^{+0.010}$ & 0n111 (0n1??) & \nodata & \nodata & \nodata & \nodata & \nodata & 0111 (01??) & 0.5$_{-0.0}^{+0.5}$ & {\sc int-g?} \\
  \enddata

  \tablenotetext{a}{Scores in parenthesis are the scores determined
    with the \citet{2013ApJ...772...79A} classification
    scheme. Objects with index values corresponding to {\sc int-g} but
    are within 1~$\sigma$ of the {\sc fld-g} value are classified with
    a score of ?.}

  \tablenotetext{b}{The overall gravity classification value and the
    68\% confidence limits calculated using our modified version of
    the \citet{2013ApJ...772...79A} classification scheme
    (Section~\ref{sec:results}).}

  \tablenotetext{c}{Although classified as {\sc fld-g}, the spectral
    indices show hints of {\sc int-g}.}

\end{deluxetable}
\clearpage
\end{landscape}

\newpage
\clearpage
\begin{landscape}
\LongTables
\renewcommand{\tabcolsep}{3pt}
  \begin{deluxetable}{lccccccccccc}
    \tablecaption{Substellar Members of the AB Dor Moving Group \label{table:abdmembers}}
    \tabletypesize{\small}
    \tablecolumns{5}
    \tablewidth{0pt}
    \tablehead{
      \colhead{Name} &
      \colhead{SpT} &
      \colhead{Gravity} &
      \colhead{$\mu_{\alpha}cos\delta, \mu_{\delta}$} &
      \colhead{$\pi$} & 
      \colhead{d$_{phot}$\tablenotemark{a}} &
      \colhead{RV} &
      \multicolumn{2}{c}{BANYAN II Web} &
      \multicolumn{2}{c}{BANYAN II} & 
      \colhead{Ref \tablenotemark{b}} 
      \\
      \colhead{ } &
      \colhead{ } &
      \colhead{(NIR\tablenotemark{c})} &
      \colhead{(mas yr$^{-1}$)} &
      \colhead{(mas)} &
      \colhead{(pc)} &
      \colhead{(km/s)} &
      \multicolumn{2}{c}{kinematics} &
      \multicolumn{2}{c}{kinematics+SED} &
      \colhead{ } 
      \\
      \colhead{ } &
      \colhead{ } &
      \colhead{ } &
      \colhead{ } &
      \colhead{ } &
      \colhead{ } &
      \colhead{ } &
      \colhead{P$_{AB Dor}$ \tablenotemark{a,d}} &
      \colhead{P$_{others}$ \tablenotemark{a,d}} &
      \colhead{P$_{AB Dor}$ \tablenotemark{a,e}} &
      \colhead{P$_{others}$ \tablenotemark{a,e}} &
      \colhead{ } 
    }
    \startdata
    \cutinhead{Bona Fide Members \tablenotemark{f}}
    CD-35~2722B & L3 & {\sc int-g} & $-$4.6$\pm$1.9, $-$59.8$\pm$1.6 & 47$\pm$3 & \nodata & 31.4$\pm$1.0 & \ldots & \ldots & \ldots & \ldots & 1,2 \\
    2MASS~14252798$-$3650229 & L4 & {\sc int-g} & $-$268$\pm$15, $-$47.3$\pm$19 & 111$\pm$12 & \nodata & 5.37 $\pm$0.256 & \ldots & \ldots & \ldots & \ldots & 3,4,5\\
    2MASS~03552337+1133437 & L3 & {\sc vl-g} & 218$\pm$5, $-$626$\pm$5, & 109.6$\pm$1.3 & \nodata & 11.92$\pm$0.22 & \ldots & \ldots & \ldots & \ldots & 6,7 \\ 
    WISEP~J00470106+680352 & L6 & {\sc int-g} & 381$\pm$12, $-$212$\pm$12 & 82$\pm$3 & \nodata & $-$20$\pm$1.4 & \ldots & \ldots & \ldots & \ldots & 8 \\
    SDSS~J11101001+0116131 & T5.5 & low & $-$217.1$\pm$0.7, $-$280.9$\pm$0.6 & 52.1$\pm$12 & \nodata & 7.5$\pm$3.8 & \ldots & \ldots & 97\% & \ldots & 9,4 \\
    \cutinhead{Strong Candidate Members \tablenotemark{g}}
    2MASS~J00012171+1535355 & L4 & {\sc int-g},[$\beta$] & 129$\pm$4, $-$177$\pm$7 & \nodata & 28$\pm$6 & \nodata & 98\% & \ldots & 97\% & \ldots & 5,10  \\
    2MASS~J00584253$-$0651239 & L1 & {\sc int-g},[$\beta$] & 138$\pm$4, $-$123$\pm$4 & 31.4$\pm$2.5 & 29$\pm$6 & \nodata & 95\% &  & 64\% & 34\% $\beta$Pic & 5,12 \\
    GU Psc b\tablenotemark{h} & T3.5 & low & 98$\pm$15, -92$\pm$15& \nodata & 47$\pm$9 & $-$1.6$\pm$0.4 & 56\% & 11\% $\beta$Pic & 88\% & 12\% $\beta$Pic & 11 \\
    PSO~J039.6352$-$21.7746 & L2.6 & {\sc int-g} & 95.5$\pm$1.5, $-$150.4$\pm$4.3 & 27.0$\pm$5.4 & 26$\pm$5 & \nodata & 93\% & \ldots & \ldots & \ldots & 12 \\
    2MASS~J03164512$-$2848521 & L1 & {\sc int-g},[$\beta$] & 98$\pm$4, $-$99$\pm$7 & \nodata & 33$\pm$7 & \nodata & 97\% & \ldots & 97\% & \ldots & 5 \\
    2MASS~J03264225$-$2102057\tablenotemark{i} & L5 & {\sc fld-g} [$\beta$/$\gamma$] & 93$\pm$6, $-$135$\pm$6 & \nodata & 26$\pm$5 & \nodata & 98\% & \ldots & 99\% & \ldots & 5,10 \\
    PSO~J318.4243+35.1277 & L0.9 & {\sc int-g} & 109.0$\pm$1.9, $-$71.5$\pm$1.6 & 27.9$\pm$3.6 & 29$\pm$9 & \nodata & 88\% & \ldots & \ldots & \ldots & 12 \\
    2MASS~J22064498$-$4217208 & L3 & {\sc vl-g},[$\gamma$] & 132.6$\pm$4.8, $-$187.7$\pm$9.3 & \nodata & 36$\pm$7 & \nodata & 92\% & \ldots & 99\% & \ldots & 5,10 \\
    2MASS~J22443167+2043433\tablenotemark{j} & L6.5 & low & 242.6$\pm$7.3, $-$219.6$\pm$7.1 & \nodata & 24$\pm$8 & \nodata & 95\% & 3\% $\beta$Pic & 99.6\% & 10 \\
    PSO~J358.5527+22.1393 & L1.9 & {\sc int-g?} & 97.0$\pm$2.1, $-$88.3$\pm$1.9 & 22.9$\pm$3.7 & 38$\pm$8 & \nodata & 79\% & \ldots & \ldots & \ldots & 12 \\
    \cutinhead{Possible Candidate Members \tablenotemark{k}}
    \\
    2MASS~J00192626+4614078 & M8 & {\sc int-g},[$\beta$] & 125$\pm$4, $-$75$\pm$4 & \nodata & 18$\pm$5 (36$\pm$10) & \nodata & $<$0.1\% (60\%) & 67\% (9\%) $\beta$Pic & 53\% & \ldots & 5 \\
    2MASS~J00425923+1142104 & M9 & {\sc int-g},[$\beta$] & 92.7$\pm$10.0, $-$75$\pm$9 & \nodata & 40$\pm$8 (81$\pm$16) & \nodata & 1\% ($<$0.1\%) & 75\% ($<$0.1\%) $\beta$Pic & 13\% & 6\% $\beta$Pic & 5 \\
    2MASS~J06322402$-$5010349 & L3 & [[$\beta$]] & $-$96.3$\pm$4.2, 9.1$\pm$6.7 & \nodata & 28$\pm$6 & \nodata & 0.2\% & \ldots & 30\% & \ldots & 5 \\
    2MASS~J06420559+4101599 & L/T & pec & $-$4.8$\pm$4.9, $-$370.5$\pm$8.5 & \nodata & \tablenotemark{l} & \nodata & \nodata & \nodata & 49\% & 10 \\
    2MASS~J08034469+0827000 & M6 & {\sc int-g},[$\beta$] & $-$72$\pm$3, $-$201$\pm$5 & \nodata & 20$\pm$4 (40$\pm$8) & \nodata & 17\% (7\%) & 69\% ($<$0.1\%) CAR & 91\% & \ldots & 5 \\
    PSO~J232.2915+63.2151 & M7.8 & {\sc int-g},[$\beta$] & $-$125.6$\pm$3.4, 32.5$\pm$3.4 & 35.5$\pm$4.2 & 11$\pm$3 & \nodata & 37\% & \ldots & 25\% & \ldots & 5,12 \\
    PSO~J292.9372$-$06.3469 & M7.6 & {\sc int-g} & 21$\pm$3, $-$106$\pm$3 & \nodata & 39$\pm$8 (80$\pm$20) & \nodata & 7\% (0\%) & 3\% $\beta$Pic & \nodata & \nodata & 12 \\
    PSO~J306.0126+16.7969 & L2.3 & {\sc int-g?} & 64$\pm$4, $-$83$\pm$6 & \nodata & 40$\pm$8 & \nodata & 43\% & \ldots & \ldots & \ldots & 12 \\
    2MASS~J20391314$-$1126531 & M7 & {\sc int-g},[$\beta$] & 54$\pm$3, $-$100$\pm$4 & \nodata & 45$\pm$9 (92$\pm$20) & \nodata & 17\% ($<$0.1\%) & 5\% ($<$0.1\%) $\beta$Pic & 2\% & \ldots & 5 \\
    2MASS~J21572060+8340575 & M9 & [[$\gamma$]] & 116.2$\pm$1.3, 46$\pm$9 & \nodata & 29$\pm$6 (58$\pm$18) & \nodata & 53\% ($<$0.1\%) & \ldots & 31\% & \ldots & 5 \\
    PSO~J334.2624+28.9438 & L3.5 & {\sc int-g} & 77$\pm$27, $-$53$\pm$11 & \nodata & 59$\pm$12 & \nodata & 0.6\% & \ldots & \ldots & \ldots & 12 \\
    PSO~J351.3434$-$11.3514 & M6.5 & {\sc int-g} & 148.8$\pm$2.3, $-$132.3$\pm$1.9 & 22.1$\pm$6.5 & 44$\pm$10 & \nodata & 50\% & \ldots & \ldots & \ldots & 12  \\
    2MASS~J23255604$-$0259508 & L1 & {\sc int-g},[$\gamma$] & 85$\pm$6, $-$106$\pm$4 & \nodata & 63$\pm$13 & \nodata & 3\% & \ldots & 73\% & \ldots & 5,10 \\
    2MASS~J23360735$-$3541489 & M9 & {\sc vl-g},[$\beta$] & 70$\pm$8, $-$80.7$\pm$10.0 & \nodata & 39$\pm$8 (79$\pm$16) & \nodata & 0.2\% ($<$0.1\%) & 97\% ($<$0.1\%) THA & 60\% & 39\% THA & 5 \\
    2MASS~J23433470$-$3646021 & L3--L6 & {\sc vl-g},[$\gamma$] & 97$\pm$6, $-$109.4$\pm$10.1 & \nodata & 40$\pm$16 & \nodata & 52\% & 30\% THA & 46\% & 38\% $\beta$Pic, 16\% THA & 5 \\
    2MASS~J23520507$-$1100435 & M8 & {\sc int-g},[$\beta$] & 100$\pm$4, $-$121$\pm$4 & \nodata & 20$\pm$7 (42$\pm$15) & \nodata & 0.1\% (82\%) & 95\% (5\%) $\beta$Pic & 91\% & \ldots & 5,10 \\
    2MASS~J23532556$-$1844402A & M6.5 & {\sc vl-g},[$\gamma$] & 90$\pm$3, $-$78$\pm$3 & \nodata & 15$\pm$3 (30$\pm$6) & \nodata & $<$0.1\% ($<$0.1\%) & $<$0.1\% (90\%) $\beta$Pic & 20\% & 46\% THA, 34\% $\beta$Pic & 5 \\
    \cutinhead{Probable Young Field Interlopers \tablenotemark{m}}
    \\
    PSO~J004.7818+41.0231 & L0.1 & {\sc fld-g} & 101$\pm$4, $-$133$\pm$1 & \nodata & 27$\pm$4 & \nodata & 19\% & 47\% $\beta$Pic & \ldots & \ldots & 12 \\
    PSO~J035.8683$-$15.1937\tablenotemark{n} & M7.1 & {\sc int-g} & 135.1$\pm$1.9, $-$137.7$\pm$2.4 & 18.3$\pm$3.7 & 44$\pm$14 & \nodata & 49\% & \ldots & \ldots & \ldots & 12 \\
    PSO~J167.1260+68.5039 & L1.8 & {\sc int-g},[$\gamma$] & $-$221.9$\pm$3.1, $-$193.7$\pm$3.9 & 46.1$\pm$6.5 & 14$\pm$3 & $-$9.8$\pm$0.1 & 0.2\% & \ldots & \ldots & 6\% CAR & 5,12 \\
    PSO~J236.7729$-$16.4422 & M9.4 & {\sc int-g},[$\beta$] & $-$70.1$\pm$1.5, $-$148.9$\pm$3.7 & 36.6$\pm$4.1 & 26$\pm$6 & \nodata & 0.3\% & \ldots & 11\% & \ldots & 5,12 \\
    \enddata

    \tablenotetext{a}{Properties in parenthesis are calculated
      assuming the object is overluminous compared with the field
      sequence by 1.5~mag. This reflects the systematic uncertainty in
      the photometric distances for our late-M~dwarf candidates
      (Section~\ref{sec:dphot}).}

    \tablenotetext{b}{Table references:
      1--\citet{2011ApJ...729..139W}, 2--\citet{2012ApJ...758...56S},
      3--\citet{2010ApJ...723..684B}, 4--\cite{2012ApJS..201...19D}, 
      5--\citet{2015ApJS..219...33G}, 6--\citet{2013AN....334...85L},
      7--\citet{2013AJ....145....2F}, 8--\citet{2015ApJ...799..203G},
      9--\citet{2015ApJ...808L..20G}, 10--\citet{2014ApJ...783..121G},
      11--\citet{2014ApJ...787....5N}, 12--this paper}

    \tablenotetext{c}{The NIR gravity classification system used is
      from \citet{2013ApJ...772...79A}. Gravity classifications in
      single brackets are visual classifications using NIR spectra by
      \citet{2015ApJS..219...33G}. For objects without NIR spectra, we
      note the visual gravity classifications from optical spectra by
      \citet{2015ApJS..219...33G} in double brackets.}

    \tablenotetext{d}{Probability based on BANYAN~II web tool which
      only uses the kinematic information to calculate membership
      probability. Membership for other significant groups is
      included.}

    \tablenotetext{e}{Probability from BANYAN~II using kinematic and
      photometric information from \citet{2015ApJS..219...33G} and
      \citet{2015ApJ...798...73G}. The membership probability is the
      probability of being a young moving group member (with the most
      likely group being the AB~Dor Moving Group), not necessarily of
      only the AB~Dor Moving Group.}

    \tablenotetext{f}{\emph{Bona Fide} members have parallaxes, radial
      velocities, and spectroscopically confirmed low gravity.}

    \tablenotetext{g}{Objects with membership probability as
      determined by the BANYAN~II webtool of at least 75\% and
      spectroscopically confirmed low gravity.}

    \tablenotetext{h}{GU Psc b is a widely accepted strong candidate
      to the AB~Dor Moving Group.}

    \tablenotetext{i}{2MASS~J03264225$-$2102057 has discrepant NIR
      gravity classifications of $\beta$/$\gamma$ (visual) and {\sc
        fld-g} (\citet{2013ApJ...772...79A} indices) from
      \citet{2015ApJS..219...33G}. Thus because there are visual signs
      of youth and the membership probability is high, we still
      consider it to be a strong candidate member.}

    \tablenotetext{j}{CFHT parallax for this object (Liu et al.,
      submitted) also suggests that it is a strong candidate member.}

    \tablenotetext{k}{Objects with membership probability as
      determined by the BANYAN~II webtool of 15--75\% or $UVWXYZ$
      positions consistent with AB~Dor Moving Group membership.}
    
    \tablenotetext{l}{We did not compute a photometric distance
      because the object is very peculiar, thus any photometric
      distance would likely be inaccurate.}

    \tablenotetext{m}{We consider objects with BANYAN~II
        membership probabilities $<$15\% and $UVWXYZ$ positions
        inconsistent with AB~Dor Moving Group membership to be likely
        field interlopers.}

    \tablenotetext{n}{We consider PSO~J035.8$-$15 as a likely field
      interloper because its $UVWXYZ$ positions are inconsistent with
      membership (Section~\ref{ssec:indivobj})}

  \end{deluxetable}
  \renewcommand{\tabcolsep}{6pt}
\clearpage
\end{landscape}
\newpage

\clearpage
\begin{deluxetable}{lccc}
  \tablecaption{Synthesized Photometry\tablenotemark{a} \label{table:synthphot}}
  \tablewidth{0pt}
  \tablehead{
    \colhead{Name} &
    \colhead{$J_{MKO}$} &
    \colhead{$H_{MKO}$} &
    \colhead{$K_{MKO}$}
\\
    \colhead{ } &
    \colhead{(mag)} &
    \colhead{(mag)} &
    \colhead{(mag)}
  }
  \startdata
  PSO~J004.7+41    & 14.150$\pm$0.009 & 13.539$\pm$0.007 & 13.056$\pm$0.007 \\
  PSO~J035.8$-$15    & 13.75$\pm$0.01   & 13.32$\pm$0.01   & 12.92$\pm$0.01   \\
  PSO~J039.6$-$21    & 14.73$\pm$0.04   & 14.24$\pm$0.05   & 13.68$\pm$0.05   \\
  PSO~J167.1+68 & 13.044$\pm$0.009 & 12.295$\pm$0.006 & 11.57$\pm$0.005 \\
  PSO J232+63 & 11.534$\pm$0.003 & 10.992$\pm$0.003 & 10.538$\pm$0.003 \\
  PSO~J236.8$-$16 & 13.81$\pm$0.02 & 13.237$\pm$0.014 & 12.708$\pm$0.017 \\
  PSO J292$-$06 & 13.837$\pm$0.011 & 13.296$\pm$0.009 & 12.830$\pm$0.009 \\
  PSO J306+16  & 15.57$\pm$0.03 & 14.657$\pm$0.020 & 13.823$\pm$0.016 \\
  PSO~J318.4+35    & 14.23$\pm$0.02   & 13.47$\pm$0.02   & 12.77$\pm$0.02   \\
  PSO J334+28 & 16.737$\pm$0.008 & 15.821$\pm$0.005 & 15.010$\pm$0.003 \\
  PSO~J351.3$-$11    & 13.59 $\pm$0.01  & 13.09$\pm$0.01   & 12.67$\pm$0.01   \\
  PSO~J358.5+22    & 15.26$\pm$0.04   & 14.62$\pm$0.04   & 14.02$\pm$0.04   \\
  \enddata
  \tablenotetext{a}{$MKO$ magnitudes are synthesized from our NIR
    spectra (Section~\ref{sec:results}).}
\end{deluxetable}
\newpage
\begin{deluxetable}{lccc}
  \tablecaption{Physical Properties\label{table:physprop} \tablenotemark{a}\tablenotemark{b}}
  \tabletypesize{\small}
  \tablewidth{0pt}
  \tablecolumns{4}
  \tablehead{
    \colhead{Name} &
    \colhead{M$_{bol}$} &
    \colhead{Mass} &
    \colhead{T$_{eff}$} 
    \\
    \colhead{ } &
    \colhead{(mag)} &
    \colhead{(M$_{Jup}$)} &
    \colhead{(K)}
  }
  \startdata
  \cutinhead{Our AB Dor Moving Group Candidates}
  PSO~J004.7+41 & [14.1~$\pm$~0.4] & [40$^{+11}_{-13}$] & [1950$^{+200}_{-200}$] \\
  PSO~J035.8$-$15 & 12.1~$\pm$~0.7 & 80$^{+40}_{-30}$  & 2700$^{+300}_{-400}$  \\
  PSO~J039.6$-$21 & 14.1~$\pm$~0.4 & 37$^{+5}_{-5}$ & 1950$^{+130}_{-150}$ \\
  PSO~J167.1+68 & 13.3~$\pm$~0.2 & 52$^{+14}_{-16}$ & 2300$^{+150}_{-190}$ \\
  PSO~J232.2+63 & 11.3~$\pm$~0.2 & 130$^{+20}_{-40}$ & 3050$^{+80}_{-150}$ \\
  PSO~J236.8$-$16 & 13.7~$\pm$~0.3 & 44$^{+12}_{-15}$ & 2090$^{+160}_{-180}$  \\
  PSO~J292.9$-$06 & [12.9~$\pm$~0.4 (11.4~$\pm$~0.5)] & [55$^{+11}_{-9}$ (110$^{+30}_{-30}$)] & [2450$^{+180}_{-200}$ (3000$^{+120}_{-170}$)] \\
  PSO~J306.0+16 & [14.3~$\pm$~0.5] & [34$^{+5}_{-6}$] & [1850$^{+170}_{-190}$] \\
  PSO~J318.4+35 & 13.4~$\pm$~0.2 & 45$^{+5}_{-5}$ & 2220$^{+110}_{-110}$ \\
  PSO~J334.2+28 & [14.6~$\pm$~0.5] & [31$^{+6}_{-5}$] & [1740$^{+180}_{-180}$] \\
  PSO~J351.3$-$11 & 12.3~$\pm$~1.4 & 70$^{+50}_{-30}$  & 2700$^{+400}_{-700}$ \\
  PSO~J358.5+22 & 14.1~$\pm$~0.4 & 36$^{+5}_{-6}$ & 1930$^{+140}_{-170}$ \\
  \cutinhead{Bona Fide Members}
  CD-35 2722B & 13.7~$\pm$~0.3 & 40$^{+4}_{-4}$ & 2090$^{+80}_{-80}$ \\
  WISEP~J00470106+680352 & 15.7~$\pm$~0.2 & 20$^{+3}_{-7}$ & 1340$^{+40}_{-40}$ \\
  2MASS~J03552337+1133437 & 15.4~$\pm$~0.2 & 23$^{+2}_{-5}$ & 1430$^{+40}_{-40}$ \\
  2MASS~J14252798$-$3650229 & 15.4~$\pm$~0.4 & 22$^{+3}_{-8}$ & 1420$^{+80}_{-100}$ \\
  SDSS~J11101001+0116131\tablenotemark{c} & 17.30$\pm$~0.05 & 10--12 & 940$\pm$20 \\
  \cutinhead{Previously Identified Candidates}
   2MASS~J00011217+1535355 & [14.9~$\pm$~0.5] & [28$^{+5}_{-4}$] & [1630$^{+160}_{-150}$] \\
   2MASS~J00192626+4614078 & [13.2~$\pm$~0.6 (11.7~$\pm$~0.6)] & [49$^{+12}_{-11}$ (90$^{+30}_{-30}$)] & [2300$^{+200}_{-300}$ (2890$^{+190}_{-200}$)] \\
   2MASS~J00425923+1142104 & [13.7~$\pm$~0.5 (12.2~$\pm$~0.3)] & [41$^{+7}_{-7}$ (72$^{+13}_{-11}$)] & [2090$^{+180}_{-200}$ (2720$^{+110}_{-150}$)] \\
   2MASS~J00584253$-$0651239 & 13.9~$\pm$~0.5 & 39$^{+7}_{-7}$ & 2040$^{+180}_{-200}$ \\
   Gu Psc b & [16.9~$\pm$~0.3] & [11.9$^{+2}_{-1.5}$] & [1060$^{+100}_{-110}$] \\
   2MASS~J03164512$-$2848521 & [13.9~$\pm$0.5] & [38$^{+6}_{-7}$] & [2010$^{+180}_{-200}$] \\
   2MASS~J03264225$-$2102057 & [14.9~$\pm$0.5] & [28$^{+4}_{-5}$] & [1610$^{+150}_{-190}$] \\
   2MASS~J06322402$-$5010349 & [14.5~$\pm$0.5] & [31$^{+5}_{-5}$] & [1750$^{+180}_{-180}$] \\
   2MASS~J06420559+4101599 \tablenotemark{d} & \nodata & [11--12] & \nodata \\
   2MASS~J08034469+0827000 & [12.4~$\pm$0.5 (10.8~$\pm$~0.5)] & [67$^{+15}_{-13}$ (150$^{+30}_{-30}$)] & [2650$^{+160}_{-200}$ (3160$^{+70}_{-110}$)] \\
   2MASS~J15291017+6312539 & [13.2~$\pm$0.6 (11.7~$\pm$~0.6)] & [50$^{+12}_{-11}$ (100$^{+30}_{-30}$)] & [2300$^{+200}_{-300}$ (2910$^{+180}_{-200}$)] \\
   2MASS~J20391314$-$1126531 & [12.4~$\pm$0.6 (10.9~$\pm$~0.5)] & [67$^{+17}_{-15}$ (150$^{+30}_{-40}$)] & [2660$^{+160}_{-300}$ (3150$^{+80}_{-190}$)] \\
   2MASS~J21572060+8340575 & [13.5~$\pm$0.5 (12.0~$\pm$~0.3)] & [44$^{+7}_{-7}$ (81$^{+15}_{-13}$)] & [2190$^{+150}_{-200}$ (2800$^{+100}_{-120}$)] \\
   2MASS~J22064498$-$4217208 & [14.4~$\pm$0.5] & [32$^{+6}_{-6}$] & [1810$^{+180}_{-190}$] \\
   2MASS~J22443167+2043433 & [22.0~$\pm$0.5] & [22$^{+4}_{-9}$] & [1380$^{+150}_{-150}$] \\
   2MASS~J23255604$-$0259508 & [13.7~$\pm$0.5] & [41$^{+8}_{-7}$] & [2110$^{+180}_{-200}$] \\
   2MASS~J23360735$-$3541489 & [13.6~$\pm$0.5 (12.1~$\pm$~0.3)] & [42$^{+8}_{-6}$ (78$^{+14}_{-12}$)] & [2140$^{+180}_{-180}$ (2770$^{+90}_{-120}$)] \\
   2MASS~J23433470$-$3646021 & [14.1~$\pm$0.9] & [37$^{+9}_{-8}$] & [1950$^{+280}_{-310}$] \\
   2MASS~J23520507$-$1100435 & [13.3~$\pm$0.6 (11.8~$\pm$~0.6)] & [47$^{+12}_{-12}$ (90$^{+30}_{-30}$)] & [2280$^{+240}_{-310}$ (2880$^{+190}_{-200}$)] \\
   2MASS~J23532556$-$1844402 & [12.4~$\pm$0.5 (10.8~$\pm$~0.5)] & [68$^{+16}_{-14}$ (150$^{+30}_{-40}$)] & [2660$^{+160}_{-200}$ (3170$^{+60}_{-160}$)] \\
  \enddata

  \tablenotetext{a}{Properties in parenthesis are calculated assuming
    the object is overluminous compared with the field sequence by
    1.5~mag. This reflects the systematic uncertainty in the
    photometric distances for our late-M~dwarf candidates.}

  \tablenotetext{b}{Properties in brackets are calculated using
    photometric distances for objects where no parallax was available.}

  \tablenotetext{c}{Properties from \citet{2015ApJ...808L..20G}}

  \tablenotetext{d}{Because of the spectral peculiarities, this object
    is classified only as an L/T dwarf and does not have a precise
    spectral type, thus we do not estimate the M$_{bol}$, mass, or
    T$_{eff}$. The mass range estimate is taken from
    \citet{2014ApJ...783..121G}.}

\end{deluxetable}
\newpage
\clearpage
\twocolumngrid

\end{document}